\documentclass[3p,twocolumn,times]{elsarticle}

\usepackage{lineno,hyperref}
\modulolinenumbers[5]

\usepackage{bm}
\usepackage{graphicx}
\usepackage{comment}
\usepackage{amsmath,amssymb,amsthm} 
\usepackage{verbatim}
\usepackage{float}
\usepackage{color}
\journal{Journal of Non-Crystalline Solids}

\begin{document}

\begin{frontmatter}

\title{Roles of liquid structural ordering in glass transition, crystallization, \\ and water's anomalies}

\author{Hajime Tanaka}
\cortext[mycorrespondingauthor]{Corresponding author}
\ead{tanaka@iis.u-tokyo.ac.jp}
\address{Research Center for Advanced Science and Technology, University of Tokyo, 4-6-1 Komaba, Meguro-ku, Tokyo 153-8904, Japan}
\address{Department of Fundamental Engineering, Institute of Industrial Science, University of Tokyo, 4-6-1 Komaba, Meguro-ku, Tokyo 153-8505, Japan}

\begin{abstract}
The liquid state is one of the fundamental and essential states of matter, but its physical understanding is far behind the other states, such as the gas and solid states, due to the difficulties associated with the high density causing many-body correlations and the lack of long-range order. Significant open problems in liquid science include glass transition, crystallization, and water's anomalies. Austen Angell has contributed tremendously to these problems and proposed many new concepts of fundamental importance. In this article, we review how these concepts have influenced our work on liquid physics, focusing on the roles of liquid structural ordering in glass transition, crystallization, and water's anomalies. 
\end{abstract}


\end{frontmatter}


\section{Introduction}

Liquids are one of the fundamental states of matter around us, and a physical understanding of the liquid state is of vital importance. The physics of an equilibrium state of simple liquids like hard spheres has been rather well established~\cite{hansen1990theory}. However, there are many open problems concerning liquids not that simple. They include (1) metastable liquids below the meting points (i.e., supercooled liquids)~\cite{DebenedettiB} and its dynamic arrest (i.e, glass transition)~\cite{Angel863_1988,AngellR,ediger1996supercooled,Angel6463_2000,Debenedetti2001,berthier2011theoretical,tanaka2012bond,royall2015role}, (2) liquid-to-crystal transition~\cite{kelton2010nucleation,sosso2016crystal}, (3) physical properties of liquids with complex directional interactions such as water~\cite{eisenberg2005structure,DebenedettiB,Angell1983,AngellCR,Debenedetti2003,Angell_glass,angell2008insights,gallo2016water}, and (4) liquid and amorphous polymorphs~\cite{poole1997polymorphic,mcmillan2004polyamorphic,mcmillan2007polyamorphism,tanaka2020liquid}. 

Austen Angell has carried out a number of pioneering studies in these topics and has been at the forefront of research moving forward over the years. I mainly studied critical phenomena and soft matter physics until 1995, and my primary interests were rather far from these topics. When I first saw the so-called Angell plot, which classifies the types of slowing down of liquid dynamics towards the glass-transition point $T_{\rm g}$ in terms of liquid fragility. Fragile liquids show super-Arrhenius-type dynamic slowing down upon cooling, whereas strong liquids exhibit Arrhenius-type one. I was fascinated by this classification and became very interested in what physical factors control the fragility of liquids. This is how I got started with the research of liquid science. I first met Austen and his family with excitement in the International Discussion Meeting on Relaxations in Complex Systems held at Vigo, Spain, in 1997. Since then, I have been very fortunate to keep communicating with him closely over the years. 

This article will review our research on glass transition, crystallization, and water's anomalies in connection to many new concepts and ideas brought by Angell. Since I recently wrote a lengthy review article on liquid-liquid transition and polyamorphism~\cite{tanaka2020liquid}, I will not discuss this topic much in this article.  
In Sec. \ref{sec:glass}, we discuss our study on glass transition initiated from my interest in the Angell plot. In Sec.~\ref{sec:crystal}, we discuss crystal nucleation from a supercooled liquid and glass-forming ability. In Sec.~\ref{sec:water}, we discuss the physical origin of water's anomalies. In Sec.~\ref{sec:summary}, we summarize.

\section{Glass transition} \label{sec:glass}

\paragraph{Fragility and glass-forming ability: Two-order-parameter model}
Motivated by Angell's classification of glass-forming liquids, I wonder what controls the liquid fragility. I came up with the idea that the frustration on crystallization may be a critical factor. 

\begin{figure}[t]
\begin{center}
\includegraphics[width=8.cm]{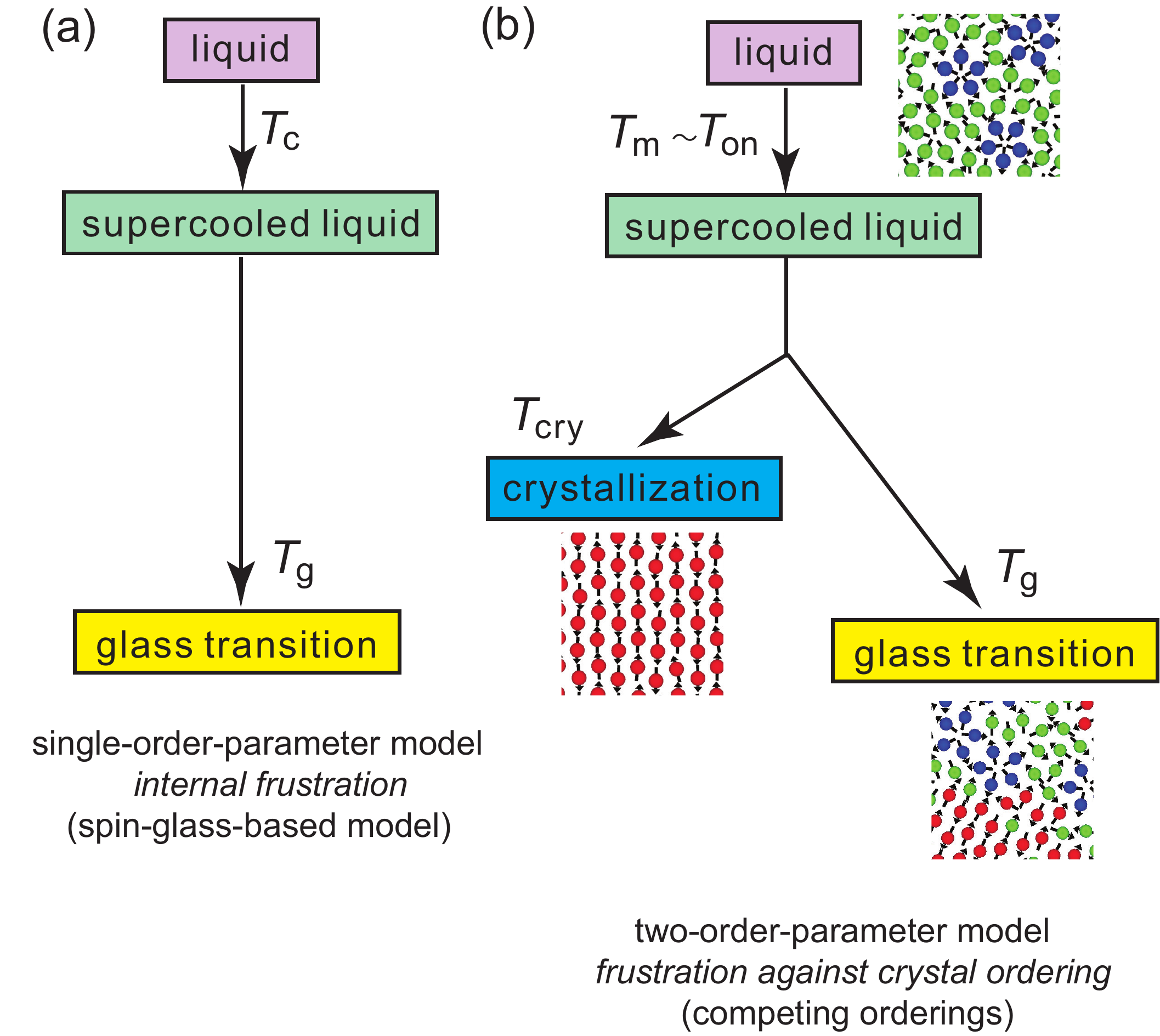}
\end{center}
\caption{(a) Single-order-parameter model of glass transition based on spin-glass-like scenarios, where frustration is internally embedded in the order parameter itself. In this scenario, the key temperatures are the mode-coupling critical temperature  $T_{\rm c}$, the glass transition temperature $T_{\rm g}$, and the ideal glass transition temperature $T_0$. 
(b) Two-order-parameter model of glass transition based on frustration against crystallization (or compering orderings), where frustration is between a part of the potential compatible with the crystal and the other part incompatible with it. In this scenario, the key temperatures are the onset temperature of the cooperativity $T_{\rm on}$, which is located near the melting point $T_{\rm m}$, $T_{\rm g}$, and $T_0$. 
}
\label{fig:models}
\end{figure}

The role of local icosahedral ordering in frustrating crystallization was first pointed out by Frank~\cite{frank1952supercooling}.
Subsequently, the concept of geometrical frustration associated with icosahedral ordering has been popularised in the community of structural glasses because of its analogy to frustration in spin glasses and has become a mainstream theoretical concept. For example, Steinhardt et al. proposed a theoretical model based on the icosahedral ordering, which suffers from geometrical frustration of the 6-fold bond orientational order parameter $Q_6$~\cite{steinhardt1983bond}. Similarly, Tarjus and Kivelson and their coworkers~\cite{tarjus2005frustration,ferrer1998supercooled} proposed a frustration-limited domain theory in which the critical point of structural ordering is avoided by internal frustration in the order parameter. In both approaches, the frustration originates from the ``internal'' geometrical frustration of the order parameter itself. 
These approaches can be regarded as a single-order-parameter description of glass transition, similarly to the spin-glass theory described by a spin variable alone (see Fig.~\ref{fig:models}(a)). Spins on a triangular lattice, for example, suffer from intrinsic geometrical frustration. 
The link between a structural glass and a spin glass model, particularly Potts glass~\cite{kirkpatrick1987dynamics}, was also the basis of the random-first-order transition (RFOT) theory~\cite{kirkpatrick1989scaling,parisi2010mean,berthier2011theoretical,kirkpatrick2015colloquium}. Its connection to the mode-coupling theory~\cite{gotze2008complex} was also be shown~\cite{kirkpatrick1989random}. We may say that all these models are based on frustration in a spin-glass-like sense (see Fig.~\ref{fig:models}(a)). 

\begin{figure*}
\begin{center}
\includegraphics[width=14.cm]{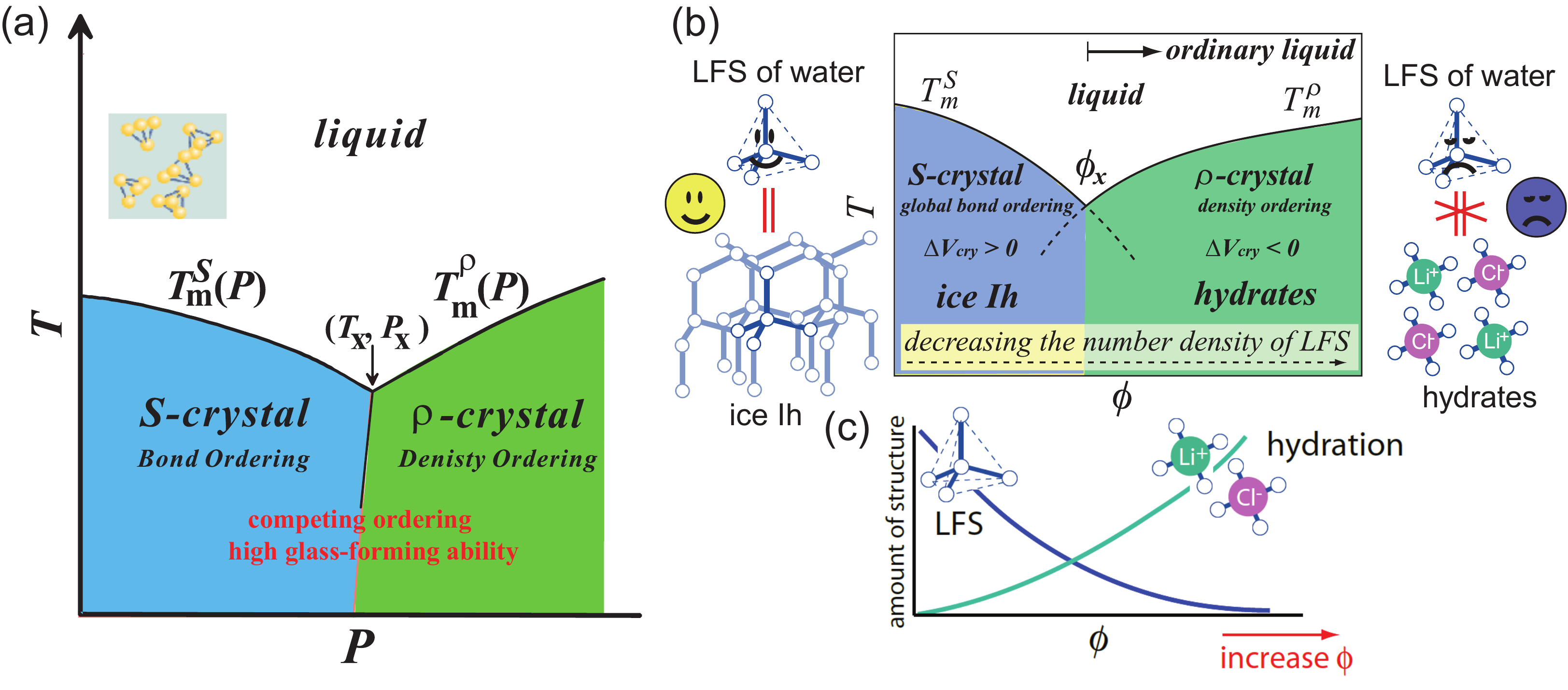}
\end{center}
\caption{(a) V-shaped phase diagram of water-type liquids~\cite{tanaka2002simple}. Stable crystals are $S$-type crystals with tetrahedral symmetry that expand upon crystallization at low pressure, whereas they are $\rho$-type crystals that shrink upon crystallization at high pressure. We proposed that water-type liquids include water, five atomic elements (Si, Ge,
Sb, Bi, and Ga), and some group III-V (e.g., InSb, GaAS, and GaP) and II-VI compounds (e.g., HgTe, CdTe, and CdSe)~\cite{tanaka2002simple}.
The melting point has a minimum, $T_{\rm X}$, at $P_{\rm X}$, around which the $S$- and $\rho$-type orderings compete and the degree of frustration on crystallization is maximized. 
(b) Schematic phase diagram of water/LiCl mixtures~\cite{Angel1058_1970,kobayashi2011possible,kobayashi2011relationship}. Locally favored structures have symmetry consistent with ice crystals ($S$-crystal) but not with hydrate crystals ($\rho$-crystal). If we replace the salt concentration $\phi$ with pressure, this figure represents the phase diagram of water or other water-type liquids (see (a)). Although both adding salt and applying pressure decrease locally favored tetrahedral structures, their roles should differ since the former has local effects whereas the latter has global effects. (c) Schematic representation of the $\phi$ dependence of the fraction of locally favored structures and hydration structures. We note that hydrated structures and locally favored tetrahedral structures are the sources of frustration against tetrahedral crystals ($S$-crystal) and hydrate crystals ($\rho$-crystal), respectively. 
Panels (b) and (c), which are courtesy of Mika Kobayashi, are reproduced from Fig.~30 of Ref.~\cite{tanaka2013importance}.
}
\label{fig:competing}
\end{figure*}

\begin{figure}[t]
\begin{center}
\includegraphics[width=8.cm]{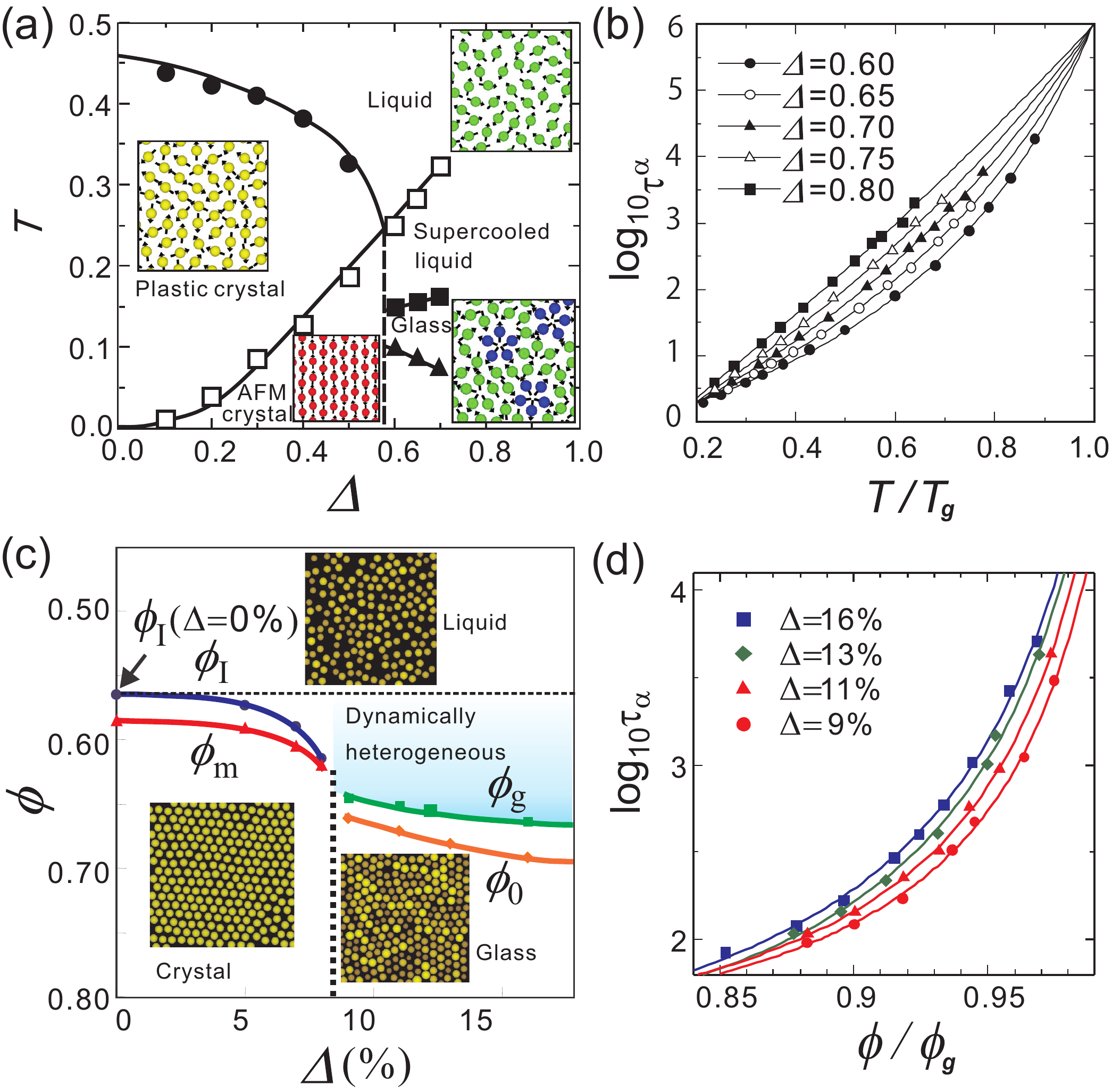}
\end{center}
\caption{(a) The phase diagram of 2D spin liquid as a function of $T$ and the strength of energetic frustration $\Delta$~\cite{shintani2006frustration}. 
An increase in $\Delta$ changes the stable equilibrium crystal from a hexagonal plastic crystal, where spins can rotate on the lattice, to an antiferromagnetic crystal. 
For $\Delta \geq 0.6$, the system forms glass with a specific cooling rate. 
(b) The Angell plot for 2D spin liquids with various $\Delta$. A liquid with larger $\Delta$ is less fragile, i.e., stronger, and has a better glass-forming ability~\cite{shintani2006frustration}.  
(c) The phase diagram of 2D polydisperse hard-sphere-like liquid as a function of $T$ and the variance of size polydispersity $\Delta$ (\%) that is a measure of the frustration strength against crystallization~\cite{kawasaki2007correlation}. 
The stable equilibrium crystal can be formed up to $\Delta=9$\%, but above it, the system tends to form glass without crystallization.  
(d) The Angell plot for liquids with various polydispersity $\Delta$. A liquid with larger $\Delta$ is less fragile, i.e., stronger, and has a better glass-forming ability~\cite{kawasaki2007correlation}, as in the 2D spin liquids.  
Panels (a) and (b) are reproduced from Figs. 2 and 6b of Ref.~\cite{shintani2006frustration}. Panel (d) is reproduced from Fig. 1(b) of Ref.~\cite{kawasaki2007correlation}.
}
\label{fig:spin}
\end{figure}

In the above models based on a single order parameter, crystallization is out of consideration, i.e., the kinetic avoidance of crystallization is implicitly assumed. Thus, particular forms of free energy were introduced to describe glass transition alone. 
Contrary to this, I thought that the same free energy should describe crystallization and vitrification and that for modeling glass transition, it is necessary to consider squarely that a liquid always crystallizes upon slow enough cooling. There should be a link between crystallization and glass transition as long as crystallization does not involve other phenomena such as phase separation. Thus, unlike models based on the internal frustration in a single order parameter, I proposed a ``two-order parameter model''~\cite{tanaka1999two,tanaka1999two1,tanaka1999two2,tanaka2005two1,tanaka2005two2, tanaka2005two3}, where frustration against crystallization caused by another type of ordering plays a critical role in avoiding crystallization and causing vitrification (see Fig.~\ref{fig:models}(b)). In other words, we focus on the frustration embedded in the free energy of a system~\cite{tanaka2012bond}: A part of the free energy tends to form a crystal, whereas the other part tends to prevent it. Our model based on ``frustration against crystallization'' or ``competing orderings'' is essentially different in physics from the other frustration models mentioned above, although it might not have been recognized so clearly in the glass community. 
For example, our model can discuss the ability to form glass depending on the strength of the frustration, whereas the models based on internal frustration of a single amorphous order parameter cannot because crystallization is out of consideration.

Based on this model, I speculated that stronger frustration against crystallization makes a glass-former stronger and increases the glass-forming ability~\cite{tanaka2005relationship}. 
This view is consistent with the link between the kinetic and thermodynamic fragilities pointed out by Martinez and Angell~\cite{martinez2001thermodynamic}.
For example, it was predicted that metallic glass-formers with a stronger tendency of icosahedral ordering should be less fragile (i.e., stronger) and better glass-formers, as long as icosahedral ordering is not too strong to form quasi-crystals~\cite{tanaka2003roles,tanaka2005relationship}. 
I also focused on the fact that tetrahedral liquids such as water, silicon, germanium, and bismuth, generally have a V-shaped temperature ($T$)-pressure ($P$) phase diagram~\cite{tanaka2002simple} (see Fig.~\ref{fig:competing}(a)). 
A liquid expands upon crystallization into a low-pressure crystal, which we call $S$-type crystal. In contrast, above the triple point $P_{\rm m}^{\rm min}$, it shrinks upon crystallization into a high-pressure crystal, which we call $\rho$-type crystal. Our model predicts that the glass-forming ability and fragility should be maximized near the triple point since the competition (i.e., frustration) between the two types of orderings ($S$-ordering and $\rho$-ordering) is maximized there.    

Later we performed two-dimensional (2D) numerical simulation to confirm our claim by introducing a spin to each particle with an angle-dependent three-body potential (strength $\Delta$) favoring pentagonal symmetry in addition to an isotropic Lennard-Jones potential~\cite{shintani2006frustration,shintani2008universal} (see Fig.~\ref{fig:spin}(a)). Pentagonal local structures that cannot tile the space induce strong frustration against crystallization. 
We indeed found that a liquid with a stronger tendency (larger $\Delta$) of local pentagonal ordering (i.e., stronger frustration against crystallization) becomes stronger and has a better glass-forming ability (see Fig.~\ref{fig:spin}(b)). 
This liquid forms an antiferromagnetic crystal upon crystallization at high $\Delta$, and the antiferromagnetic orientational ordering competes with the pentagonal spin ordering in a supercooled state. 
We note that the increase in the anisotropic part of the potential leads to the increase in the activation energy for particle motion at high temperatures above the onset temperature $T_{\rm on}$ below which cooperative motion emerges, also making liquids stronger. 

We also performed simulation studies on 2D weakly polydisperse hard-sphere-like liquids, where the degree of size polydispersity $\Delta$ is a measure of the strength of frustration against crystallization into a hexatic crystal~\cite{kawasaki2007correlation}. 
This system has a $\phi$-$\Delta$ phase diagram similar to the above 2D spin liquids (see Fig. \ref{fig:spin}(c)), showing the generality of our frustration-based scenario. 
For this system, we have also found that a liquid suffering stronger frustration against crystallization, i.e., larger $\Delta$ becomes stronger and has a better glass-forming ability (see Fig.~\ref{fig:spin}(d)).

Molinero, Sastry, and Angell~\cite{molinero2006tuning} also performed numerical simulations of modified Stillinger-Weber potential, which is made of an isotropic part and anisotropic part favoring tetrahedral symmetry. 
They found that the glass-forming ability and the fragility are maximized around the triple point, consistent with the prediction of our two-order-parameter model~\cite{tanaka2002simple}. 
I vividly remember a fascinating discussion on this topic with Angell and Molinero in the International Discussion Meeting on Relaxations in Complex Systems held at Lille, France, in 2005. 
Later, Angell and his coworkers even proved a high glass-forming ability of atomic germanium near the triple point by performing high-pressure experiments~\cite{bhat2007vitrification}. They successfully produced a metallic amorphous state of single-component atomic germanium by temperature quench.  

A similar $V$-shaped melting point minimum is also observed in binary mixtures. Such melting point minimum is widely known as a eutectic point. 
Angell also did pioneering experiments with Sare on the glass-forming ability of a water-salt mixture~\cite{Angell_Sare,AngellCR}. It is one of the critical experiments that was the basis of developing our two-state model. 
We also did experiments on the same system (a water-LiCl mixture) to test the prediction of a three-state model, which includes hydration structures as the third state to our two-state model of water (see Fig.~\ref{fig:competing}(b)). This model phenomenologically describes the dependence of the viscosity on the salt concentration~\cite{kobayashi2011possible,kobayashi2011relationship}. In the course of these works, Angell kindly provided us with many valuable comments. 

Recently, we have used the model with tunable tetrahedrality $\lambda$ developed by Molinelo, Sastry, and Angell~\cite{molinero2006tuning} to study the link between the degree of competing orderings and the glass-forming ability in a system with a V-shaped phase diagram (see Fig.~\ref{fig:competing}(a))~\cite{russo2018glass}. We will discuss this problem and provide a more detailed physical origin of the phenomenon in Sec.~\ref{sec:crystal}.

\paragraph{Structural origin of slow glassy dynamics} 

\begin{figure*}[t]
\begin{center}
\includegraphics[width=16.cm]{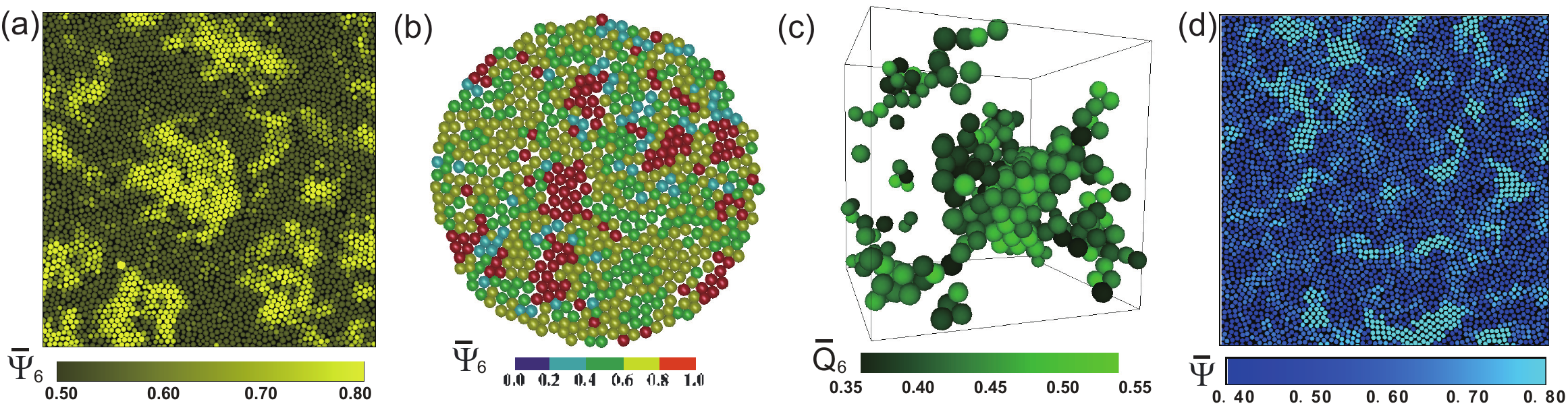}
\end{center}
\caption{Spatial fluctuations of angular orders developed in various supercooled glass-forming liquids. We stress that these angular orders are not coupled to the density field and thus cannot be detected by a two-body density correlator, such as 
the radial distribution function $g(r)$ and structure factor $S(q)$~\cite{tanaka2010critical}.  
(a) Spatial distribution of hexatic order parameter $\bar{\Psi}_6$ time-averaged over $\tau_\alpha$ in a 2D supercooled polydisperse hard-sphere-like liquid (polydispersity $\Delta$=9\%)~\cite{kawasaki2007correlation}. 
(b) Spatial distribution of hexatic order parameter $\bar{\Psi}_6$ time-averaged over $\tau_\alpha$ in a 2D supercooled polydisperse driven granular liquid (polydispersity $\Delta$=9\%)~\cite{watanabe2008direct}. 
(c) Spatial distribution of 6-fold bond orientational order parameter $\bar{Q}_6$ time-averaged over $\tau_\alpha$ in a 3D supercooled polydisperse hard-sphere-like liquid (polydispersity $\Delta$=6\%)~\cite{tanaka2010critical,kawasaki2010structural}. 
(d) Spatial distribution of spin alignment order parameter $\bar{\Psi}$ time-averaged over $\tau_\alpha$ in a 2D supercooled spin liquid (the frustration parameter $\Delta=0.6$)~\cite{shintani2006frustration}. 
}
\label{fig:order}
\end{figure*}

\begin{figure*}[h!]
\begin{center}
\includegraphics[width=16.cm]{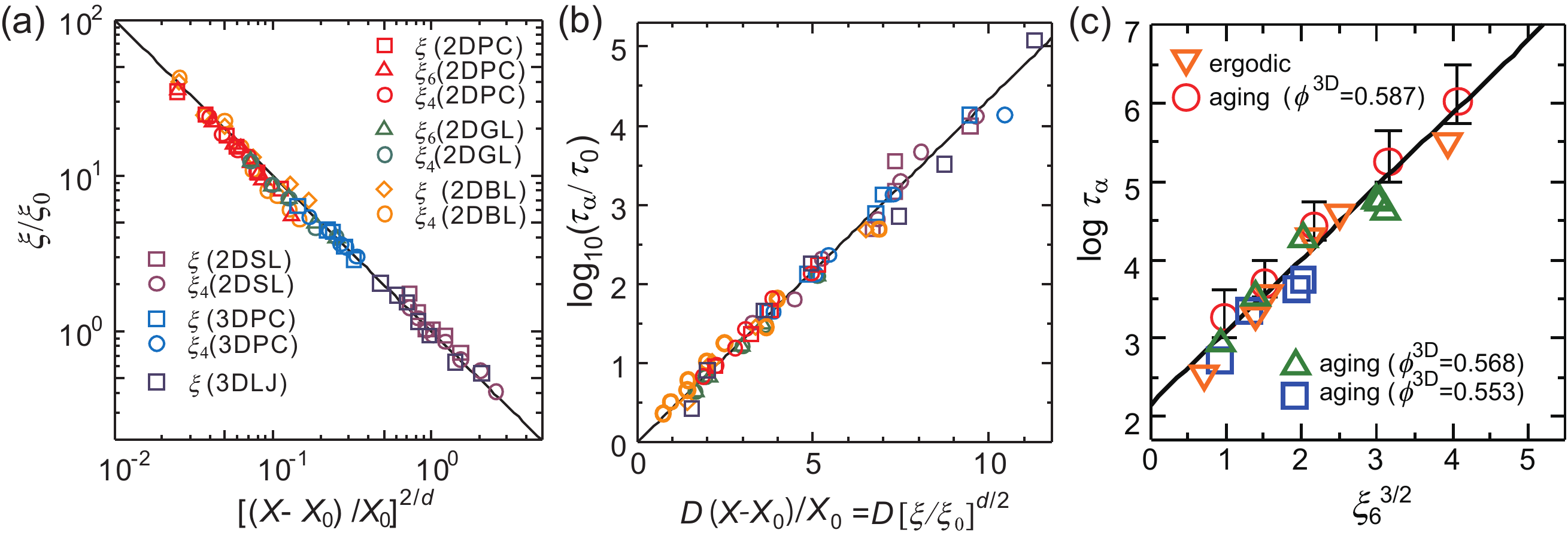}
\end{center}
\caption{(a) Power-law relation between $\xi/\xi_0$ and $(X-X_0)/X_0$~\cite{tanaka2010critical}. $X$ is the volume fraction $\Phi$ for 2D and 3D polydisperse hard-sphere-like liquids (2DPC and 3DPC), 2D granular liquids (2DGL), and 2D binary hard-sphere-like liquids (2DBL), whereas the temperature $T$ for 2D spin (2DSL) and 3D Lennard-Jones liquids (3DLJ). $\xi$ and $\xi_6$ are the static correlation lengths, and $\xi_4$ is the dynamic correlation length. 
$\xi_6$ is the correlation length of hexatic order $\Psi_6$ for 2DPC and 2DGL, whereas the correlation length of 6-fold bond orientational order $Q_6$ for 3DPC and 3DLJ. 
(b) Relation between $\log (\tau_\alpha/\tau_0)$ and $D_{\rm F} (X-X_0)/X_0=D_{\rm F} (\xi/\xi_0)^{d/2}$~\cite{tanaka2010critical}, where $D_{\rm F}$ is the fragility index and $d$ is the spatial dimensionality. 
(c) Relationship of $\tau_\alpha$ to $\xi_6$ during the aging process of 3DPC (polydispersity: 6\%) for the volume fractions $\Phi$ = 0.550, 0.568, and 0.587, together with the relationship of $\tau_\alpha$ to $\xi_6$ in the equilibrium liquid as a function of $\Phi$~\cite{kawasaki2014structural}. Note that during aging, both $\tau_\alpha$ and $\xi$ increase with the aging time. 
Panels (a) and (b) are reproduced from Figs. 4a and b of Ref.~\cite{tanaka2010critical}, respectively, and panel (c) from Fig. 4(f) of Ref.~\cite{kawasaki2014structural}.
}
\label{fig:scaling}
\end{figure*}

One of the central questions of glass transition is the physical origin of drastic slowing down of the dynamics towards the glass transition point. 
The extreme case is the fragile limit according to the Angell classification. Many simulation models of spherical particles belong to the fragile-limit glass formers~\cite{tanaka2019revealing}.
Our two-order-parameter model predicts that a system only weakly frustrated against crystallization, which is sitting in the border between crystallization and vitrification, should belong to the fragile-limit glass formers~\cite{tanaka2012bond}. 
Based on this idea, we studied weakly frustrated liquids and found that crystal-like orientational order grows with an increase in the degree of supercooling. 
We show such examples of growing angular orders upon cooling in various supercooled liquids, as shown in Fig.~\ref{fig:order}, indicating the generality in fragile liquids. 
We emphasize that two-body density correlators, such as the radial distribution function $g(r)$ and structure factor $S(q)$, are hard to detect these angular orders. This fact also implies little coupling between the density and the angular orders for fragile-limit liquids (note that this is not the case for strong liquids, such as silica). In these liquids, steric repulsion is the primary source of structural ordering. Noting that an angular order requires at least three-body correlations, we may say that glassy amorphous order must be intrinsically due to many-body correlations, casting doubt on scenarios based on the two-body correlation~\cite{tanaka2012bond,tanaka2019revealing,tanaka2020role}. 

We also found that these spatial fluctuations of static structural order parameters are strongly correlated with the mobility fields, i.e., dynamic heterogeneities~\cite{shintani2006frustration,kawasaki2007correlation,watanabe2008direct,tanaka2010critical,kawasaki2010structural,leocmach2012roles}. 
These results not only support a structure-dynamics link revealed by an isoconfigurational sampling method developed by Harrowell and his coworkers~\cite{widmer2004reproducible} but also show the importance of angular order, i.e., many-body interactions, at the origin of structure-dynamics correlation. 
 
Furthermore, the correlation length $\xi$ has been revealed to diverge towards the ideal glass transition 
point $X_0$ as~\cite{shintani2006frustration,kawasaki2007correlation,watanabe2008direct,tanaka2010critical,kawasaki2010structural,leocmach2012roles} 
\begin{equation}
\xi=\xi_0 \left(\frac{|X-X_0|}{X_0} \right)^{-\nu}, \label{eq:xi}
\end{equation}
where $\nu=2/d$, $d$ is the spatial dimensionality, and $X$ is temperature $T$ or volume fraction $\phi$. 
We also found that the static correlation length grows with an increase in the degree of supercooling, in proportion to the dynamic correlation length $\xi_4$~\cite{shintani2006frustration,kawasaki2007correlation,watanabe2008direct,tanaka2010critical,kawasaki2010structural,
leocmach2012roles}. This suggests that the dynamic heterogeneity is a consequence of the spatial fluctuation of static structural order. 
Moreover, we found that the exponent $\nu$ is the same as the critical exponent for the correlation length for the $d$-dimensional Ising universality class of critical phenomena~\cite{tanaka2010critical}: $\nu \sim 2/d$.  
We show this relation suggesting the Ising criticality for various liquids is shown in Fig.~\ref{fig:scaling}(a). Similar Ising-like criticality was also reported by Mosayebi et al.~\cite{mosayebi2010probing} and Zheng et al.~\cite{zheng2021translational}.

Furthermore, we obtained the following relation between $\xi$ and the structural relaxation time $\tau_\alpha$ (see Fig.~\ref{fig:scaling}(b)):
\begin{equation}
\tau_\alpha=\tau_0 \exp \left( \frac{K (\xi/\xi_0)^{d/2}}{k_{\rm B}T} \right), \label{eq:tau}
\end{equation}
where $\tau_0$ is the microscopic time, $K$ is related to the fragility index $D_{\rm F}$ as $K=D_{\rm F} k_{\rm B}T$ (larger $D_{\rm F}$ indicates a stronger character), and $k_{\rm B}T$ is the thermal energy.   
This relation indicates that the activation energy for particle motion increases in proportion to $(\xi/\xi_0)^{d/2}$. 
Furthermore, we found~\cite{tanaka2010critical} that the dynamics of the structural order parameter belongs to the dynamic universality of the kinetic Ising model, i.e., model A~\cite{hohenberg1977theory,OnukiB}. 
We note that our angular structural order parameter is non-conserved since it can change locally at each point independently from other points (as spin can flip independently without any constraint).
This non-conserved nature of the order parameter is consistent with the model A-type dynamics.

\paragraph{Generalized angular order parameter: Packing capability $\Psi$}
In the above examples showing the Ising criticality, angular ordering has a connection to the crystal symmetry (see, e.g., Fig.~\ref{fig:order}) because these systems suffer from weak frustration against crystallization.  
However, such a connection is not expected for binary mixtures that involve phase separation upon crystallization. The phase separation cuts the link between the liquid-state and crystal-state free energies.   
To overcome this difficulty, we have recently introduced a new structural order parameter, i.e., local packing capability $\Psi$, which is a measure of local vibrational entropy, i.e., local free energy. 
This structural order parameter $\Psi$ can detect sterically favored topologies in an order-agnostic manner.  
It characterizes the angular order of the configuration of neighboring particles around a particle, thus intrinsically reflecting many-body correlations. 
We may regard $\Psi$ as a generalized bond orientational order parameter that does not rely on specific symmetry. 

Using this order parameter, we have confirmed that relations (\ref{eq:xi}) and (\ref{eq:tau}) hold universally for any particulate glass-formers, including binary mixtures, in which the interaction potentials are isotropic and additive~\cite{tong2018revealing,tong2019structural,tong2020role}. 
Furthermore, we have found that a link between the amplitude of fast $\beta$ motion and slow structural relaxation~\cite{dyre2006colloquium,starr2002we,widmer2006predicting,kawasaki2007correlation,larini2008universal,betancourt2015quantitative} is through local structures identified as the local packing capability~\cite{tong2018revealing}. 

\begin{figure*}[t!]
\begin{center}
\includegraphics[width=14.cm]{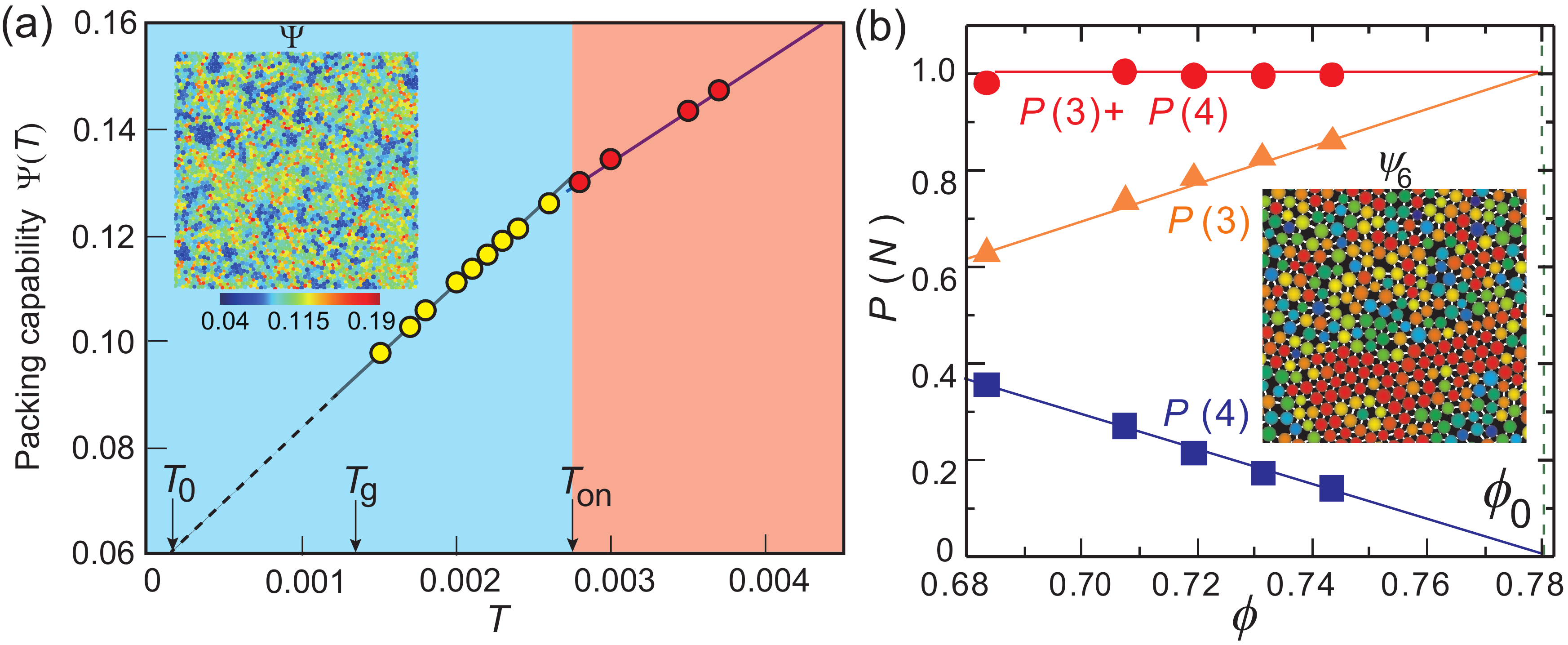}
\end{center}
\caption{(a) Evolution of structural order $\Psi$ in instantaneous liquid states (circles) in a 2D polydisperse harmonic systems (the polydispersity $\Delta=13\%$). The onset temperature ($T_{\rm on}=0.00276$) and dynamical glass transition temperature for the slowest cooling rate ($T_g=0.00135$) are indicated, allowing us to separate the simple-liquid (light red), supercooled and glass regimes (light blue).  For instantaneous liquid states, the temperature dependence of $\Psi$ can be fitted with a linear function in the supercooled regime $(\Psi-\Psi_0)/\Psi_0 \propto (T-T_0)/T_0$  (solid line; dashed line below $T_g$). Note that smaller $\Psi$ means higher order. The inset image is the spatial distribution of $\Psi$ at $T=0.0018$. See Ref.~\cite{tong2019structural} for the details.
(b) $\phi$-dependence of the fraction of triangles and squares, which are averaged over $10 \tau_\alpha$, for 2D polydisperse hard-sphere-like system ($\Delta=$9\%). Squares (geometrical defects) decrease, or transform to compact  triangles, with an increase in $\phi$ and tend to disappear around $\phi_0$ completely. The inset image shows the correlation of the instantaneous hexatic order parameter, $\psi_ 6$, to geometrical defects (i.e., voids) at $\phi=0.73$. Bonds are shown as thin white lines. $\psi_6$ is evidently anti-correlated with the geometrical defects (black voids). See Ref.~\cite{tanaka2010critical} for the details.
}
\label{fig:defect}
\end{figure*}

We have also found that the average local packing capability defined in an instantaneous liquid state, $\Psi$, under thermal noise have the following relation to the system temperature ($T$) or density ($\rho$)~\cite{tong2019structural} (see Fig.~\ref{fig:defect}(a)):
\begin{equation}
(\Psi-\Psi_0)/\Psi_0=(T-T_0)/T_0=(\rho_0-\rho)/\rho_0, \label{eq:Psi}
\end{equation}  
where $\Psi_0$ and $\rho_0$ are the values of $\Psi$ and $\rho$ at the ideal glass-transition point. This relation indicates that $\Psi$ is an essential control variable of the slow glassy dynamics not only locally but also globally~\cite{tong2019structural}. 
For example. we have confirmed a general relationship between $\Psi$ and $\tau_\alpha$:
\begin{equation}
\tau_\alpha=\tau_0 \exp \left (\frac{B \Psi_0}{\Psi-\Psi_0} \right), \label{eq:Psi_tau}
\end{equation}  
where $B$ is a constant.

It has been known~\cite{berthier2010critical} that liquids interacting with the Weeks-Andersen-Chandler (WCA) and Lennard-Jones (LJ) potentials, which have an identical $g(r)$ but very different dynamics at the same temperature. 
Recently, there have been some efforts to save this failure of the two-body description of slow glassy dynamics by its modification~\cite{landes2020attractive,nandi2021microscopic}.   
However, we have shown that these liquids actually have very different local packing capability $\Psi$ at the same temperature, responsible for the different dynamics~\cite{tong2020role}. We have confirmed that both systems obey relation (\ref{eq:Psi_tau}). This finding indicates that a two-body correlation such as $g(r)$ is not enough to detect the amorphous order responsible for slow dynamics of a supercooled state, and many-body angular correlations play an essential role in slow glassy dynamics ~\cite{tanaka2012bond,tanaka2019revealing,tanaka2020role}. 

\paragraph{Link between glassy order parameter and slow dynamics}
Now we consider how local liquid structures determine slow glassy dynamics.  
For a weakly polydisperse system, the spatial correlation function of the coarse-grained crystal-like bond orientational order parameter, a complex order parameter that has both amplitude and phase information, can directly detect the correlation length of growing structural order. Here, the coarse-graining up to neighboring particles is necessary to remove the disturbance due to icosahedral ordering in hard-sphere-like systems. Thus, it is hard to detect a proper correlation length without coarse-graining (see, e.g., Ref.~\cite{charbonneau2013decorrelation}).

What is critical here is that the phase (angular) coherence of the bond orientational order parameter plays an essential role in detecting the static correlation length $\xi$~\cite{kawasaki2007correlation,tanaka2010critical,kawasaki2010structural,tanaka2012bond} (see below).   
On the other hand, the amplitude correlation function fails in detecting $\xi$.  
For a scalar structural order parameter, the spatial coarse-graining over the correlation length $\xi$~\cite{tong2018revealing,tong2019structural} or the time averaging over $\tau_\alpha$~\cite{shintani2006frustration,kawasaki2007correlation,tanaka2010critical,kawasaki2010structural,leocmach2012roles} are crucial for seeing nearly one-to-one structure-dynamic correlation ($\xi \sim \xi_4$) in supercooled glass-forming liquids.  
This fact means that the dynamics of a particle cannot be determined locally but depends on its environment; in other words, it is determined in a coarse-graining manner~\cite{berthier2007structure,tong2018revealing,tong2019structural}.

We have shown that the size of the environment affecting the motion of a particle determines the static correlation length $\xi$ as well as the dynamic correlation length $\xi_4$~\cite{tong2018revealing}.
Using isoconfigurational averaging, we have also studied how the mobility field develops from an inherent structure with time. 
We have found that particle motion starts to evolve from most defective parts with low local packing capability (i.e., large $\Psi(\bm{r})$)  and gradually spreads towards regions with higher packing capability (smaller $\Psi(\bm{r})$), following the $\Psi(\bm{r})$ field. 
The strongest correlation is seen between the mobility field pattern at the maximum dynamic heterogeneity (its characteristic length=$\xi_4$) and the static order parameter field with the coarse-grained length $\xi$, i.e., $\xi_4 \sim \xi$. 
Since this process proceeds statistically, there is no one-to-one relationship between the local scalar order parameter such as $\Psi(\bm{r})$ and the mobility field. 
Here we note that since $\Psi(\bm{r})$ is inversely correlated with the local Debye-Waller factor (solidity) (or, positively correlated with the amplitude of fast $\beta$ motion)~\cite{tong2018revealing}, this result is consistent with the correlation between the Debye-Waller factor and the slow dynamics~\cite{scopigno2003fragility,dyre2006colloquium,widmer2006predicting,kawasaki2007correlation,larini2008universal,simmons2012generalized,zhang2021fast}. 

Here we consider the role of the phase (angular) coherence of the complex bond orientational order parameter. 
Considering the above finding, we speculate that the phase coherence controls how the mobility field statistically develops, following the structural order parameter field. 
This fact again strongly suggests the critical importance of {\it angular} structural order in determining slow glassy dynamics. 
We stress that the angular coherence is tightly linked to the spatial extendability of sterically favored topologies without voids.  

Finally, we also emphasize that Eq.~(\ref{eq:tau}) holds even in out-of-equilibrium situations during aging~\cite{kawasaki2014structural} (see Fig.~\ref{fig:scaling}(c)) and for systems under spatial confinements~\cite{watanabe2011structural}. 
These results strongly suggest that liquid dynamics at a specific location is controlled by a structural order parameter of many-body origin there, irrespective of whether the liquid is in equilibrium or out of equilibrium or whether it is confined or not confined. 

\paragraph{The origin of the Ising-like criticality}
These studies indicate that structural fluctuation of particulate glass-formers interacting with isotropic potentials shows Ising-like criticality, suggesting a discrete symmetry of glassy structural ordering. Langer proposed an Ising model of glass transition, focusing on the two-state feature of sterically favored configurations~\cite{langer2013ising,langer2014theories}. This is an intriguing idea. We might be able to say that frustration generally changes the nature of the order parameter from the continuous to discrete symmetry~\cite{tanaka2010critical} since similar phenomena are also observed in spin systems~\cite{chandra1990ising,weber2003ising}. Below we consider this problem from a different angle.

In the above, we have emphasized the importance of the phase coherency of orientational order parameter and the spatial coherence of packing capability. Such local coherence of order between neighboring particles is favored sterically because a particle with high orientational order (or packing capability) can gain an extra free volume (or vibrational entropy) if its neighbor also coherently has high order. This situation is essentially the same as the favored parallel alignment of neighboring spins in the Ising model.

Here, we point out an essential difference in the character of the order-parameter fluctuations between 2D monodisperse and glass-forming polydisperse disks. In the former, ordered patches with some phase coherence are formed in a liquid state, and its coherence length $\xi_{\rm coh}$ (i.e., the patch size) exponentially toward the hexatic-ordering point. On the other hand, in the latter, frustration creates low ordered regions, leading to critical-like fluctuations of the order parameter amplitude in addition to the phase, as shown in Figs.~\ref{fig:order}(a) and (b). The correlation length diverges, obeying the Ising-like power law instead of the exponential growth. Similar behaviors are also observed in other systems, such as 2D spin liquid (see Fig.~\ref{fig:order}(d)~\cite{shintani2006frustration} and glass-forming binary mixtures~\cite{tong2018revealing, tong2019structural}. We speculate that the introduction of this disorder due to frustration and the resulting order-parameter-amplitude fluctuations may be the origin of the discrete twofold symmetry, i.e., the Ising-like criticality.

In relation to the above speculation, we focus on the nature of the disorder. The above order parameter $\Psi$ is scalar, representing the local packing capability around a central particle. Thus, its decrease with decreasing $T$ or increasing $\phi$ indicates the decrease of the degree of disorder. 
We also define voids for weakly 2D polydisperse hard-sphere-like systems as follows~\cite{tanaka2010critical}. We first apply Delaunay triangulation to a spatial pattern of the centers of mass of particles. When the length of a side of a triangle connecting particle $i$ and $j$ is larger than a critical value $\sigma^c_{ij}$, which we set $\sigma^c_{ij}=1.5 (\sigma_i+\sigma_j)/2$, we cut that bond. In this way, we can identify geometrical defects, which are polygons whose number of sides are more than 4 (squares, pentagons, $\cdots$). The driving force of geometrical ordering is steric repulsion, which maximizes entropy by gaining vibrational entropy while sacrificing configurational entropy. The inset image of Fig.~\ref{fig:defect}(b) illustrates the spatial distribution of geometrical defects (squares, pentagons, $\cdots$) for 2D polydisperse hard-sphere-like system~\cite{tanaka2010critical}, which is anti-correlated with the degree of hexatic order (i.e., tiling with rather regular triangles). We also plot the $\phi$-dependence of $P(3)$ and $P(4)$ in Fig.~\ref{fig:defect}(b).  
Triangles become more and more dominant with an increase in $\phi$, and squares (i.e., geometrical defects) tend to disappear completely at the ideal glass-transition volume fraction $\phi_0$. The fraction of voids, which is roughly the fraction of squares, $P(4)$, follows $P(4) \propto (\phi_0-\phi)$.

Interestingly, the $T$-dependence of voids, $P(4)$, is essentially the same as that of the packing capability, $\Psi$, Eq.~(\ref{eq:Psi}) (compare Figs.~\ref{fig:defect}(a) and (b)). We stress that since relation~(\ref{eq:Psi}) holds in any systems interacting with isotropic interactions, including both 2D and 3D systems~\cite{tong2019structural}, the behaviors of these two quantities characterizing local packing are universal. 
The packing capability $\Psi$ and the fraction of voids $P(4)$ act as control parameters of the dynamics as the temperature does. We may regard these quantities as `free volume'. This result supports the above speculation on the importance of frustration-induced disorder for the Ising-like criticality. However, since the discussion is speculative, identifying the origin of discrete symmetry remains a task for future theoretical investigation.

\paragraph{Ising-criticality scenario or RFOT scenario}

\begin{figure*}
\begin{center}
\includegraphics[width=11.cm]{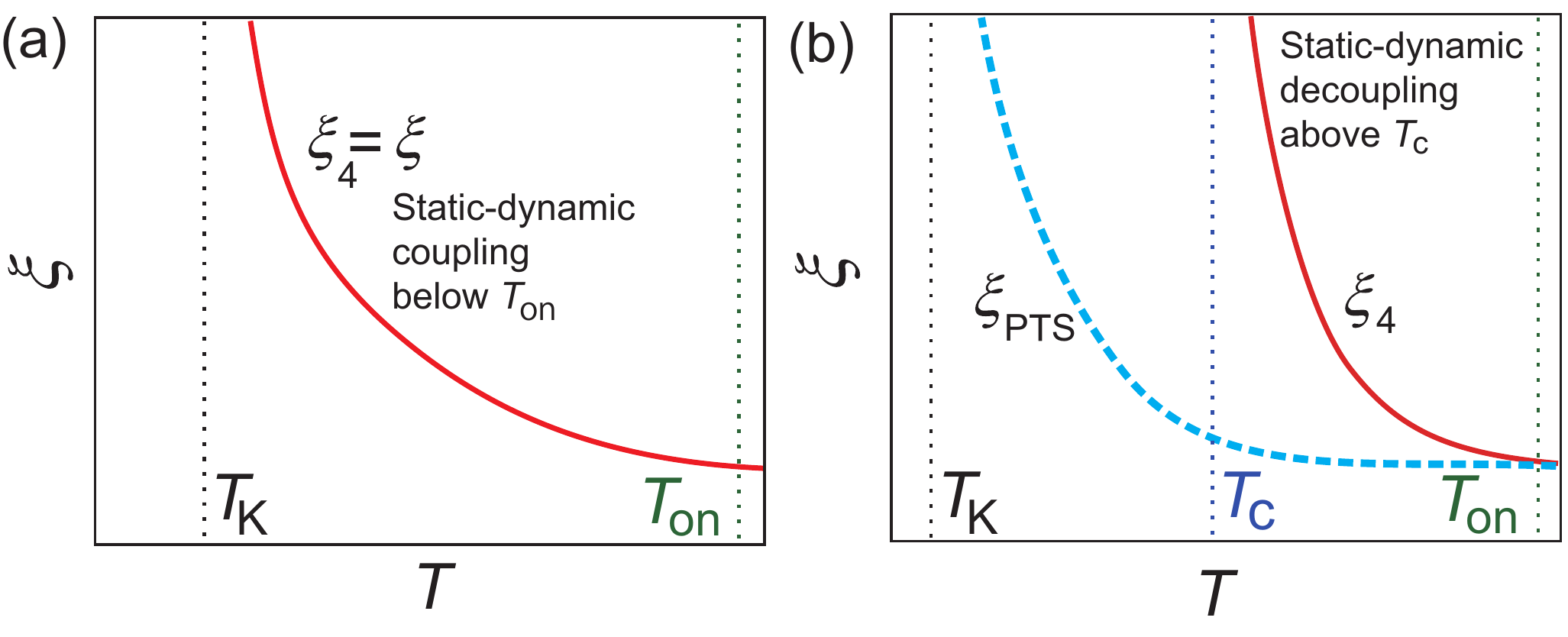}
\end{center}
\caption{
Comparison of the temperature dependences of the static ($\xi$ and $\xi_{\rm PTS}$) and dynamic correlation lengths $\xi_4$ between the Ising criticality (a) and RFOT (b) scenarios. 
$T_{\rm on}$ is the onset temperature, $T_{\rm c}$ is the mode-coupling critical temperature, and $T_{\rm K}$ is the Kauzmann temperature equal to the ideal glass transition temperature $T_0$.
}
\label{fig:xi}
\end{figure*}

\begin{figure*}[h]
\begin{center}
\includegraphics[width=11.cm]{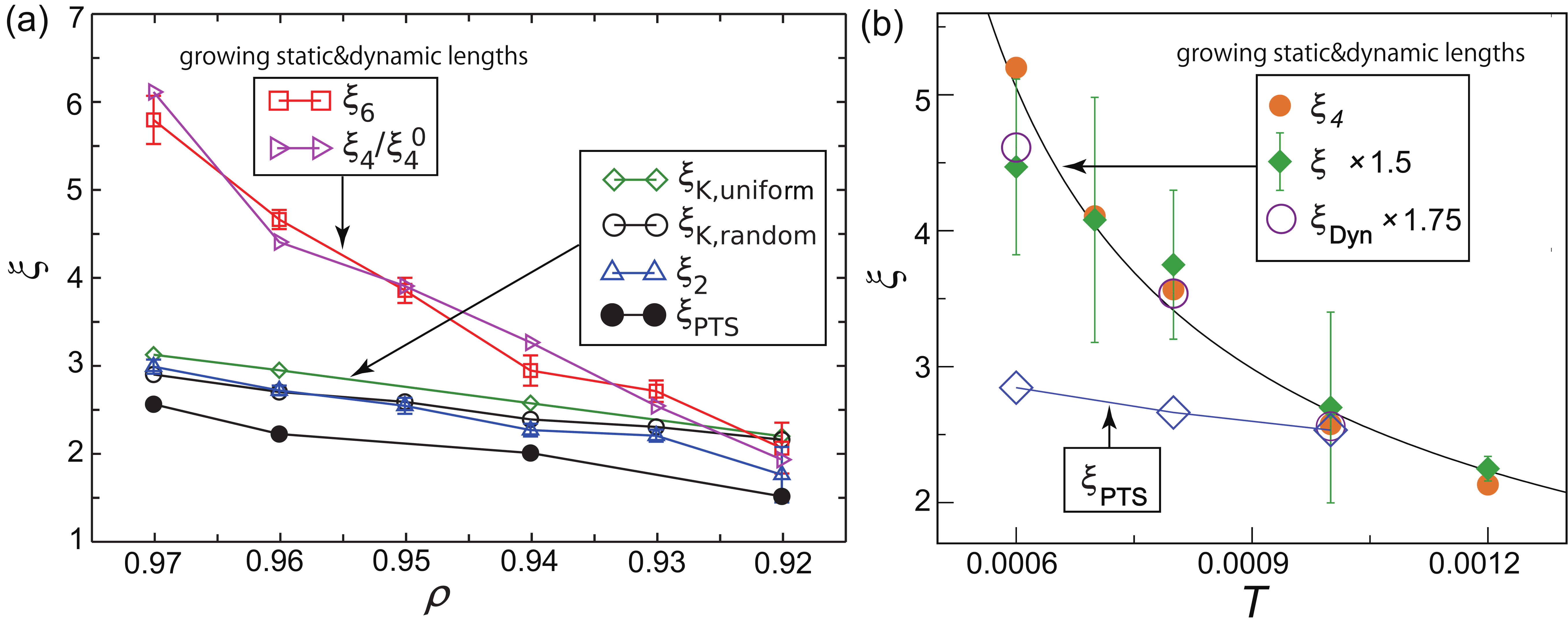}
\end{center}
\caption{(a) Various correlation lengths as a function of density $\rho$ for 2D polydisperse hard-sphere-like liquid (polydispersity $\Delta=$11\%)~\cite{russo2015assessing}: red squares for bond orientational order $\xi_6$, blue triangles for the two-body correlations $\xi_2$, black circles for $\xi_K$ for random pinning, green diamonds for $\xi_K$ for uniform pinning, black filled circles for $\xi_{\rm PTS}$, and pink triangles for the dynamic correlation
length $\xi_4$. The dynamical correlation length is scaled with $\xi_4^0 \sim 1.7$ to ease the visual comparison with the static length scales.
(b) Temperature dependences of the dynamic correlation length $\xi_4$ and the static correlation length $\xi$ for 3D soft binary mixture ($T_0 = 3.5 \times 10^{-4}$)~\cite{tong2018revealing}. The solid line is a power-law fit with $\xi = \xi_0((T - T_0)/T_0)^{-2/3}$. Here, we include the dynamic $\xi_{\rm Dyn}$ and static PTS $\xi_{\rm PTS}$ lengths of the same system reported in Ref.~\cite{kob2012non} for comparison: the dynamic length $\xi_{\rm Dyn} \times 1.75$ (open circles) and the static length $\xi_{\rm PTS} \times 4.6$ (open diamonds). 
Panel (a) is made based on Fig. 1D of Ref.~\cite{russo2015assessing}, and panel (b) is reproduced from Fig.~5(f)  of Ref.~\cite{tong2018revealing}.
}
\label{fig:PTS}
\end{figure*}

Our Ising-criticality scenario~\cite{tanaka2010critical,tanaka2012bond} and the RFOT scenario~\cite{kirkpatrick1989scaling,parisi2010mean,berthier2011theoretical,kirkpatrick2015colloquium} both predict the same relations for $\xi$ and $\tau_\alpha$ like Eqs.~(\ref{eq:xi}) and (\ref{eq:tau}) (for the former, below the onset temperature $T_{\rm on}$, whereas for the latter, only near $T_0$ below the mode-coupling $T_{\rm c}$). 
However, there is a fundamental difference concerning the nature of the order-parameter fluctuations.

In our scenario, spatial fluctuations of the order parameter are critical-like (i.e., obeying the Ornstein-Zernike correlation~\cite{OnukiB}) for our scenario, and, upon cooling, their size exceeds the particle size below the onset temperature $T_{\rm on}$. Thus, $T_{\rm on}$ marks a crossover from the simple Arrhenius to Vogel-Fulcher-Tammann law (see. e.g., Ref.~\cite{tanaka2005two2}). 
Below $T_{\rm on}$, the static ($\xi$) and dynamic ($\xi_4$) correlation lengths grow in a coupled manner since the growth of structural ordering is the origin of slow dynamics in this scenario (see Fig.~\ref{fig:xi}(a)). 

In contrast, mosaic-like metastable islands are a prominent player for the RFOT scenario and emerge only below the mode-coupling $T_{\rm c}$, below which the free energy is supposed to have a metastable minimum. 
Thus, the static ($\xi$) and dynamic ($\xi_4$) correlation lengths are strongly decoupled far above the Kauzmann (or ideal glass-transition) temperature $T_{\rm K}=T_0$, more precisely, above $T_{\rm c}$ (see Fig.~\ref{fig:xi}(b)).
The crucial point is that $T_{\rm on}$ is generally located much higher temperature than $T_{\rm c}$, and thus, the relation between the static and dynamic correlation lengths in a temperature region between $T_{\rm on}$ and $T_{\rm c}$ should provide critical information on which scenario is more appropriate for the description of slow glassy dynamics.  

Our molecular dynamics simulations showed that the relations given by Eqs.~(\ref{eq:xi}) and (\ref{eq:tau}) 
hold even in the temperature region between $T_{\rm on}$ and $T_{\rm c}$ (or slightly below it). 
We have confirmed for 2D polydisperse hard-sphere liquid (Fig.~\ref{fig:PTS}(a))~\cite{russo2015assessing} and 3D soft binary mixture liquid (Fig.~\ref{fig:PTS}(b))~\cite{tong2018revealing} that 
the static correlation length $\xi$ is directly coupled with the dynamic correlation length $\xi_4$ even above $T_{\rm c}$.  
According to the RFOT scenario, there should exist no rapidly growing static length above $T_{\rm c}$, where mosaic structures are absent (see Fig.~\ref{fig:xi}(b)). Thus, these results support our Ising-criticality scenario (compare Figs.~\ref{fig:xi}(a) and (b)). 

We also note that the standard point-to-set length $\xi_{\rm PTS}$, which was developed to pick up the mosaic length based on the RFOT scenario~\cite{biroli2008thermodynamic,charbonneau2012geometrical,kob2012non}, cannot detect our angular order correlation length $\xi$~\cite{russo2015assessing,tong2018revealing}. 
This fact seems to indicate that the standard point-to-set (PTS) length that is inherently designed to detect translational order cannot pick up orientational order~\cite{russo2015assessing} unless it is specially designed to detect angular order~\cite{yaida2016point} (note that the PTS defined in this way is no more order-agnostic since the knowledge of the order type is required in advance). 
The ROFT theory is exact for hard-sphere liquids at the infinite dimension ($d=\infty$), where many-body effects disappear entirely, and the two-body-level description becomes exact~\cite{parisi2020theory}.  
However, at low dimensions such as $d=2$ and 3 relevant for real systems, many-body angular correlations that originate from entropy gain associated with high particle packing capability lead to `entropy-driven structural ordering' in a supercooled liquid~\cite{tanaka2012bond,tanaka2019revealing,tanaka2020role}. 

The physics behind our scenario can be relatively easily understood by noting that (1) crystallization of hard spheres discovered by Alder and Wainwright~\cite{alder1957phase} is purely entropically (or sterically) driven and (2) hexatic ordering plays a critical role in the ordering of hard disks in 2D; i.e., an intermediate hexatic phase with quasi-long-range orientational order emerges between liquid and crystal states~\cite{nelson2002defects}.  
Thus, we may conclude that in liquids made of particles where interparticle steric repulsions play a critical role in structural ordering, slow glassy dynamics is caused by {\it entropy-driven angular ordering, i.e., enhancement of packing capability}. 
Recently, Han and his coworkers~\cite{zheng2021translational} showed that liquids made of hard ellipsoids also exhibits Ising-criticality near the glass transition, indicating that our scenario is valid as long as 
structural ordering is entropically driven. 

We emphasize that entropically driven structural order can grow its correlation length without limitation upon cooling until geometrical frustration embedded in the structural order parameter prevents it. 
We note that, unlike crystalline order, amorphous order cannot tile the space thoroughly and thus intrinsically suffers from internal frustration in the structural order itself. For example, polydispersity above a particular level prevents infinite growth of crystal-like bond orientational order. Thus, the critical-like power-law growth of the correlation length $\xi$ and the resulting VFT-like divergence of $\tau_\alpha$ may eventually stop upon cooling. 

Finally, we note that the situation should be markedly different for energetically driven structural ordering such as tetrahedral ordering in water and silica. 
Locally favored structures formed energetically by directional bonds must accompany a sizeable entropic loss upon the ordering (see below)~\cite{Tanaka2000,tanaka2019revealing,tanaka2020role}. 
Thus, these structures cannot extend spatially and remain localized without growing, increasing only their fraction upon cooling.   

\begin{figure*}[t]
\begin{center}
\includegraphics[width=12.cm]{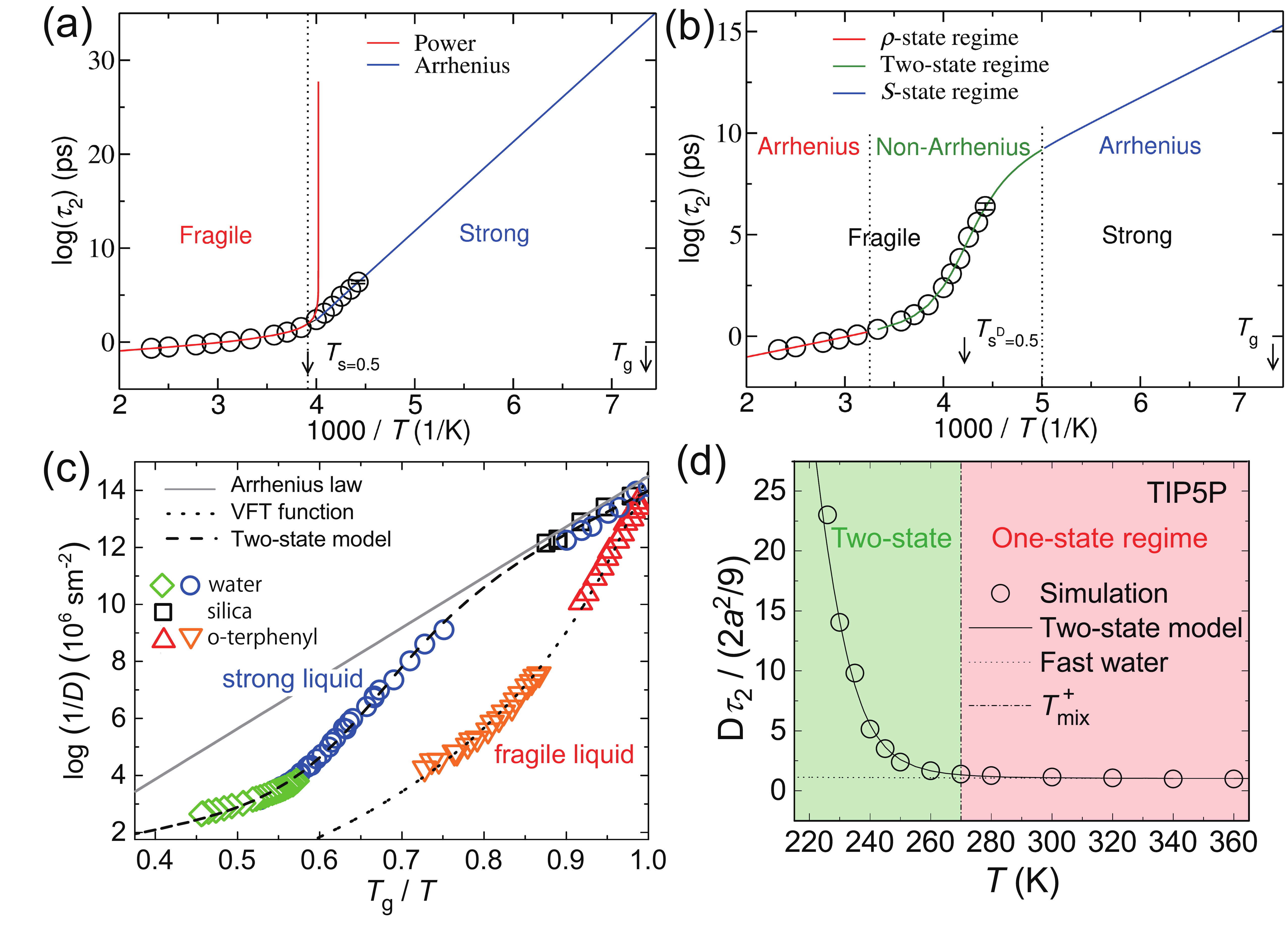}
\end{center}
\caption{Dynamical crossovers in TIP5P water and real water at 1 bar. (a) The Widom-line interpretation~\cite{xu2005relation} of dynamical crossovers in TIP5P water: a crossover from power-law above the Widom line ($s=1/2$ line here) to Arrhenius law below the Widom line. This scenario results in unphysically slow dynamics at $T_{\rm g}$. (b) Our hierarchical two-state scenario for dynamical crossovers in TIP5P water. In the pure $\rho$-state regime, water shows an Arrhenius behavior. When entering the two-state regime upon cooling, water shows the first dynamic crossover from the Arrhenius to non-Arrhenius (the so-called ``fragile'') behavior; When leaving the two-state regime by further cooling, water exhibits the second dynamic crossover from the non-Arrhenius to Arrhenius (the so-called ``strong'') behavior. (c) Two-state fitting to the temperature dependence of the experimentally measured diffusion constants of water and silica, together with a fit of our hierarchical two-state model. We also show the data of a typical fragile liquid, o-terphenyl, with a fit of the Vogel-Fulcher-Tammann law. On the details of the experimental data, see Ref.~~\cite{shi2018origin}.
(d)  Breakdown of the Stokes-Einstein-Debye relation in TIP5P water. The solid and dotted lines represent the two-state and dynamic $\rho$-state contributions, respectively, indicating that the growth of dynamic $S$-state upon cooling causes the breakdown of the Stokes-Einstein-Debye relation in supercooled water. The effective hydrodynamic radius $a = 1.3$~\AA was estimated from the high-temperature data for TIP5P water. 
These figures (a)-(d) are reproduced from Figs. 15(a) and (b) of Ref.~\cite{shi2018common}, Fig.~5 and Fig. 4A of Ref.~\cite{shi2018origin}, respectively. See Refs.~\cite{shi2018origin,shi2018common} for the details.
} 
\label{fig:water_dyn}
\end{figure*}

\paragraph{Fragile-to-strong transition}
Angell and his coworkers~\cite{ito1999thermodynamic} discovered that some liquids like water exhibit fragile behavior at a high temperature, whereas strong behavior near $T_{\rm g}$. This unusual behavior has been widely known as ``fragile-to-strong transition''. Later, this transitional behavior has been reported for various liquids such as silica~\cite{saika2001fragile}, metallic glass-formers~\cite{zhang2010fragile,zhou2015structural}, and chalcogenide glasse-formers~\cite{wei2015phase}, and has attracted considerable attention in the glass community. It motivated us to consider the physical origin of the fragile-to-strong transition of water~\cite{Tanaka2003}. 
This unusual phenomenon has been claimed to be explained by a crossover from the power-law-type mode-coupling anomaly to a strong (Arrhenius) behavior upon crossing the Widom line of liquid-liquid transition~\cite{xu2005relation}. 
This scenario has become very popular. 

I focused on the fact that the fragile behavior of water is observed even at a temperature almost twice of $T_{\rm g}$, which is usually thought free from the glass transition or far above the onset temperature of cooperative slow glassy dynamics, $T_{\rm on}$. 
I proposed~\cite{Tanaka2003} that this transition may not be related to glass transition, but due to the crossover between the two states from a high-temperature disordered liquid without locally favored structures  (tetrahedral structures in the case of water and icosahedral structures for metallic liquids), which we call ``$\rho$-liquid'', to a low-temperature liquid full of locally favored structures, which we call ``$S$-liquid''. Since $\rho$- and $S$-liquids have different activation energies, 
a liquid with the fraction of locally favored structures $s$ should have the activation energy of $E(s)=E_\rho (1-s)+E_S s$, where $E_\rho$ and $E_S$ are the activation energy for pure $\rho$-liquid ($s=0$) and $S$-liquid ($s=1$), leading to an Arrhenius-to-Arrhenius crossover. The assumption behind this simple additive form of $E(s)$ is that the lifetime of locally favored structures is much shorter than $\tau_\alpha$~\cite{Tanaka2003}, which was later confirmed by numerical simulations~\cite{shi2018common}. 
It was predicted that the structural relaxation time $\tau_{\alpha}$ is described as a function of $s$ by the following phenomenological relation~\cite{Tanaka2003}:
\begin{equation}
\tau_\alpha(s)=\tau_0 \exp \left(\frac{E_\rho (1-s)+E_S s}{k_{\rm B}T} \right),
\end{equation}
where $\tau_0$ is the microscopic time. 

Recently, we have confirmed the validity of this phenomenological relation based on numerical simulations of realistic classical water models~\cite{shi2018origin,shi2018common}, but at the same time found that 
$s$ is not the fraction of locally favored structures, but its spatial average up to neighbors $s^D$. This is because the dynamics of a molecule cannot be determined by itself but also affected by its neighbors.  
The crucial point is that diffusion occurs via the exchange between neighboring molecules, and hydrogen-bond reconnections are necessary for the diffusion of a water molecule. 
This consideration has led us to develop a hierarchical two-state model~\cite{shi2018origin,shi2018common}. 

Figure~\ref{fig:water_dyn} shows the analyses of the dynamics of TIP5P water based on the Widom-line scenario (a) and our hierarchical two-state scenario (b), and the hierarchical two-state model analysis of the dynamics of actual water (c). 
We also note that our two-state model can describe the breakdown of the Stokes-Einstein-Debye relation without any adjustable parameters (Fig.~\ref{fig:water_dyn}(d) (see Refs.~\cite{shi2018origin,shi2018common} on the details). 
This indicates that the breakdown of the Stokes-Einstein-Debye relation in water does not come from glassiness but simply from the two-state feature. 
We also discuss this problem in more detail in Sec.~\ref{sec:water}. 

We argue that this Arrhenius-Arrhenius crossover scenario based on the two-state model may be valid also for silica, metallic, and chalcogenide liquids since these liquids tend to form distinct locally favored structures. 
For silica, we have evidence for the two-state feature, similarly to water~\cite{shi2018impact,shi2019distinct}. For metallic glass-formers such as CuZr, it was shown that local structural orderings such as icosahedral ordering are linked to the fragile-to-strong transition, although local structures may not be one type~\cite{zhou2015structural}.  
We note that Angell and his coworkers pointed out the importance of medium-range order and liquid-liquid transition in the fragile-to-strong transition in phase-change materials used for memory~\cite{wei2017structural,wei2017glass,lucas2020liquid}. 

\paragraph{Perspective}

The above consideration has led us to classify glass-forming liquids based on which of entropy or energy is dominant in structural ordering~\cite{tanaka2019revealing,tanaka2020role}. Liquids with entropy-driven structural ordering, such as hard spheres, can have amorphous structural (angular) order whose correlation length tends to diverge towards the ideal glass-transition point but may eventually stop growing due to geometrical frustration in the amorphous order parameter. In these liquids, the packing capability is the essential order parameter. On the other hand, for liquids with energy-driven structural ordering, such as water and silica, the size of locally favored structures cannot grow and be localized due to a significant entropy loss upon structural ordering. Lowering temperature increases only the fraction of locally favored structures ($s$) (without increasing their sizes), leading to an Arrhenius-Arrhenius crossover discussed above. In these liquids, the fraction of locally favored structures ($s$) is the order parameter. 
Finally, the intermediate behavior is expected for liquids in which both entropy and energy contribute to the free energy. This case of liquids with intermediate fragility remains unclear, and further research is necessary to reveal the origin of the slow glassy dynamics in such liquids. 

We have also found that entropy-driven structural ordering in a supercooled liquid exhibits critical-like fluctuations, whose growth leads to the increased activation energy for particle motion.  However, the origin of the Ising-like criticality and its link to slow dynamics remain elusive. This problem is a challenging but interesting subject for future theoretical study.

\section{Crystallization}~\label{sec:crystal}

\paragraph{Role of preordering of supercooled liquid in crystal nucleation}

\begin{figure*}[t]
\begin{center}
\includegraphics[width=16.cm]{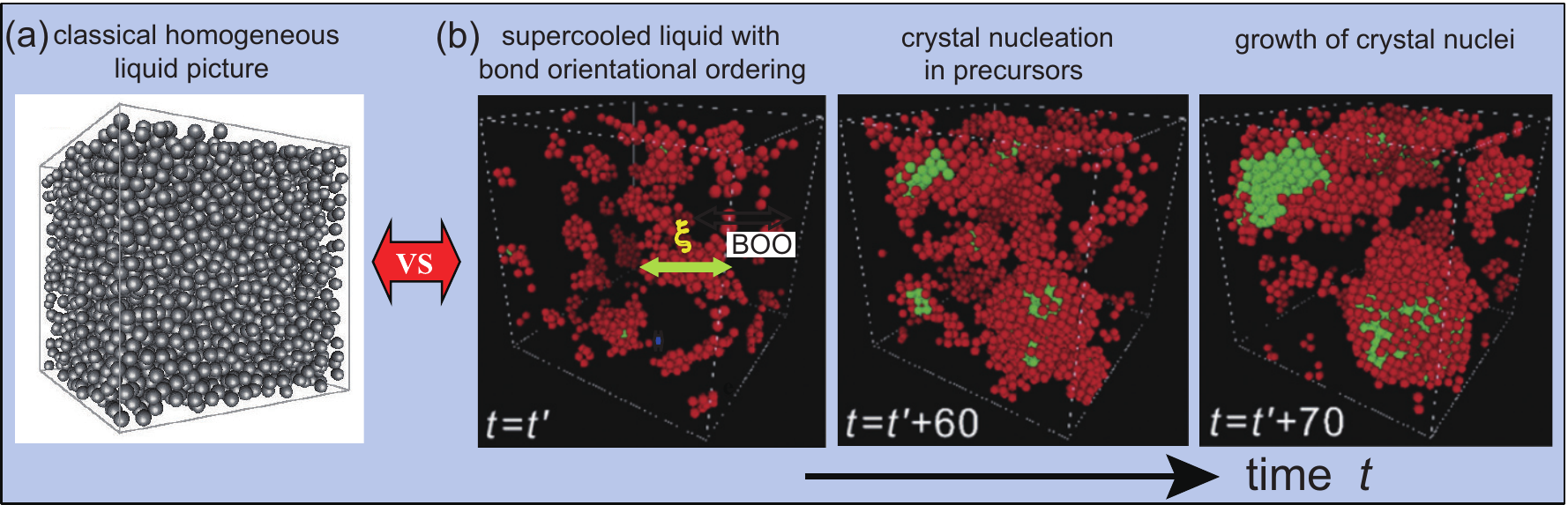}
\end{center}
\caption{
(a) A homogeneous liquid state before crystal nucleation, assumed in the classical nucleation theory and density functional theory.  
(b) A supercooled liquid state of a monodisperse hard-sphere-like liquid with transient high 6-fold bond orientational order ($Q_6$) regions (red particles)~\cite{kawasaki2010formation}, which act as precursors for crystal nucleation (left panel). As times go on, crystal nuclei (green particles) are formed exclusively in red precursor structures (middle panel) and grow with time (right panel). This figure is reproduced from Fig.~2 of Ref.~\cite{kawasaki2010formation}.
}
\label{fig:NG}
\end{figure*}

Crystallization is another important topic in the field of liquid science~\cite{kelton2010nucleation,sosso2016crystal}. 
Angell also made many pioneerinng experimental studies on crystal nucleation~\cite{dupuy1982controlled,kadiyala1984separation,angell1986crystallization,angell1988structural}.
As discussed above, we view glass transition as a consequence of frustration against crystallization. Then, as described above, we have found that in liquids suffering from weak frustration against crystallization,  
crystal-like orientational order develops in a supercooled state~\cite{kawasaki2007correlation,tanaka2010critical,kawasaki2010structural,leocmach2012roles}. 
This finding led us to the idea that preordering in the form of crystal-like orientational order acts as precursors for crystal nucleation~\cite{kawasaki2010formation,kawasaki2010structural} (see Fig.~\ref{fig:NG}(b)). 

On the other hand, in the classical nucleation theory and density functional theory of crystallization~\cite{DebenedettiB,kelton2010nucleation}, a supercooled liquid has been regarded as homogeneous (see Fig.~\ref{fig:NG}(a)), and thus, crystal nuclei have been assumed to be formed randomly in space. 
Contrary to this, we have confirmed in hard spheres~\cite{kawasaki2010formation,kawasaki2010structural,russo2012microscopic} that crystal nuclei are always formed in regions of high crystal-like bond orientational order formed in a supercooled state~\cite{russo2012microscopic,tanaka2012bond,russo2016crystal}. We emphasize that this structural ordering is an intrinsic feature of a supercooled liquid state not suffering from strong frustration against crystallization. 

Crystal nucleation begins with increasing the phase (angular) coherence of the crystal-like bond orientational order without accompanying density change. Later, the density change, i.e., translational ordering, comes into play only when the crystal nucleus becomes large enough to have its spatial periodicity. 
Such crystal nucleation behavior was directly confirmed by confocal microscopy experiments in colloidal systems by Tan et al.~\cite{tan2014visualizing}. 
We have found that preordering also plays a critical role in the polymorph selection by simulations~\cite{russo2012microscopic} and confocal microscopy experiments~\cite{li2020revealing}. 
Furthermore, we have confirmed that this scenario of crystal nucleation is valid for soft spheres~\cite{russo2012selection} and water~\cite{russo2014new}, indicating the universality of this scenario of crystal nucleation~\cite{russo2016crystal}.

The above scenario suggests that preordering compatible with the symmetry of a crystal to be formed promotes crystal nucleation since preordering lowers the liquid-crystal interface tension due to wetting effects~\cite{kawasaki2010formation}. 
We emphasize that the wetting effects are different from those in a scalar order parameter~\cite{tanaka2001interplay} since the relevant order parameter responsible for the effects is orientational. The effects are a consequence of symmetry matching between the precursor and crystal structures, leading to the polymorph selection due to preordering~\cite{russo2012microscopic,li2020revealing}.  
Such preordering also plays a critical role in heterogeneous crystal nucleation on the substrate~\cite{arai2017surface}.  
As mentioned above, we pointed out a possibility that competition between two types of crystals leads to the minimum melting point at the triple point~\cite{tanaka1998simple}, making a liquid structure more disordered and increasing the frustration against crystallization and thus glass-forming ability~\cite{tanaka2005relationship}.  
Angell and coworkers indeed found that this is the case for liquids with a modified Stillinger-Weber potential~\cite{molinero2006tuning}. 

\paragraph{Impact of the preorder-crystal similarity on glass-forming ability}

\begin{figure*}[t]
\begin{center}
\includegraphics[width=12.cm]{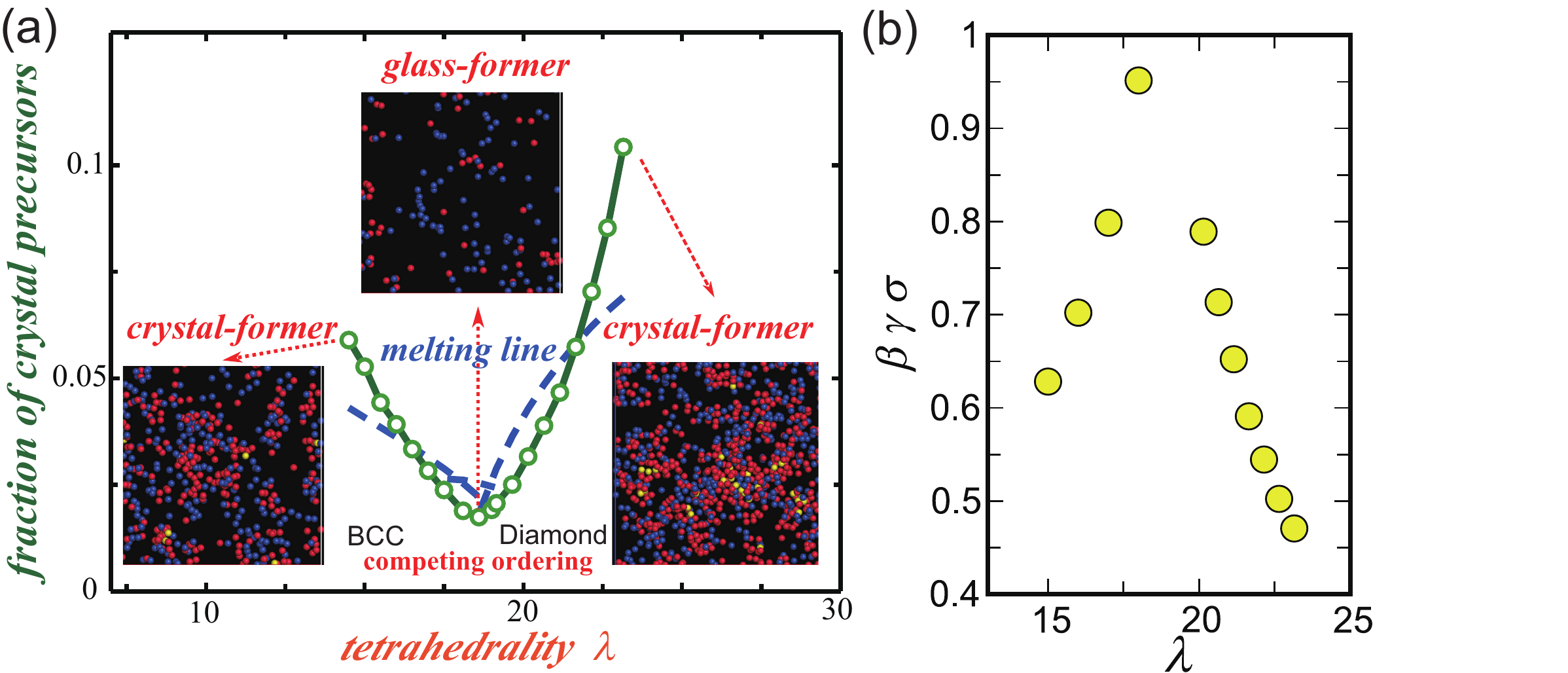}
\end{center}
\caption{(a) V-shaped phase diagram of liquids with the modified Stillinger-Weber potential, where the tetrahedrality $\lambda$ can be changed continuously. For low and high $\lambda$, body-centered cubic (BCC) crystal ($\rho$-crystal) 
diamond-cubic (DC) crystal ($S$-crystal) are formed, respectively. The melting point (blue dashed line) has a minimum at the triple point.  Here we also show the dependence of the fraction of crystal precursors (.i.e., crystal-like orientationally ordered particles) on $\lambda$ (green curve). The fraction of crystal precursors has a distinct minimum around the triple point, leading to high glass-forming ability. We also show three snapshots of crystal precursors from simulations. 
(b) The dependence of the scaled interface tension $\beta \gamma \sigma$ ($\gamma$: interface tension; $\sigma$: particle size). It has a steep peak at the minimum of the 
fraction of crystal precursors around the triple point (see (a)). 
Panels (a) and (b) are made from Fig. 4(a) and Fig. 3(a) of Ref.~\cite{russo2018glass}, respectively.
}
\label{fig:triple}
\end{figure*}

Recently, we have studied the physical origin of enhanced glass-forming ability near the melting point minimum, focusing on the degree of preordering in a supercooled state~\cite{russo2018glass}. 
According to the classical nucleation theory~\cite{kelton2010nucleation}, the crystal nucleation rate $I$ is obtained as 
\begin{equation}
I=f\frac{D}{D_0} \exp(-\beta \Delta F^\ast)\:,
\end{equation}
where $f$ is a numerical factor (which depends on terms like the Zeldovich factor~\cite{kelton2010nucleation}), $D/D_0$ is the adimensional diffusion constant,
$\Delta F^\ast$ is the free energy barrier that the system has to overcome in order to crystallize, 
and $\beta=1/k_{\rm B}T$. 
Here we note that the crystallization kinetics is controlled by the diffusion constant $D$ and not by the viscosity $\eta$,
which are decoupled for $T \leq T_{\rm m}$ (see, e.g., Refs.~\cite{tanaka2003possible,tanaka2012bond,ediger2008crystal}). 
It should be noted that the Stokes-Einstein relation is violated below the onset temperature $T_{\rm on}$. 
The fact that crystal nucleation is controlled by $D$ and not by $\eta$ can also resolve Kauzmann's paradox~\cite{tanaka2003possible,tanaka2012bond}. 

Now we consider the crystal nucleation based on our two-order-parameter model. 
The free-energy cost to form a crystal nucleus, of volume $V_n$ and interface area $S_n$, can be expressed in the adimensional form by scaling it with the thermal energy as follows~\cite{russo2018glass}:
\begin{eqnarray}
\beta\Delta F=-V_n(\beta \Delta \mu)+S_n (\beta \gamma),   \label{eq:Delta F}
\end{eqnarray}
where $\gamma$ is the liquid-crystal interfacial tension, and $\Delta \mu=\mu_{\rm liquid}-\mu_{\rm solid}$ is the chemical potential difference between solid and liquid, i.e., the driving force of crystallization.
It is important to note that the terms competing in the dimensionless free-energy cost are $\beta\Delta \mu$ and $\beta\gamma$, and not bare $\Delta \mu$ and $\gamma$ themselves (see Eq.~(\ref{eq:Delta F})). Thus, the nucleation rate should not be discussed by the absolute values of $\Delta \mu$ and $\gamma$ but by those scaled by the available thermal energy. 
This also allows us to make a unified description of the glass-forming ability for thermal and athermal (hard-sphere) systems. 
In this picture, $\beta \Delta F^\ast$ is expressed as 
\begin{equation}
\beta \Delta F^\ast = c (\beta \gamma)^d/(\beta \Delta \mu)^{d-1}, 
\end{equation}
where $d$ is the dimensionality and $c$ is a numerical factor depending on the shape of the nucleus.

We have shown~\cite{russo2018glass} that the macroscopic definition of the interfacial tension can be directly linked to the structural difference between the two phases via the general two-order-parameter Ginzburg-Landau theory, where the interfacial tension is obtained from the gradients of the order parameter fields~\cite{tanaka2012bond}:
\begin{equation}
\beta \gamma=\int dx \left[K_1 \left( \frac{d \rho}{dx} \right)^2+
K_Q \left( \frac{d Q}{dx} \right)^2 \right], \label{eq:grad}
\end{equation}
where $\rho$ and $Q$ are density and structural order parameters, respectively, $x$ is the coordinate perpendicular to the interface, and $K_\rho$ and $K_Q$ are positive coefficients associated with the spatial gradients of $\rho$ and $Q$ respectively. 
Since $Q$ is not scalar, the above relation is approximate but may be enough for a qualitative explanation.  
The above relation shows that not only $|\nabla \rho|$ but also $|\nabla Q|$ are essential in determining the interface tension. It also directly shows that the order parameter gradient, an intrinsic physical quantity characterizing the penalty associated with the structural difference between the two phases, is characterized by $\beta \gamma$ and not by $\gamma$. 

The nucleation rate $I$ is determined by $D/D_0$, $\beta \Delta \mu$, and $\beta \gamma$.  We have examined which of these three factors is dominant in determining $I$ and found that $\beta \gamma$ is predominant~\cite{russo2018glass}. 
Furthermore, $\beta \gamma$ is largely determined by the degree of crystal-like preordering $Q$ in a supercooled liquid (compare Figs.~\ref{fig:triple}(a) and (b)). That is, the suppression of crystal-like preorder due to competition between the two types of crystal orderings is responsible for a higher nucleation barrier, i.e., a higher glass-forming ability near the melting-point minimum. 
In the case of multi-component mixtures such as metallic alloy glass-formers, the composition in preorder also strongly affects the liquid-crystal interface tension and thus the glass-forming ability (see below)~\cite{hu2020physical}. 
This finding not only confirms the validity of the basic idea behind our two-order parameter model~\cite{tanaka1999two,tanaka1999two1,tanaka1999two2,tanaka2012bond} but also provides a simple physical explanation for the finding of the enhanced glass-forming ability near a eutectic point by Angell and coworkers~\cite{Angel1058_1970,kanno1977homogeneous,AngellR,angell2000glass,angell2008insights}. It is the crystal-liquid (preorder) similarity that controls the glass-forming ability. 
The generalization of the above theory incorporating the non-scalar nature of the angular order parameter $Q$, which is necessary to reveal the role of precursors in polymorph selection upon nucleation, remains for future research.  

\paragraph{Crystal nucleation and growth in metallic systems}
As an interesting example, we discuss the crystallization and glass-forming ability of monoatomic and multi-component metallic systems. In the field of metallic glasses, the glass-forming ability of metallic alloys is a critical issue~\cite{egami1997universal,johnson1999bulk,inoue2000stabilization}. 
As shown above, frustration against crystallization, or competing ordering, is crucial for controlling the fate of a liquid upon cooling. Concerning local atomic arrangements, there are typically two types of structural orderings in these metallic liquids. One is local icosahedral ordering (ICO), whose symmetry is incompatible with crystal. Its importance was first pointed out by Frank~\cite{frank1952supercooling}. The other is crystal-like bond orientational ordering (CRYO), which has the same local orientational symmetry as the equilibrium crystal(s).  CRYO tends to promote the formation of long-range density ordering, whereas ICO acts as the source of frustration, or impurities, against crystallization~\cite{shintani2006frustration,kawasaki2010formation}. The strength of the competition between ICO and CRYO or between different CRYO's determines the ease of crystallization. However, a too strong tendency of ICO may lead to the formation of a quasi-crystal~\cite{tanaka2003roles,tanaka2005relationship}. For Zr, for example, local icosahedral and bcc orders were reported to be competing by ab initio MD simulations~\cite{jakse2003local}. Therefore, both ICO and CRYO and their relationship should play a critical role in determining the glass-forming ability~\cite{tanaka2003roles,tanaka2005relationship,shintani2006frustration}. The presence of ICO in metallic liquids and alloys has been shown by many experimental observations~\cite{kelton_first_2003,xi_correlation_2007,hirata_geometric_2013}, and its role in glass formation has been widely studied for metallic glasses~\cite{jakse2003local,luo2004icosahedral,sheng2006atomic,shen_icosahedral_2009,cheng2011atomic,ding2014full,gonzalez2017competition}. 
However, the role of CRYO in determining the glass-forming ability and the relationship between ICO and CRYO has been elusive.

\begin{figure}[t!]
\centering
\includegraphics[width=0.45\textwidth]{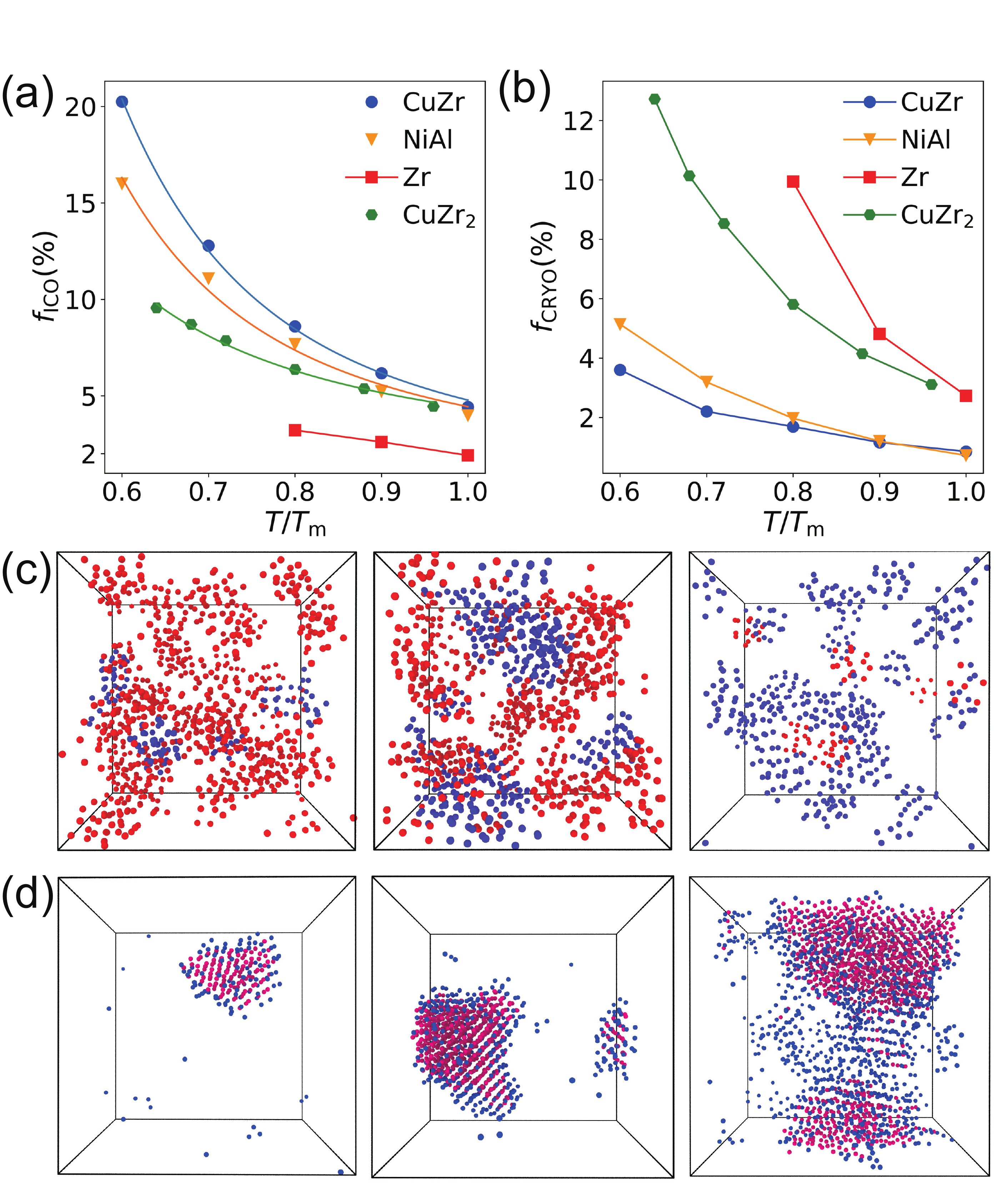}
\caption{Topological orderings in supercooled liquids.
(a) Temperature-dependence of the fraction of atoms in the icosahedral environments (ICO), $f_{\rm ICO}$, in the four systems, CuZr, NiAl, Zr, and CuZr$_2$. The solid lines for the data of CuZr, NiAl, and CuZr$_2$ are fits of the two-order-parameter model (see Eq.~(\ref{eq:s})). (b) Temperature-dependence of the fraction of atoms involved in crystal-like environments (CRYO). The order of the quantity and growth speed of $f_{\rm CRYO}$ is opposite to that of $f_{\rm ICO}$. (c) Snapshots of supercooled states with red atoms in ICO and blue atoms in CRYO environments at $0.6T_{\rm m}$ for CuZr (left) and  NiAl (middle), and at $0.8T_{\rm m}$ for Zr (right). (d) Snapshots of the spatial distribution of the crystalline phase (magenta atoms) and the precursors (i.e., CRYO) (blue atoms) for CuZr (left), NiAl (middle), and Zr (right). The degree of wettability of the precursors to the crystal increases with a decrease in GFA.}
\label{fig:metal}
\end{figure}

To address these issues, we have studied the crystallization kinetics of CuZr, NiAl, CuZr$_2$, and Zr by numerical simulations~\cite{hu2020physical}. 
We have compared the change of the fraction of ICO particles with decreasing temperature in Fig.~\ref{fig:metal}(a). The faction of ICO particles decreases in the order of CuZr, NiAl, CuZr$_2$, and Zr and increases more slowly in the same order upon cooling. In particular, Zr has little ICO. We have also confirmed that the temperature-dependence of ICO can be well described by the two-order-parameter model (see Eq.~(\ref{eq:s}))~\cite{tanaka1998simple_g,tanaka2003roles,tanaka2012bond,tanaka2020role} (solid curves in Fig.~\ref{fig:metal}(a)), which confirms the validity of this model also to metallic systems. The spatial distributions of ICO are also shown for CuZr, NiAl, and Zr in Fig.~\ref{fig:metal}(c). These results demonstrate that ICO in these metallic systems promotes glass formation and leads to topological frustration against crystallization~\cite{tanaka2003roles}. Microscopically, a larger amount of ICO particles causes a more substantial liquid-crystal structural contrast, resulting in larger $\beta \gamma$. We note that prevention of crystallization by ICO does not come from the $\beta\Delta \mu$ decrease, but the $\beta\gamma$ increase, as shown above.

Next, we turn our attention to CRYO. 
Figure~\ref{fig:metal}(b) shows how the fraction of CRYO atoms increases with lowering the temperature. The spatial distributions of CRYO are also shown in Fig.~\ref{fig:metal}(d) together with ICO for the three systems. The order of the degree of CRYO is the opposite of that of ICO. 
We can see that the faction of CRYO particles increases in the order of CuZr, NiAl, CuZr$_2$, and Zr, and increases much faster in the same order upon cooling. Zr has the highest fraction of CRYO, and its growth upon cooling is the fastest. In contrast, CuZr only has a tiny fraction of CRYO. Moreover, with decreasing $T/T_{\rm m}$, the spatial correlation of CRYO in Zr grows quickly, whereas it almost does not change in CuZr. 
Since CRYO has the local orientational symmetry same as the equilibrium crystal, its formation reduces $\beta\gamma$ and promotes crystallization. This suggests that suppressing CRYO should help with glass formation. 

Chemical ordering is another crucial factor controlling crystal nucleation in multi-component systems. The significance of composition fluctuations in crystal nucleation has been pointed out~\cite{desgranges2014unraveling,puosi2018dynamical,desgranges2019can,ingebrigtsen2019crystallization}. We have shown that the similarity between the composition in CRYO and the crystal to be formed plays a critical role in crystal nucleation~\cite{hu2020physical}.

We have also found that the crystal-growth rate is quite different among CuZr, NiAL, and Zr~\cite{hu2020physical}. The growth rate is slower in the order of CuZr, NiAl, and Zr. 
To account for the different kinetics among the three systems, we study how the amount of the precursors defined as CRYO in the liquid state affects the crystal-growth rate. For CuZr, the amount of CRYO around the nucleus is so low that it cannot wet the crystal nucleus, as can be seen in the left panel of Fig.~\ref{fig:metal}(d). Furthermore, the chemical composition of the preorder is different from that of the crystal for this system~\cite{hu2020physical}. The crystal phase cannot quickly grow because of the substantial structural and compositional differences between the liquid and crystal phases across the interfaces. In contrast, there is much more CRYO for NiAl (the middle panel of Fig.~\ref{fig:metal}(d)). In the early stage before the crystal grows, crystal precursors are formed while surrounding the critical nucleus. The fraction of CRYO particles becomes higher than that of crystalline particles. The amount of CRYO in Zr is even higher, and the surrounding liquid has a strong tendency to form crystal-like preordering. The chemical frustration is also absent in the monoatomic system. This fact indicates that the nucleus is thoroughly wet by CRYO during its growth (the right panel of Fig.~\ref{fig:metal}(d)). These results suggest that CRYO also plays a critical role in determining not only the nucleation kinetics but also the growth kinetics by tuning the properties of the liquid-crystal interfaces.

According to CNT, after the nucleation of crystals, their growth rate $U$ is given by  
\begin{equation}
U = k_0 D \left[ 1 - \exp\left(- \beta{\Delta \mu} \right) \right], 
\label{eq2}
\end{equation}
where $k_0$ is a constant. 
Among these liquids, the diffusion constant $D$ and $\beta\Delta \mu$ are quite similar~\cite{hu2020physical}. Thus, Eq.~(\ref{eq2}) predicts that the crystal growth rate should also be similar. Contrary to this, our study clearly shows that the crystal growth rates are significantly different among them~\cite{hu2020physical}. This result shows the severe failure of CNT in predicting the crystallization kinetics of metallic liquids at large supercooling. This failure may originate from ignorance of the preordering in the liquid phase near the crystal growth front in CNT. 
For example, the enhancement of the liquid phase's wettability to the crystal through preordering has been overlooked in the previous theories of crystal growth, including CNT. 
Our finding suggests that classical theory needs fundamental modifications to take liquid structural and chemical orderings into account. It may be of substantial importance not only for metallic alloys but also for other materials, including phase-change materials~\cite{wuttig2007phase,lencer2011design,greer_new_2015,wei2017glass,wei2017structural,
wei2019phase,persch2021potential}.

Concerning the above, we mention ultra-fast crystal growth observed in single atomic systems~\cite{greer_new_2015}. Recently, Sun and Harrowell have shown that such ultra-fast growth involves little to no activation process, which is linked to the magnitude of the displacements associated with the transformation to a crystal~\cite{sun2018mechanism,sun2020displacement}. We have recently shown that this fast crystal growth occurs via a diffusionless collective process in the presence of CRYO at the growth front~\cite{gao2021fast}. This is related to the fact that the transformation of CRYO to the corresponding crystal requires only tiny but cooperative displacements since CRYO already has the bond orientational order compatible with the crystal to be formed (see the inset of Fig. 2a of Ref.~\cite{russo2012microscopic}). Furthermore, we have found that under a particular condition, the intrinsic mechanical instability of a disordered glassy state subject to the crystal growth front allows for domino-like fast crystal growth even at ultra-low temperatures~\cite{gao2021fast}.

\paragraph{Perspective}
In the above, we discuss homogeneous crystal nucleation. Preordering also plays a significant role in heterogeneous nucleation~\cite{arai2017surface} through the coupling between the surface field and structural order parameter ~\cite{watanabe2011structural} and also in crystal growth at low temperatures~\cite{gao2021fast}. Remarkably, we have recently demonstrated that the preordering in the crystal-glass interface enables crystal growth even at $T=0$ (without thermal noises), i.e., crystallization due to mechanical instability~\cite{gao2021fast}.  The mechanical aspect of a glass state~\cite{yanagishima2017common,tong2020emergent,yanagishima2021towards} and its link to crystallization is also an interesting issue for future study. 

Here we should note that identification of local structures, e.g., by using bond orientational order parameters~\cite{steinhardt1983bond} is not necessarily reliable,  
Recently, machine learning has been used to detect the symmetry of a crystal structure more accurately, which may be helpful for a better description of crystal nucleation~\cite{spellings2018machine,boattini2019unsupervised,leoni2021nonclassical,becker2021unsupervised}.  

In the above, we have shown that preordering at the crystal growth front significantly influences crystal-growth kinetics. This fact has been overlooked in classical theory. Developing a theory of crystal growth is an important task for the future.

\section{Water's anomalies} \label{sec:water}
\paragraph{Research history of water's anomalies}
Water is the essential liquid familiar to us on the earth, but its physical properties are shrouded in mystery. Water has a peculiarity not seen in other molecular liquids. For example, it is widely known that water has the anomalous temperature- and pressure-dependences of various physical quantities such as the isothermal compressibility, specific heat, diffusion constant, and viscosity, as exemplified by the maximum density at 4$^\circ$C~\cite{eisenberg2005structure}. These anomalies were proposed to be explained by ``a mixture model'' or ``a continuum model''~\cite{eisenberg2005structure}. The former model assumes the state of water to be a mixture of a disordered liquid structure and a regular structure such as ice or clathrate and focuses on the temperature and pressure variations of these fractions, as proposed by R\"ontgen and Pauling. Many models are proposed along this line (e.g., Refs.\cite{davis1965two,angell1971two,ponyatovsky1998metastable,urquidi1999origin}). 
On the other hand, the latter model assumes the structure of water to be homogeneous and focuses on the continuous change of the liquid structure with the change of temperature and pressure, as proposed by Pople and others. 

\begin{figure}[t!]
\begin{center}
\includegraphics[width=8.cm]{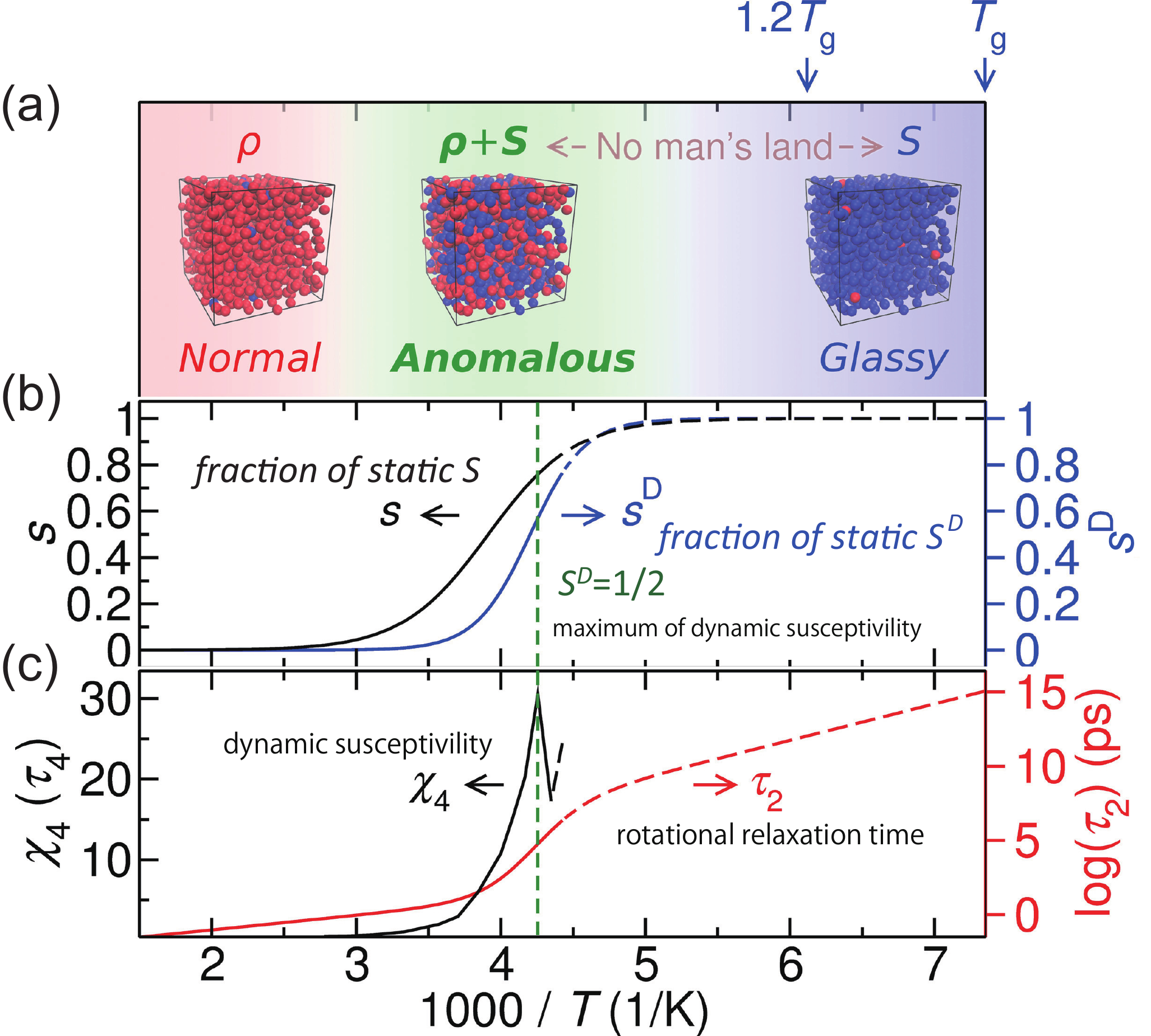}
\end{center}
\caption{Hierarchical two-state scenario of water's anomalies. (a) In the two-state scenario, the full available temperature range can be divided into three regimes: (1) At high temperature ($\rho$-state regime), water is predominated by a disordered state and behaves like normal liquids; (2) In the two-state regime, water is mainly a dynamic mixture of  $\rho$ and $S$ states, behaving anomalously; (3) Below the two-state regime ($S$-state regime), water is predominated by $S$-state and shows glassy behaviors. 
(b) Fractions, $s$ and $s^D$, of the thermodynamic (static) and dynamic $S$ states, respectively, as a function of inverse temperature. $s$ becomes 1/2 on the static Schottky line, which coincides with the Widom line possessing criticality. However, we stress that the peaks of the thermodynamic response functions, such as the isothermal compressibility and heat capacity, are produced by the two-state Schottky anomaly alone and do not require criticality. 
(c) Reorientational time $\tau_2$ in log scale and four-point susceptibility $\chi_4$ as a function of inverse temperature. At ambient pressure, dynamic $S$ state grows rapidly upon cooling, and their fraction reaches a value of 1/2 at $T_{s^{D}= 1/2}$, where $\chi_4$ maximizes, indicating the unique two-state behavior. When entering and leaving the two-state regime, dynamic quantities show two crossovers, centered by the dynamic Schottky line ($s^D =1/2$ line). The two-state regime is far from the glass transition temperature $T_{\rm g}$, and as a result, the dynamic slowing down, the dynamic heterogeneity, and the breakdown of the Stokes-Einstein-Debye relation all start far away from $T_{\rm g}$. This figure is reproduced from Fig.~5 of Ref.~\cite{shi2018origin}. 
}
\label{fig:two_state}
\end{figure}

However, this classical debate has been challenged by a new scenario introduced by Speedy and Angell~\cite{Speedy1976}, who focused on the critical-like power-law behaviors of thermodynamic quantities, such as the thermal expansion coefficient, isothermal compressibility, and heat capacity. This scenario is based on the retraction of the gas-liquid spinodal line at ambient pressure and is known as the retracting spinodal scenario. 
For confirming this scenario, it is crucial to trace the gas-liquid spinodal line at negative pressure. This is experimentally quite challenging but has been pioneered by Caupin and his coworkers~\cite{azouzi2013coherent,caupin2015escaping}.  
This physically plausible scenario has attracted renewed interest from not only chemists but also physicists. Many people have been attracted by such critical-like behavior of water, the most familiar liquid to us. 

The situation has been further changed primarily by a discovery of liquid-liquid transition, i.e., the second critical point in model water by Stanley and his coworkers~\cite{Poole1992} and also by an experimental discovery of two amorphous forms of water by Mishima and his coworkers~\cite{mishima1985apparently}, which has led to the second-critical point scenario (see, e.g., Ref.~\cite{Mishima1998}). 
These explanations of water's thermodynamic anomalies based on thermodynamic singularities have become quite popular, having contributed to the acceleration of the research of liquid science, including not only water but also liquid-liquid transition~\cite{Debenedetti2003,AngellwaterR,gallo2016water}. We note that Angell and his coworkers also developed a thermodynamic model based on a van-der Waals model, incorporating directional hydrogen bonding, to explain both water's anomalies and liquid-liquid transition~\cite{poole1994effect}.

The singularity-based scenarios are attractive and intriguing for physicists. 
The retraction of the gas-liquid spinodal line is physically possible. However, the criticality associated with a spinodal line is usually not strong. 
Only for a mean-field-like system do we expect strong criticality towards the spinodal line. We note that according to the Ginzburg criterion~\cite{OnukiB}, long-range interactions are necessary for such mean-field-like behavior. Although inter-water interactions are relatively long-range, it is unclear whether they are long enough to induce the mean-field behavior. Furthermore, once passing the gas-liquid spinodal line upon cooling, density fluctuations spontaneously should grow to form the gas phase. However, such behavior has not been reported 
at least at positive pressures. 

The second-critical point scenario for water's anomaly is also appealing. 
I thought there might be a second critical point associated with liquid-liquid transition, but I did not believe that the origin of water's anomalies is criticality. 
From my own research experience of critical anomalies of binary mixtures~\cite{tanaka1982acoustic,tanaka1985theoretical}, I learned that the critical anomalies are observed only in the vicinity of a critical point, and there should also be hyperscaling relations between the critical exponents~\cite{OnukiB}. The suggested location of the second critical point was too far away from the experimentally accessible temperature-pressure range, and there were no such hyperscaling relations between the power-law exponents obtained by power-law fittings of various physical quantities showing the anomalies. We point out that the estimation of the non-critical background contributions has not been done reliably in most previous analyses.

\paragraph{Two-order-parameter model of water}

Based on the above considerations on the weakness of the singularity-based scenarios, I thought that water's anomalies originate from local structural ordering similarly to the two-order-parameter model for glassy anomalies. Thus, I proposed a simple two-order-parameter (or two-state) model~\cite{tanaka1998simple,Tanaka2000,Tanaka2000a}. Since R\"ontgen, a mixture model has been popular, and various related models have been proposed.  Angell also proposed such a model~\cite{angell1971two}. Previous mixture models regard water as a mixture of two distinct structural units, thus ignoring a large difference in the degeneracy between the two states and the resulting large entropy change associated with their transformation. This has led to the overestimation of locally favored structures in most previous mixture models; for example, at ambient condition, the fraction of ordered structures were often estimated nearly or higher than 50\%~\cite{davis1965two,angell1971two,ponyatovsky1998metastable,urquidi1999origin}. 
We also stress that it is physically inappropriate to interpret water as a mixture of low-density (LDL) and high-density (HDL) liquids (e.g., Refs.~\cite{nilsson2015structural,johari2015thermodynamic}) or a mixture of low-density amorphous (LDA) and high-density amorphous (HDA) water (e.g., Ref.~\cite{ponyatovsky1998metastable}) (see Ref.~\cite{tanaka2020liquid} concerning the reasons). 
This overestimation of the fraction of ordered structures may partly come from the fact that if we simply decompose water into two components using isosbestic points of structural~\cite{urquidi1999origin} and spectroscopic data~\cite{walrafen1986raman}, the estimated fraction of the ordered structures becomes considerably high.

I proposed a simple two-state model to explain water's anomalies, putting a particular focus on a significant loss of configurational entropy upon the formation of tetrahedral ordering upon hydrogen-bonding~\cite{tanaka1998simple,Tanaka2000,Tanaka2000a}. 
I introduced the structural order parameter $S$ to represent tetrahedral orientational ordering and the density order parameter $\rho$ to represent density ordering (note that $S$ does not represent entropy). 
The relevance of the two-order-parameter description was supported by quantitative characterization of orientational and translational order in model water~\cite{Errington2001}. 

We also derived the equation of motion for these two order parameters while treating the order parameter $S$ as a non-conserved order parameter~\cite{tanaka1999two,tanaka2000general,takae2020role}. Note that locally favored structures can be formed and annihilated locally, and their number density is thus not conserved. The two structures exchange dynamically in a liquid state. The set of dynamic equations should be generally valid for describing liquid dynamics with locally favored structures, including liquid-liquid transition~\cite{tanaka2020liquid}. 

In this model, locally favored structures (LFTS) are tetrahedral structures stabilized by hydrogen bonding and having only small structural fluctuations, whereas normal-liquid structures (DNLS) have many disordered structures with large structural fluctuations. 
In other words, the formation of LFTS from DNLS is accompanied by a significant loss of the configurational entropy. Thus, our two-state model predicts a much smaller fraction of locally favored structures than conventional mixture models considering only the mixing entropy and neglecting the configurational entropy change associated with the transformation between the two components. For example, our model predicts that the fraction of locally favored tetrahedral structures is about 10\% at ambient conditions. We also note that LFTS cannot extend spatially because of the significant loss of the configurational entropy to form LFTS from DNLS. This rationalizes the assumption that only the fraction of LFTS increases without its size growth upon cooling. 

\begin{figure*}[tbh]
\begin{center}
\includegraphics[width=12.cm]{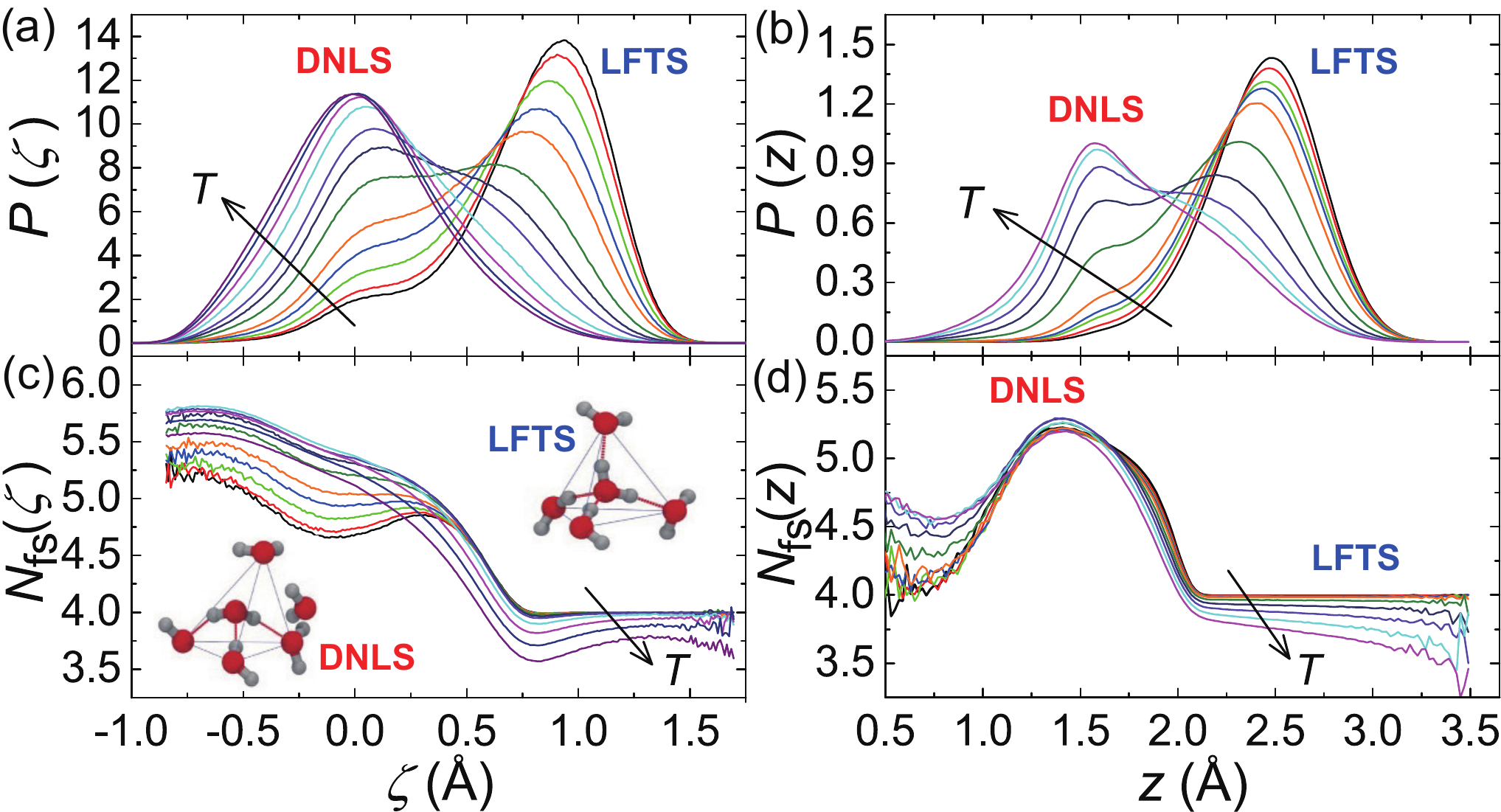}
\end{center}
\caption{Evidence of the two-state features revealed by structural index detecting the translational order of the second shell.  
(a) The $\zeta$ distribution of TIP5P water. (b) The $z$ distribution of BKS silica. 
For both (a) and (b), we can see distinct bimodal distributions, and the fraction of LFTS, $s$, increases with decreasing $T$.
(c), (d) The distributions of the number of the first shell oxygen and silicon atoms, $N_{\rm fs}$ for water and silica, respectively.  
These figures are made from Fig. 1 of Ref.~\cite{shi2018impact}.
}
\label{fig:water_silica}
\end{figure*}

The free energy of a non-ideal mixture of the two types of local structures (two states) can be written as~\cite{tanaka1998simple,Tanaka2000,Tanaka2000a}:
\begin{eqnarray}
G=G_{\rho} + s \Delta G + k_{\mathrm{B}} T \left[s \ln s + (1-s) \ln (1-s)\right] \nonumber \\
+ Js(1-s), \nonumber 
\end{eqnarray}
where $G_\rho$ is the free energy of $\rho$-liquid made of DNLS, $\Delta G$ is the free-energy difference between LFTS and DNLS, $s$ is the fraction of LFTS, $J$ is the strength of coorperativity of tetrahedral ordering, and $k_{\mathrm{B}}$ is the Boltzmann constant. $\Delta G$ can be further written as $\Delta G=\Delta E - T \Delta \sigma + P \Delta V$, where $\Delta E$, $\Delta \sigma$, and $\Delta V$ characterize the energy, entropy, and volume differences between LFTS and DNLS, respectively. The coexistence of the two states reaches equilibrium if $\partial G / \partial s=0$, which leads to the following relation:
\begin{equation}
\frac{\partial G}{\partial s}=\Delta G+k_{\mathrm{B}}T\mathrm{ln}\left(\frac{s}{1-s}\right)+J\left(1-2s\right)=0.
\label{eq:dGds}
\end{equation}
Then, the fraction $s$ of LFTS can be obtained analytically from Eq.~(\ref{eq:dGds}) if the coorperativity is negligible ($J \sim 0$; see below) or $s \ll 1$~\cite{tanaka1998simple,Tanaka2000,Tanaka2000a}:
\begin{equation}
s \cong \frac{1}{1+\exp \left( \frac{\Delta E+J - T \Delta \sigma + P \Delta V}{k_\mathrm{B}T}\right)}.
\label{eq:s}
\end{equation}
In the case of $s \ll 1$, which is roughly satisfied in the experimentally accessible temperature range at ambient and high pressures, we have the following very simple approximate expression for $s$:
\begin{equation}
s \sim \exp \left( \frac{-\Delta E -J+ T \Delta \sigma - P \Delta V}{k_\mathrm{B}T} \right).
\label{eq:Boltz}
\end{equation}

This model predicts the fraction of LFTS $s$ increases with decreasing $T$, following Eq.~(\ref{eq:s}) (see Fig.~\ref{fig:two_state}(a)). 
Liquids at high and low-temperature limits are pure $\rho$-state and $S$-state, respectively. A liquid is a dynamic mixture of $\rho$-state and $S$-states in the intermediate temperature range. 
As will be discussed later, the dynamic quantities such as the diffusion constant and viscosity cannot directly be described by the static order parameter $s$, but by the dynamical version (see Figs.~\ref{fig:two_state}(b) and (c)).

The anomalous parts of thermodynamic quantities such as the density, isothermal compressibility, and heat capacity can be commonly expressed by Eq.~(\ref{eq:s}), and further approximated by the common Boltzmann factor Eq.~(\ref{eq:Boltz}), the simplest mathematical expression for the water's anomalies, as long as $s \ll 1$. 
It has been shown that the Boltzmann factor can indeed describe the experimental results of water's anomalies well. Thus, we have argued that our two-state model can explain water's anomalies without invoking thermodynamic singularities~\cite{tanaka1998simple,Tanaka2000,Tanaka2000a}. 
Later, a similar two-state mode was used by Anisimov and his coworkers~\cite{Holten2012,Singh2016two,Biddle2017} and Nilsson and Pettersson~\cite{nilsson2015structural} to analyze the water's anomalies. However, they included the effects of the second critical point and emphasized the importance of criticality, unlike us. 
A possibility of entropy-driven cooperative ordering was also proposed for water~\cite{Holten2012}. 
Recently, Caupin and Anisimov~\cite{caupin2019thermodynamics} also discussed the relation between the two-state description and gas-liquid spinodal in supercooled and stretched (negative pressure) water in detail.  
This relation has also been discussed by Angell and Kapko~\cite{angell2016potential} and us~\cite{russo2018water}, using the modified Stillinger-Weber model. 

We have shown that the two-state model fittings nicely describe the thermodynamic anomalies of water~\cite{tanaka1998simple,Tanaka2000,Tanaka2000a}. 
However, there has been no clear experimental and numerical evidence for the existence of the two states. 
It was challenging to determine the water structure due to large thermal fluctuations characteristic of a liquid state. Therefore, the problem has been regarded as an unresolved issue in the field of liquids for a long time.

\paragraph{Numerical evidence for the two-state feature}

\begin{figure*}[tb]
\begin{center}
\includegraphics[width=11.cm]{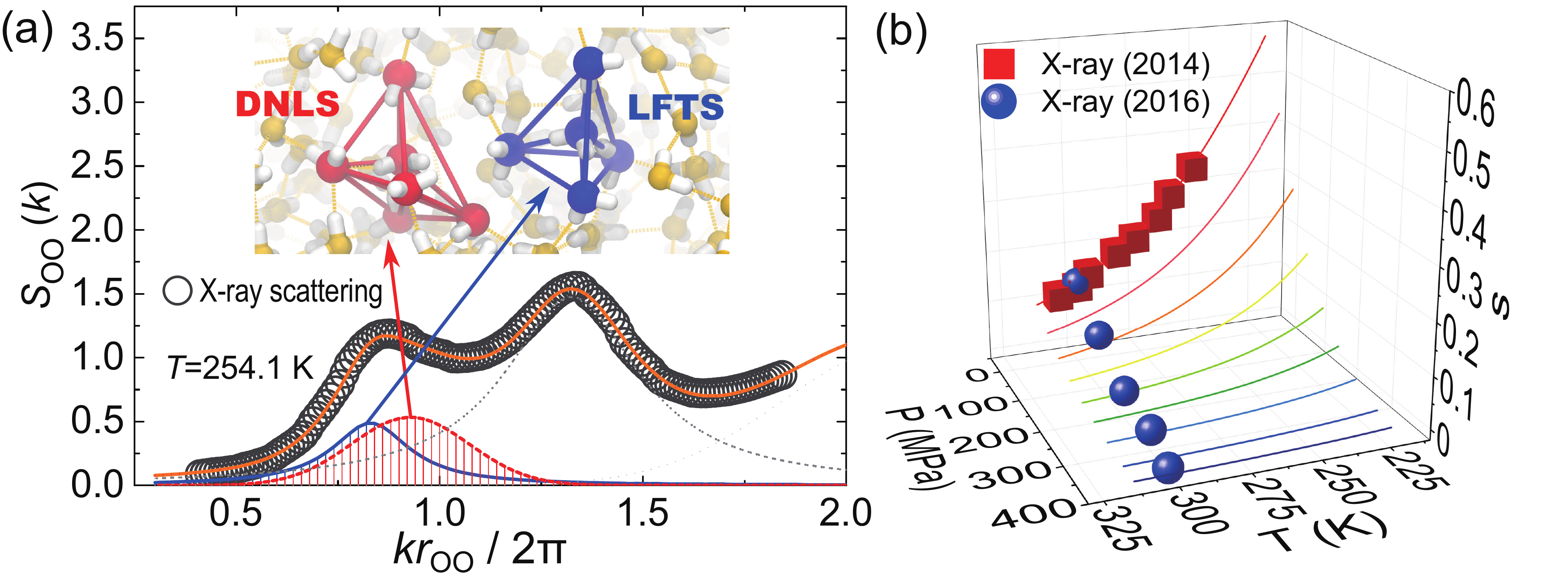}
\end{center}
\caption{(a) Decomposition of the first peak of the oxygen-oxygen partial structure factor $S_{\rm OO}(k)$ of water to the FSDP from locally favored tetrahedral structures (LFTS), i.e. $S$-state, and the peak from disordered norlmal liquid structures 
(DNLS), i.e., $\rho$-state. (b) The fraction of LFTS, $s$, of liquid water directly estimated from the experimental X-ray scattering data as a function of temperature $T$ and pressure $P$. The solid lines are the two-state fits by Eq.~(\ref{eq:s}). 
Panels (a) and (b) are made from the cover image and Fig. 1B of Ref.~\cite{shi2020direct}, respectively. 
}
\label{fig:FSDP}
\end{figure*}

\begin{figure*}[thb]
\begin{center}
\includegraphics[width=11.cm]{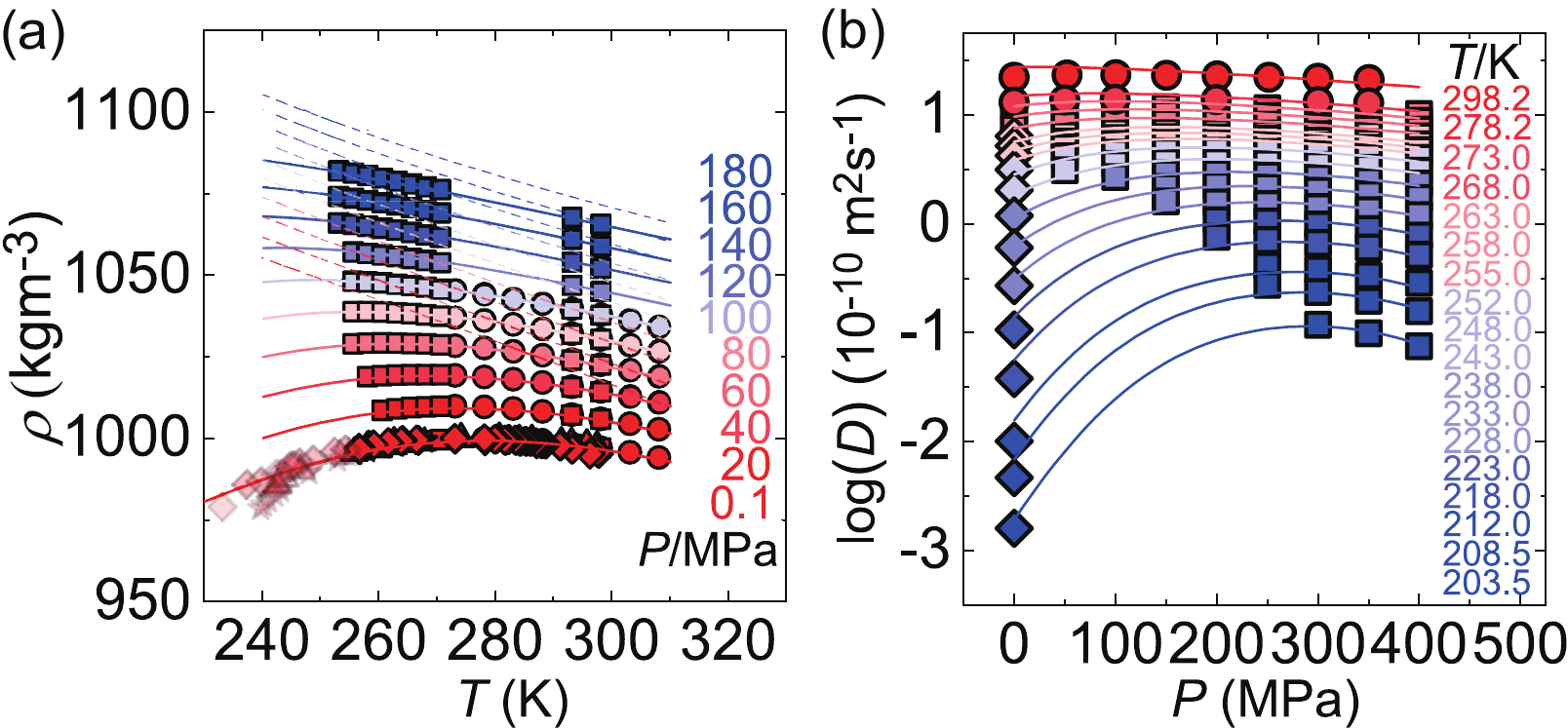}
\end{center}
\caption{Two-state-model analysis of experimentally measured thermodynamic and dynamic data of liquid water. (a) $T$-dependence of density $\rho$. The symbols represent experimental data, and the solid curves are the two-state model predictions. The dash curves denote the background contributions from DNLS. The blue and red colors represent higher and lower pressures, respectively. The pressure of each isobar is shown on the right side.
In this fitting, the value of $s$ is determined independently from the FSDP analysis of X-ray scattering data (see Fig.~\ref{fig:FSDP}).  
(b) The $P$-dependence of the diffusion coefficient $D$ for various $T$. Symbols are experimental data, and the curves are fits by the hierarchical two-state model.  
The figures (a) and (b) are reproduced from Fig. 2A and Fig. 3B of Ref.~\cite{shi2020anomalies}, respectively. See Ref.~\cite{shi2020anomalies} on the details.
}
\label{fig:fit}
\end{figure*}

Under these circumstances, we have introduced a new structural index $\zeta$ for the structural analysis of simulated classical water models, which measures the local translational order in the second shell~\cite{Russo2014}. 
Many structural descriptors have been developed to characterize water's local structure. Among them, focusing on the second shell of nearest neighbors~\cite{soper2000structures}, descriptors such as the local structure
index (LSI)~\cite{shiratani1998molecular} and $d_5$~\cite{cuthbertson2011mixturelike} have been introduced. However, these descriptors did not consider hydrogen bonding. 
The key to developing the structural index $\zeta$ was to recognize the importance of hydrogen bonding to analyze the water arrangement in the neighboring shells around a water molecule (see Ref.~\cite{shi2018Microscopic} on the detailed comparison between these descriptors). 
We have also introduced a similar structural index $z$ to analyze the liquid structure of silica~\cite{shi2018impact}.
We have shown that the liquid structures of water and silica show distinct bimodal distributions of these structural indices $\zeta$ and $z$ (see Fig.~\ref{fig:water_silica}(a) and (b)) and that the temperature and pressure dependences of the fractions of the two structures are in almost perfect agreement with the theoretical predictions of the two-state model mentioned above~\cite{tanaka1998simple,Tanaka2000,Tanaka2000a}. 
Thus, we have succeeded in revealing the tetrahedral structure hidden in the fluctuating liquid structure of water~\cite{Russo2014,shi2018Microscopic} and silica~\cite{shi2018impact}.

Angell and Kanno~\cite{angell1976density} pointed out the similarity between water and silica, but at the same time showed that the sharpness of the density maximum of water is much sharper than that of silica. 
We also note that silica also shows polyamorphism (see, e.g., Ref.~\cite{huang2004amorphous}) as water does~\cite{mishima1985apparently,loerting2006amorphous,bachler2021experimental}.
We have revealed~\cite{shi2018impact} that the similarity comes from the common two-state feature associated with the formation of  LFTS (see Fig.~\ref{fig:water_silica}(a) and (b)), whereas the difference stems from a much higher degree of the internal structural order of LFTS in water than silica. This difference originates from the following reason: The H-O-H angle in water is almost fixed by intramolecular covalent bonds much stronger than intermolecular hydrogen bonds, whereas the O-Si-O angle in silica largely fluctuates. In other words, H$_2$O is a rigid molecule, but SiO$_2$ is just a mixture of atoms in reality.   
We have also shown that this difference in the degree of the internal structural order of LFTS may also be responsible for the drastic difference in glass-forming ability between these two systems~\cite{shi2018impact}. 
The high order of LFTS makes the liquid structure more similar to the crystal, lowering the liquid-crystal interface tension. 

As shown in Figs.~\ref{fig:water_silica}(c) and (d), LFTS is characterized by four first neighbors bonded to the central atom, whereas DNLS has five or six neighbors in the first shell for both water and silica. 
The thermodynamic anomaly of water has been considered to be described by the power-law, $(T-T_{\rm s})^{-a}$, associated with the singularity at $T_{\rm s}$ in the second critical point scenario of Stanley et al.~\cite{Mishima1998} and the retracting spinodal scenario by Speedy and Angell~\cite{Speedy1976}. On the other hand, we have argued that the water's anomalies are explained by a two-state model and described by an exponential function (i.e., the Boltzmann factor)~\cite{tanaka1998simple,Tanaka2000,Tanaka2000a}. 
The above studies support our scenario~\cite{tanaka1998simple,Tanaka2000,Tanaka2000a}. For example, the density decrease below 4$^\circ$C with decreasing temperature can be quantitatively explained by the increase in the fraction of LFTS ($s$) of the bulky tetrahedral structure of a large specific volume (see below). We can explain the thermodynamic anomalies such as the isothermal compressibility and specific heat similarly. For example, the transformability between LFTS and DNLS gives water extra-softness, which is maximized when their fractions are equal (i.e., $s=1/2$), leading to the peak of the isothermal compressibility: the Schottky anomaly characteristic of the two-state model. 

We have also shown that kinetic anomalies such as a steep increase in the viscosity with decreasing temperature and the violation of the Stokes-Einstein-Debye relation can also be explained in a unified manner by the coarse-grained structural order parameter $s_{\rm D}$, in which the structural order parameter $s$ is averaged up to the nearest neighbor particles~\cite{shi2018origin,shi2018common}, as discussed in Sec.~\ref{sec:glass}. The reason why coarse-graining is necessary for the description of dynamical properties is that the motion of molecules requires the exchange of configurations with neighboring particles, which involves the recombination of hydrogen bonds. We have shown that a water molecule locally having a tetrahedral structure around it can move with the reconnection of hydrogen bonds only when at least two water molecules without tetrahedral order (i.e., defects in our context) exist in its neighbors~\cite{shi2018origin,shi2018impact}. 
The popular theory of dynamical anomalies claimed that the anomalies are described by the power-law predicted by the mode-coupling theory based on the glass transition~\cite{xu2005relation} or by criticality~\cite{Mishima1998}. In contrast, the fact that the dynamic fluctuation (dynamic susceptibility) exhibits its maximum at the temperature where the fraction of the two dynamic states becomes 1/2, i.e., the dynamic Schottky line ($s^D=1/2$), and not at the Widom line ($s=1/2$) (see Fig.~\ref{fig:two_state}) indicates that the dynamic anomaly and the violation of the Stokes-Einstein law are caused by neither glass transition nor criticality, contrary to the popular scenarios. 

\paragraph{Experimentally accessible evidence for the two-state feature}
As described above, we have shown convincing numerical evidence for the two-state feature in model waters through introducing the structure index $\zeta$. 
However, since the structure index $\zeta$ is defined by microscopic local structures, it cannot be applied to experimental systems. Accordingly, the controversy concerning both thermodynamic and dynamic anomalies of water could not be settled experimentally. The key is to find a way to determine the existence of the two states and their fractions `experimentally'. 

To do so, we have focused on the so-called First Sharp Diffraction Peak (FSDP), which is a scattering peak that appears in the structure factor $S(k)$ of many tetrahedral liquids (e.g., silica and silicon) at wavenumbers corresponding to distances longer than the average interatomic distance. The origin of this peak has remained unresolved and a matter of controversy over the years. 
Recently, we have discovered for silica that FSDP is caused by the density wave with a wavelength corresponding to the height of the tetrahedral structure~\cite{shi2019distinct}. This finding strongly suggests the existence of a similar FSDP in the structure factor of water.
We found through molecular dynamics simulations that the density wave associated with the tetrahedral structure leads to the appearance of FSDP also in water~\cite{shi2020direct}, as in silica~\cite{shi2019distinct}. Specifically, we have proven that there is a peak (FSDP) whose wavelength corresponds to the height of a tetrahedral structure, in the partial structure factor of oxygen (see Fig.~\ref{fig:FSDP}). We stress that this peak can be directly identified by X-ray and neutron scattering. We have also shown by numerical simulations of model waters that the intensity of FSDP is proportional to the fraction $s$ of the locally favored tetrahedral structures, which we have independently determined using the molecular-level structure index $\zeta$~\cite{shi2020direct}. These results provide firm evidence for the validity of our two-state model based on both experimental and simulation data. Thus, we conclude that liquid water is a dynamic mixture of DNLS and LFTS. 

Decomposing the first peak of the O-O partial structure factor obtained by X-ray scattering experiments into FSDP and the normal-liquid first peak, we have determined the fraction of LFTS, $s$, and found that $s$ indeed follows Eq.~(\ref{eq:s}) for actual water. This allows us to explain water's anomalies without any adjustable parameters to express $s$. We have shown that our two-state model in which $s$ is determined experimentally can explain water's thermodynamic and dynamic anomalies satisfactorily (see Fig.~\ref{fig:fit})~\cite{shi2020anomalies}. 

The existence of FSDP in water was not noticed until now because (1) FSDP of water is not observed as an independent peak but hidden in the first peak (see Fig.~\ref{fig:FSDP}(a)) and (2) the origin of FSDP itself was not known until we clarified it. 

We expect that this discovery of FSDP may close more than a century of a long-standing debate over the liquid water structure between the mixture and continuum models~\cite{eisenberg2005structure}, which involved Nobel laureates, R\"ontgen, Pauling, and Pople.

\paragraph{The second critical point and its influence on water's properties}

\begin{figure}[t]
\begin{center}
\includegraphics[width=7.0cm,clip]{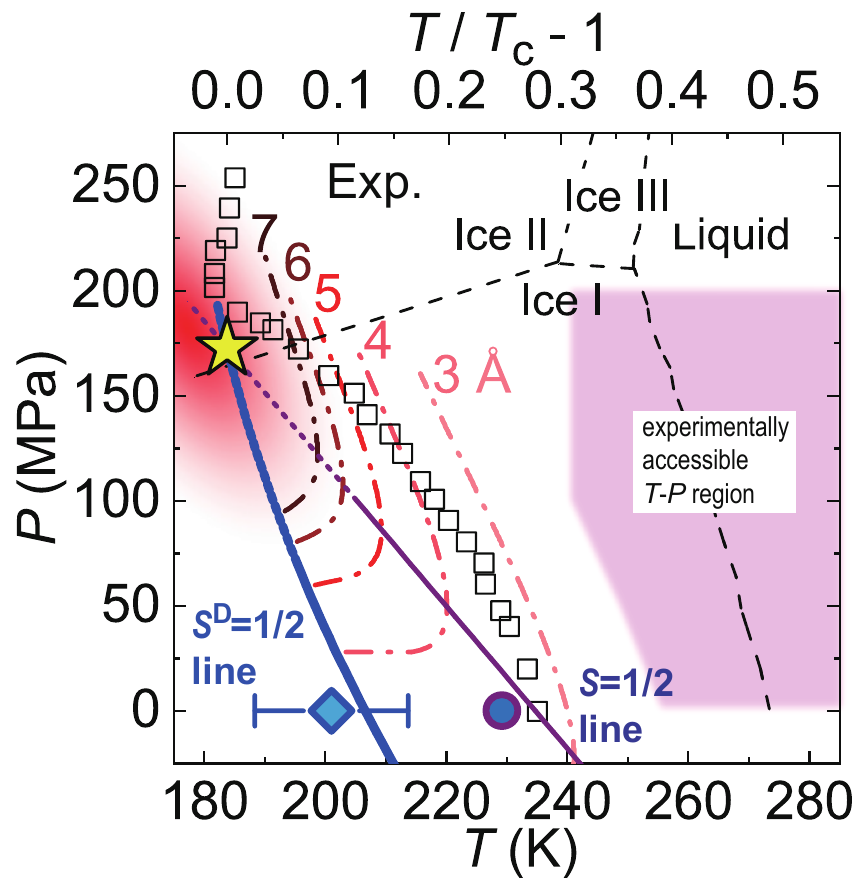}
\end{center}
\caption{The state diagram of real liquid water determined from the experimental data. The static and dynamic Schottky lines, the lines of $s(T,P)=1/2$ (thin
purple curve) and $s^D(T,P)=1/2$ (thick blue curve), respectively, meet at the liquid-liquid critical point (LLCP) (yellow star). Its location is estimated to be around 184~K and 173~MPa. The circle is the isothermal compressibility maximum obtained by X-ray scattering, and the diamond is the maximum of the rate of dynamic slowing down~\cite{shi2020anomalies}. The black dash curves represent the phase boundaries, and the squares are homogeneous nucleation temperatures. Red dot-dash curves display the contour lines of the correlation length of 3-7~\AA. The magenta region represents the $T$-$P$ range, where the experimental thermodynamic data are available without ice nucleation. The correlation length exceeds twice of the molecule size only in the vicinity of the LLCP (red region), and rapidly decays to the molecule scale at $(T-T_c)/T_c > 0.3$, and thus, becomes negligible in the experimentally accessible $T$-$P$ region (magenta region).
This figure is reproduced from Fig. 5C of Ref.~\cite{shi2020anomalies}. 
}
\label{fig:critical}
\end{figure}

Recently, the presence of the second critical point has been shown convincingly by Debenedetti and his coworkers for ST2 water~\cite{palmer2014metastable}, two realistic water models, TIP4P/2005 and TIP4P/Ice~\cite{debenedetti2020second}, and an ab initio deep neural network model~\cite{gartner2020signatures} (see also Ref.~\cite{palmer2018advances} for review). 
There have also been reported experimental indications for the second critical point~\cite{handle2017supercooled, kim2017maxima, kim2020experimental,pathak2021enhancement}. So, the presence of the second critical point in actual water has become more convincing. However, it should be noted that there have been some debates on the interpretation of recent X-ray scattering experiments on density fluctuations in supercooled water~\cite{caupin2018comment,kim2018response}. 
On this issue, we stress that the peak of the isothermal compressibility is not specific to the Widom line (i.e., criticality), but can be simply a result of a two-state feature that we call the Schottky anomaly~\cite{tanaka2012bond}: Even without any criticality, the isothermal compressibility and the heat capacity can have maxima on the Schottky line solely from the two-state feature. 
Thus, an analysis without considering the contribution of the Schottky anomaly should inevitably result in a wrong conclusion. 
We also note that there have been debates on the nature of the two forms of amorphous water, LDA and HDA (see, e.g., a recent review~\cite{handle2017supercooled}).

We have also proposed an original method to identify the existence and location of the second critical point of water~\cite{shi2020anomalies}. First, we systematically analyzed structural, thermodynamic, and dynamical experimental data based on our hierarchical two-state model~\cite{shi2018origin,shi2018common}. This model predicts that if a second critical point exists, the static and dynamic Schottky lines (see the red and blue lines in Fig.~\ref{fig:critical}) where the static and dynamic fluctuations are maximized, respectively, on the temperature-pressure phase diagram will intersect at the second critical point (see Fig.~\ref{fig:critical}). We have confirmed the validity of this prediction by simulations for a few classical models of water. Furthermore, based on experimental measurements of the isothermal compressibility and diffusivity at low temperatures, it was shown that the above two lines intersect at a temperature of 184~K and a pressure of 173~MPa in actual water, i.e., the second critical point is likely to exist in this vicinity (see the yellow star symbol in Fig.~\ref{fig:critical})~\cite{shi2020anomalies}. This prediction is expected to provide an essential guideline for future experimental searches for the second critical point. 

This result, together with the recent numerical and experimental results mentioned above, strongly suggests the presence of the second critical point in the water. However, at the same time, from the knowledge of basic critical phenomena~\cite{OnukiB}, the experimentally reachable liquid region of water (the pink region in Fig.~\ref{fig:critical}) is too far from the second critical point. Thus, the criticality associated with the second critical point can be almost neglected even if it exists (see the correlation length (single-dotted line) in Fig.~\ref{fig:critical}, which is comparable to the bare correlation length $\xi_0$, i.e., the size of LFTS, in the experimentally reachable liquid region)~\cite{shi2020anomalies}. This fact, together with the above finding for FSDP~\cite{shi2020direct}, strongly suggests that the water's anomalies can be quantitatively described by our two-order-parameter (two-state) model regarding water as a dynamic mixture of two exchangeable microscopic states~\cite{tanaka1998simple,Tanaka2000,Tanaka2000a}. The existence of the two critical points in water is also naturally explained by the existence of the two scalar order parameters, $\rho$ and $S$, for the gas-liquid and liquid-liquid critical points respectively~\cite{tanaka1999two,tanaka2000general,Tanaka2000a,tanaka2020liquid}. Concerning the liquid-liquid transition of water, we can understand it quite naturally as a consequence of the cooperative formation of locally favored tetrahedral structures (LFTS) represented by the $J$ term in Eq.~(\ref{eq:Delta F}) in the framework of our two-order-parameter model~\cite{Tanaka2000a,tanaka2000general}. However, the microscopic origin of the cooperativity remains a task for future investigation. 

\begin{figure}[t!]
\begin{center}
\includegraphics[width=7.0 cm]{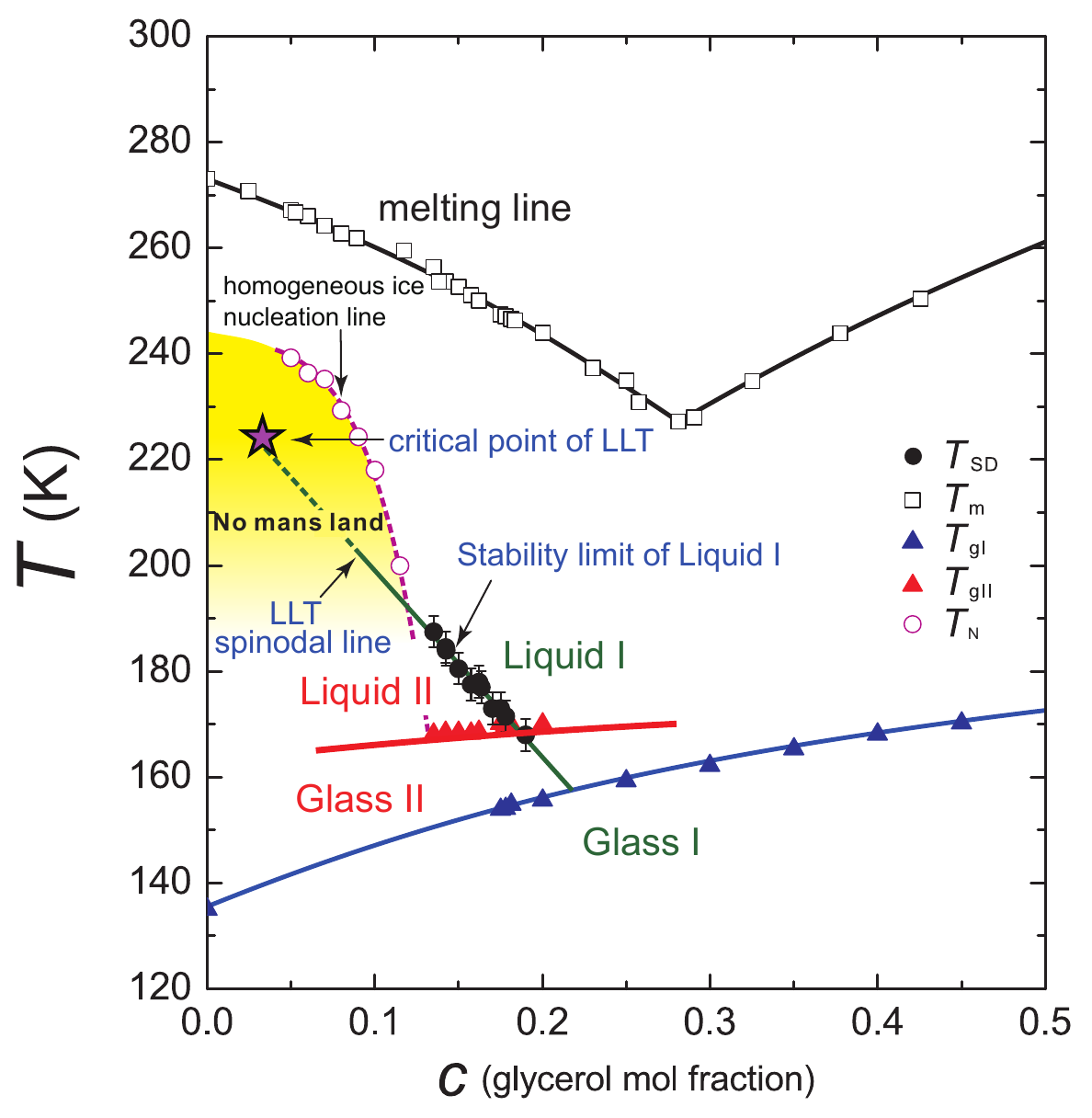}
\end{center}
\caption{Glycerol concentration versus temperature ($c$-$T$) state diagram of water-glycerol mixtures. $T_{\rm SD}$: LLT spinodal temperature (black filled circles; green line); $T_{\rm gI}$: the glass transition temperature of liquid I (blue filled triangles; blue curve). For pure water ($c = 0$), we use the widely accepted value of 136~K for  $T_{\rm gI}$;  $T_{\rm gII}$: the glass transition temperature of liquid II (red filled triangles). The red curve indicates $T_{\rm gII}$ of pure liquid II without ice cubic ice; $T_{\rm H}$: the homogeneous nucleation temperature (violet open circles) measured for the cooling rate of 100~K min$^{-1}$; $T_{\rm m}$: the melting (liquidus) temperature (black open squares; black curve). We make a linear extrapolation of $T_{\rm SD}$ (green dashed line) to estimate the position of a hypothetical critical point (CP) (magenta star) of the liquid-liquid transition (LLT). This extrapolation is necessary because we cannot access $T_{\rm SD}$ for $c < 0.13$ owing to rapid nucleation of hexagonal ice before reaching the final target temperature in the quenching process. For $c > 0.19$, on the other hand, the kinetics of LLT drastically slows down, which also prevents us from accessing LLT during the observation time. 
This figure is reproduced from Fig. 4a of Ref.~\cite{murata2012liquid}. See Ref.~\cite{murata2012liquid} for the details.
}
\label{fig:LLT}
\end{figure}

\paragraph{Liquid-liquid transition} 
Angell have also tremendously contributed to the understanding of liquid-liquid transition in water~\cite{poole1994effect,angell2000water}, aqueous solutions~\cite{angell2003hyperquenching,zhao2016apparent,woutersen2018liquid}, 
and water-like liquids such as silicon~\cite{sastry2003liquid}, silica~\cite{lascaris2014search}, model systems~\cite{xu2009monatomic}, and phase-change materials~\cite{lucas2020liquid}. 

Inspired by these pioneering works, we have also experimentally studied liquid-liquid transitions in aqueous solutions, including a glycerol-water mixture~\cite{murata2012liquid} and aqueous organic solutions~\cite{murata2013general}. 
Figure ~\ref{fig:LLT} shows the state diagram of water-glycerol mixtures with the spinodal line of liquid-liquid transition obtained experimentally~\cite{murata2012liquid}. 
These studies suggest the presence of liquid-liquid transition in pure water.  
However, the liquid-liquid transitions in aqueous solutions always occur in a state metastable against ice nucleation. This has inevitably created controversies concerning whether the observed phenomena are truly liquid-liquid transitions or not. We have expressed our opinion about these controversies in Ref.~\cite{tanaka2020liquid}, and thus do not discuss this issue here.  
These experimental results on liquid-liquid transition, including pattern evolution dynamics during the transformation, support the validity of the two-order-parameter description of liquid-liquid transition~\cite{tanaka1999two,tanaka2000general,tanaka2012bond,takae2020role,tanaka2020liquid}.
About liquid-liquid transitions in various liquids and a theoretical description, please refer to our recent review article~\cite{tanaka2020liquid}. 

\paragraph{Perspective}
One of the most important properties of water is its ability to adapt to the environment by changing its state freely depending on temperature, pressure, ionic concentration, and others. The key to this ability is that it contains a degree of freedom, the fraction of two states, that does not exist in other simple liquids. Since water is an essential liquid for us, the results of this research will help us understand the unique properties of water itself and have a significant ripple effect on a wide range of fields, including physics, chemistry, and life sciences, and earth sciences.
Furthermore, silica, silicon, and germanium, which are essential in the earth science and semiconductor industries, also have tetrahedral structures~\cite{poole1997comparison}, and the knowledge obtained from water's study is expected to help understand these materials as well~\cite{tanaka2002simple,shi2018impact}.

\section{Summary} \label{sec:summary}
In this article, we have mainly reviewed our studies about the impact of local structural ordering on liquid-glass transition, crystallization,  and water's anomalies, which have been significantly affected by Angell's pioneering works. 
Austen Angell made a number of experimental and theoretical findings and kept creating main research streams in the fields by proposing new ideas and views. 
Most scientists in the liquid science field, including myself, have been greatly affected and inspired by these intriguing ideas.   
We believe that these research streams will continue to contribute to the further future development of liquid science.

\section*{Acknowledgments}
I would like to express my sincere gratitude to the late Austin Angell for his continuous kind support and encouragement since we first met him in 1997. My research activity in the liquid science field has continued to be motivated by new concepts and ideas brought by Austen, such as fragility, fragile-to-strong transition, the relation between phase diagram and glass-forming ability, anomalous physical behaviors of water, and liquid-liquid transition. If there were no such inputs, many of our works described here might not have been made. He was also very kind personally to me and also to my family. Meeting and talking to Austen at international liquid-related conferences and communications via emails were always a treasure for me. I sincerely dedicate this paper to Austen Angell in heaven and wish him all the best.
The content of this article is based on the results obtained in fruitful collaboration with many colleagues, including Shunto Arai, Takeaki Araki, Akira Furukawa, Yuan-Chao Hu, Takeshi Kawasaki, Mika Kobayashi, Rei Kurita, Mathieu Leocmach, Alex Malins, Ken-ichiro Murata, Flavio Romano, Paddy Royall, John Russo, Trond S. Ingebrigtsen, Rui Shi, Ryotaro Shimizu, Shiladitya Sengupta, Hiroshi Shintani, Peng Tan, Hua Tong, Stefan Williams, Taiki Yanagishima, and Keiji Watanabe (alphabetical order). The author warmly thanks all of them for very fruitful collaborations and discussions. 
Finally, I acknowledge receipt of Grants-in-Aid for Specially Promoted Research (JSPS KAKENHI Grant Nos. JP20H05619 and JP25000002) and Scientific Research (A) (JSPS KAKENHI Grant No. JP18H03675) from the Japan Society for the Promotion of Science (JSPS).


\begin{thebibliography}{100}
\expandafter\ifx\csname url\endcsname\relax
  \def\url#1{\texttt{#1}}\fi
\expandafter\ifx\csname urlprefix\endcsname\relax\def\urlprefix{URL }\fi
\expandafter\ifx\csname href\endcsname\relax
  \def\href#1#2{#2} \def\path#1{#1}\fi

\bibitem{hansen1990theory}
J.-P. Hansen, I.~R. McDonald, Theory of simple liquids, Elsevier, 1990.

\bibitem{DebenedettiB}
P.~G. Debenedetti, {Metastable Liquids Concepts and Principles}, Princeton
  University Press, Princeton, 1996.

\bibitem{Angel863_1988}
C.~A. Angell, Perspective on the glass transition, J. Phys. Chem. Solids 49
  (1988) 863.

\bibitem{AngellR}
C.~A. Angell, Formation of glasses from liquids and biopolymers, Science 267
  (1995) 1924--1935.

\bibitem{ediger1996supercooled}
M.~D. Ediger, C.~A. Angell, S.~R. Nagel, Supercooled liquids and glasses, J.
  Phys. Chem. 100~(31) (1996) 13200--13212.

\bibitem{Angel6463_2000}
C.~A. Angell, Ten questions on glassformers, and a real space 'excitations'
  model with some answers on fragility and phase transitions, J. Phys.:
  Condens. Matter 12 (2000) 6463.

\bibitem{Debenedetti2001}
P.~G. Debenedetti, F.~H. Stillinger, {Supercooled liquids and the glass
  transition.}, Nature 410~(6825) (2001) 259--267.

\bibitem{berthier2011theoretical}
L.~Berthier, G.~Biroli, Theoretical perspective on the glass transition and
  amorphous materials, Rev. Mod. Phys. 83~(2) (2011) 587--645.

\bibitem{tanaka2012bond}
H.~Tanaka, Bond orientational order in liquids: Towards a unified description
  of water-like anomalies, liquid-liquid transition, glass transition, and
  crystallization, Eur. Phys. J. E 35~(10) (2012) 113.

\bibitem{royall2015role}
C.~P. Royall, S.~R. Williams, The role of local structure in dynamical arrest,
  Phys. Rep. 560 (2015) 1--75.

\bibitem{kelton2010nucleation}
K.~Kelton, A.~L. Greer, Nucleation in condensed matter: applications in
  materials and biology, Elsevier, 2010.

\bibitem{sosso2016crystal}
G.~C. Sosso, J.~Chen, S.~J. Cox, M.~Fitzner, P.~Pedevilla, A.~Zen,
  A.~Michaelides, Crystal nucleation in liquids: Open questions and future
  challenges in molecular dynamics simulations, Chem. Rev. 116~(12) (2016)
  7078--7116.

\bibitem{eisenberg2005structure}
D.~S. Eisenberg, W.~Kauzmann, {The Structure and Properties of Water}, Oxford
  University Press, Oxford, 2005.

\bibitem{Angell1983}
C.~A. Angell, {Supercooled water}, Annu. Rev. Phys. Chem. 34 (1983) 593--630.
\newblock \href {http://dx.doi.org/10.1146/annurev.pc.34.100183.003113}
  {\path{doi:10.1146/annurev.pc.34.100183.003113}}.

\bibitem{AngellCR}
C.~A. Angell, Liquid fragility and the glass transition in water and aqueous
  solutions, Chem. Rev. 102 (2002) 2627--2650.

\bibitem{Debenedetti2003}
P.~G. Debenedetti, {Supercooled and glassy water}, J. Phys.: Condens. Matter 15
  (2003) 1669--1726.
\newblock \href {http://dx.doi.org/10.1088/0953-8984/15/45/R01}
  {\path{doi:10.1088/0953-8984/15/45/R01}}.

\bibitem{Angell_glass}
C.~A. Angell, Amorphous water, Annu. Rev. Phys. Chem. 55 (2004) 559--583.

\bibitem{angell2008insights}
C.~A. Angell, Insights into phases of liquid water from study of its unusual
  glass-forming properties, Science 319~(5863) (2008) 582--587.

\bibitem{gallo2016water}
P.~Gallo, K.~Amann-Winkel, C.~A. Angell, M.~A. Anisimov, F.~Caupin,
  C.~Chakravarty, E.~Lascaris, T.~Loerting, A.~Z. Panagiotopoulos, J.~Russo,
  J.~A. Sellberg, H.~E. Stanley, H.~Tanaka, C.~Vega, L.~Xu, L.~G.~M.
  Pettersson, Water: A tale of two liquids, Chem. Rev. 116 (2016) 7463--7500.

\bibitem{poole1997polymorphic}
P.~H. Poole, T.~Grande, C.~A. Angell, P.~F. McMillan, Polymorphic phase
  transitions in liquids and glasses, Science 275 (1997) 322.

\bibitem{mcmillan2004polyamorphic}
P.~F. McMillan, Polyamorphic transformations in liquids and glasses, J. Mater.
  Chem. 14~(10) (2004) 1506--1512.

\bibitem{mcmillan2007polyamorphism}
P.~F. McMillan, M.~Wilson, M.~C. Wilding, D.~Daisenberger, M.~Mezouar, G.~N.
  Greaves, Polyamorphism and liquid--liquid phase transitions: challenges for
  experiment and theory, J. Phys.: Condens. Matter 19~(41) (2007) 415101.

\bibitem{tanaka2020liquid}
H.~Tanaka, Liquid--liquid transition and polyamorphism, J. Chem. Phys. 153~(13)
  (2020) 130901.

\bibitem{frank1952supercooling}
F.~C. Frank, Supercooling of liquids, Proc. R. Soc. Lond. A 215 (1952) 43--46.

\bibitem{steinhardt1983bond}
P.~J. Steinhardt, D.~R. Nelson, M.~Ronchetti, Bond-orientational order in
  liquids and glasses, Phys. Rev. B 28~(2) (1983) 784--805.

\bibitem{tarjus2005frustration}
G.~Tarjus, S.~A. Kivelson, Z.~Nussinov, P.~Viot, The frustration-based approach
  of supercooled liquids and the glass transition: a review and critical
  assessment, J. Phys.: Condens. Matter 17~(50) (2005) R1143.

\bibitem{ferrer1998supercooled}
M.~L. Ferrer, C.~Lawrence, B.~G. Demirjian, D.~Kivelson, C.~Alba-Simionesco,
  G.~Tarjus, {Supercooled liquids and the glass transition: Temperature as the
  control variable}, J. Chem. Phys. 109~(18) (1998) 8010--8015.

\bibitem{kirkpatrick1987dynamics}
T.~R. Kirkpatrick, D.~Thirumalai, Dynamics of the structural glass transition
  and the $p$-spin-interaction spin-glass model, Phys. Rev. Lett. 58~(20)
  (1987) 2091.

\bibitem{kirkpatrick1989scaling}
T.~R. Kirkpatrick, D.~Thirumalai, P.~G. Wolynes, Scaling concepts for the
  dynamics of viscous liquids near an ideal glassy state, Phys. Rev. A 40~(2)
  (1989) 1045.

\bibitem{parisi2010mean}
G.~Parisi, F.~Zamponi, Mean-field theory of hard sphere glasses and jamming,
  Rev. Mod. Phys. 82~(1) (2010) 789.

\bibitem{kirkpatrick2015colloquium}
T.~R. Kirkpatrick, D.~Thirumalai, Colloquium: Random first order transition
  theory concepts in biology and physics, Rev. Mod. Phys. 87~(1) (2015) 183.

\bibitem{gotze2008complex}
W.~G{\"o}tze, Complex dynamics of glass-forming liquids: A mode-coupling
  theory, Vol. 143, Oxford Univ. Press, Oxford, 2008.

\bibitem{kirkpatrick1989random}
T.~R. Kirkpatrick, D.~Thirumalai, Random solutions from a regular density
  functional hamiltonian: a static and dynamical theory for the structural
  glass transition, J. Phys. A 22~(5) (1989) L149.

\bibitem{tanaka2002simple}
H.~Tanaka, Simple view of waterlike anomalies of atomic liquids with
  directional bonding, Phys. Rev. B 66~(6) (2002) 064202.

\bibitem{Angel1058_1970}
C.~A. Angell, E.~J. Sare, Glass-forming composition regions and glass
  transition temperatures for aqueous electrolyte solutions, J. Chem. Phys. 52
  (1970) 1058--1068.

\bibitem{kobayashi2011possible}
M.~Kobayashi, H.~Tanaka, {Possible link of the V-shaped phase diagram to the
  glass-forming ability and fragility in a water-salt mixture}, Phys. Rev.
  Lett. 106~(12) (2011) 125703.

\bibitem{kobayashi2011relationship}
M.~Kobayashi, H.~Tanaka, {Relationship between the phase diagram, the
  glass-forming ability, and the fragility of a water/salt mixture}, J. Phys.
  Chem. B 115~(48) (2011) 14077--14090.

\bibitem{tanaka2013importance}
H.~Tanaka, Importance of many-body orientational correlations in the physical
  description of liquids, Faraday Discuss. 167 (2013) 9--76.

\bibitem{shintani2006frustration}
H.~Shintani, H.~Tanaka, Frustration on the way to crystallization in glass,
  Nature Phys. 2~(3) (2006) 200.

\bibitem{kawasaki2007correlation}
T.~Kawasaki, T.~Araki, H.~Tanaka, Correlation between dynamic heterogeneity and
  medium-range order in two-dimensional glass-forming liquids, Phys. Rev. Lett.
  99~(21) (2007) 215701.

\bibitem{tanaka1999two}
H.~Tanaka, Two-order-parameter description of liquids: critical phenomena and
  phase separation of supercooled liquids, J. Phys.: Condens. Matter 11~(15)
  (1999) L159.

\bibitem{tanaka1999two1}
H.~Tanaka, {Two-order-parameter description of liquids. I. A general model of
  glass transition covering its strong to fragile limit}, J. Chem. Phys.
  111~(7) (1999) 3163--3174.

\bibitem{tanaka1999two2}
H.~Tanaka, {Two-order-parameter description of liquids. II. Criteria for
  vitrification and predictions of our model}, J. Chem. Phys. 111~(7) (1999)
  3175--3182.

\bibitem{tanaka2005two1}
H.~Tanaka, {Two-order-parameter model of the liquid--glass transition. I.
  Relation between glass transition and crystallization}, J. Non-Cryst. Solids
  351~(43-45) (2005) 3371--3384.

\bibitem{tanaka2005two2}
H.~Tanaka, {Two-order-parameter model of the liquid--glass transition. II.
  Structural relaxation and dynamic heterogeneity}, J. Non-Cryst. Solids
  351~(43-45) (2005) 3385--3395.

\bibitem{tanaka2005two3}
H.~Tanaka, {Two-order-parameter model of the liquid--glass transition. III.
  Universal patterns of relaxations in glass-forming liquids}, J. Non-Cryst.
  Solids 351~(43-45) (2005) 3396--3413.

\bibitem{tanaka2005relationship}
H.~Tanaka, Relationship among glass-forming ability, fragility, and short-range
  bond ordering of liquids, J. Non-Cryst. Solids 351~(8-9) (2005) 678--690.

\bibitem{martinez2001thermodynamic}
L.~M. Martinez, C.~A. Angell, A thermodynamic connection to the fragility of
  glass-forming liquids, Nature 410~(6829) (2001) 663--667.

\bibitem{tanaka2003roles}
H.~Tanaka, Roles of local icosahedral chemical ordering in glass and
  quasicrystal formation in metallic glass formers, J. Phys.: Condens. Matter
  15~(31) (2003) L491.

\bibitem{shintani2008universal}
H.~Shintani, H.~Tanaka, Universal link between the boson peak and transverse
  phonons in glass, Nature Mater. 7~(11) (2008) 870.

\bibitem{molinero2006tuning}
V.~Molinero, S.~Sastry, C.~A. Angell, Tuning of tetrahedrality in a silicon
  potential yields a series of monatomic (metal-like) glass formers of very
  high fragility, Phys. Rev. Lett. 97 (2006) 075701.

\bibitem{bhat2007vitrification}
M.~H. Bhat, V.~Molinero, E.~Soignard, V.~C. Solomon, S.~Sastry, J.~L. Yarger,
  C.~A. Angell, Vitrification of a monatomic metallic liquid, Nature 448 (2007)
  787--790.

\bibitem{Angell_Sare}
C.~A. Angell, E.~J. Sare, Liquid-liquid immiscibility in common aqueous salt
  solutions at low temperatures, J. Chem. Phys. 49 (1968) 4713--4714.

\bibitem{russo2018glass}
J.~Russo, F.~Romano, H.~Tanaka, Glass forming ability in systems with competing
  orderings, Phys. Rev. X 8~(2) (2018) 021040.

\bibitem{tanaka2010critical}
H.~Tanaka, T.~Kawasaki, H.~Shintani, K.~Watanabe, Critical-like behaviour of
  glass-forming liquids, Nature Mater. 9~(4) (2010) 324--331.

\bibitem{watanabe2008direct}
K.~Watanabe, H.~Tanaka, Direct observation of medium-range crystalline order in
  granular liquids near the glass transition, Phys. Rev. Lett. 100~(15) (2008)
  158002.

\bibitem{kawasaki2010structural}
T.~Kawasaki, H.~Tanaka, Structural origin of dynamic heterogeneity in
  three-dimensional colloidal glass formers and its link to crystal nucleation,
  J. Phys.: Condens. Matter 22~(23) (2010) 232102.

\bibitem{kawasaki2014structural}
T.~Kawasaki, H.~Tanaka, Structural evolution in the aging process of
  supercooled colloidal liquids, Phys. Rev. E 89~(6) (2014) 062315.

\bibitem{tanaka2019revealing}
H.~Tanaka, H.~Tong, R.~Shi, J.~Russo, Revealing key structural features hidden
  in liquids and glasses, Nat. Rev. Phys. (2019) 1.

\bibitem{tanaka2020role}
H.~Tanaka, Role of many-body correlation in slow dynamics of glass-forming
  liquids: Intrinsic or perturbative, J. Stat. Mech. Theory Exp. 2020~(3)
  (2020) 034003.

\bibitem{leocmach2012roles}
M.~Leocmach, H.~Tanaka, Roles of icosahedral and crystal-like order in the hard
  spheres glass transition, Nat. Commun. 3 (2012) 974.

\bibitem{widmer2004reproducible}
A.~Widmer-Cooper, P.~Harrowell, H.~Fynewever, How reproducible are dynamic
  heterogeneities in a supercooled liquid?, Phys. Rev. Lett. 93~(13) (2004)
  135701.

\bibitem{mosayebi2010probing}
M.~Mosayebi, E.~Del~Gado, P.~Ilg, H.~C. {\"O}ttinger, Probing a critical length
  scale at the glass transition, Phys. Rev. Lett. 104~(20) (2010) 205704.

\bibitem{zheng2021translational}
Z.~Zheng, R.~Ni, Y.~Wang, Y.~Han, Translational and rotational critical-like
  behaviors in the glass transition of colloidal ellipsoid monolayers, Sci.
  Adv. 7~(3) (2021) eabd1958.

\bibitem{hohenberg1977theory}
P.~C. Hohenberg, B.~I. Halperin, Theory of dynamic critical phenomena, Rev.
  Mod. Phys 49~(3) (1977) 435.

\bibitem{OnukiB}
A.~Onuki, {Phase Transition Dynamics}, Cambridge Univ. Press, Cambridge, 2002.

\bibitem{tong2018revealing}
H.~Tong, H.~Tanaka, Revealing hidden structural order controlling both fast and
  slow glassy dynamics in supercooled liquids, Phys. Rev. X 8~(1) (2018)
  011041.

\bibitem{tong2019structural}
H.~Tong, H.~Tanaka, Structural order as a genuine control parameter of dynamics
  in simple glass formers, Nature Commun. 10 (2019) 5596.

\bibitem{tong2020role}
H.~Tong, H.~Tanaka, Role of attractive interactions in structure ordering and
  dynamics of glass-forming liquids, Phys. Rev. Lett. 124~(22) (2020) 225501.

\bibitem{dyre2006colloquium}
J.~C. Dyre, Colloquium: The glass transition and elastic models of
  glass-forming liquids, Rev. Mod. Phys. 78~(3) (2006) 953.

\bibitem{starr2002we}
F.~W. Starr, S.~Sastry, J.~F. Douglas, S.~C. Glotzer, What do we learn from the
  local geometry of glass-forming liquids?, Phys. Rev. Lett. 89~(12) (2002)
  125501.

\bibitem{widmer2006predicting}
A.~Widmer-Cooper, P.~Harrowell, Predicting the long-time dynamic heterogeneity
  in a supercooled liquid on the basis of short-time heterogeneities, Phys.
  Rev. Lett. 96~(18) (2006) 185701.

\bibitem{larini2008universal}
L.~Larini, A.~Ottochian, C.~De~Michele, D.~Leporini, Universal scaling between
  structural relaxation and vibrational dynamics in glass-forming liquids and
  polymers, Nat. Phys. 4~(1) (2008) 42--45.

\bibitem{betancourt2015quantitative}
B.~A.~P. Betancourt, P.~Z. Hanakata, F.~W. Starr, J.~F. Douglas, Quantitative
  relations between cooperative motion, emergent elasticity, and free volume in
  model glass-forming polymer materials, Proc. Natl. Acad. Sci. 112~(10) (2015)
  2966--2971.

\bibitem{berthier2010critical}
L.~Berthier, G.~Tarjus, Critical test of the mode-coupling theory of the glass
  transition, Phys. Rev. E 82~(3) (2010) 031502.

\bibitem{landes2020attractive}
F.~P. Landes, G.~Biroli, O.~Dauchot, A.~J. Liu, D.~R. Reichman, Attractive
  versus truncated repulsive supercooled liquids: The dynamics is encoded in
  the pair correlation function, Phys. Rev. E 101~(1) (2020) 010602.

\bibitem{nandi2021microscopic}
M.~K. Nandi, S.~M. Bhattacharyya, Microscopic theory of softness in supercooled
  liquids, Phys. Rev. Lett. 126~(20) (2021) 208001.

\bibitem{charbonneau2013decorrelation}
P.~Charbonneau, G.~Tarjus, Decorrelation of the static and dynamic length
  scales in hard-sphere glass formers, Phys. Rev. E 87~(4) (2013) 042305.

\bibitem{berthier2007structure}
L.~Berthier, R.~L. Jack, Structure and dynamics of glass formers:
  Predictability at large length scales, Phys. Rev. E 76~(4) (2007) 041509.

\bibitem{scopigno2003fragility}
T.~Scopigno, G.~Ruocco, F.~Sette, G.~Monaco, Is the fragility of a liquid
  embedded in the properties of its glass?, Science 302~(5646) (2003) 849--852.

\bibitem{simmons2012generalized}
D.~S. Simmons, M.~T. Cicerone, Q.~Zhong, M.~Tyagi, J.~F. Douglas, Generalized
  localization model of relaxation in glass-forming liquids, Soft Matter 8~(45)
  (2012) 11455--11461.

\bibitem{zhang2021fast}
H.~Zhang, X.~Wang, H.-B. Yu, J.~F. Douglas, Fast dynamics in a model metallic
  glass-forming material, J. Chem. Phys. 154~(8) (2021) 084505.

\bibitem{watanabe2011structural}
K.~Watanabe, T.~Kawasaki, H.~Tanaka, Structural origin of enhanced slow
  dynamics near a wall in glass-forming systems, Nat. Mater. 10~(7) (2011) 512.

\bibitem{langer2013ising}
J.~S. Langer, Ising model of a glass transition, Phys. Rev. E 88~(1) (2013)
  012122.

\bibitem{langer2014theories}
J.~S. Langer, Theories of glass formation and the glass transition, Rep. Prog.
  Phys. 77~(4) (2014) 042501.

\bibitem{chandra1990ising}
P.~Chandra, P.~Coleman, A.~Larkin, Ising transition in frustrated heisenberg
  models, Phys. Rev. Lett. 64~(1) (1990) 88.

\bibitem{weber2003ising}
C.~Weber, L.~Capriotti, G.~Misguich, F.~Becca, M.~Elhajal, F.~Mila, Ising
  transition driven by frustration in a 2d classical model with continuous
  symmetry, Phys. Rev. Lett. 91~(17) (2003) 177202.

\bibitem{russo2015assessing}
J.~Russo, H.~Tanaka, Assessing the role of static length scales behind glassy
  dynamics in polydisperse hard disks, Proc. Natl. Acad. Sci. USA 112 (2015)
  6920--6924.

\bibitem{kob2012non}
W.~Kob, S.~Rold{\'a}n-Vargas, L.~Berthier, Non-monotonic temperature evolution
  of dynamic correlations in glass-forming liquids, Nature Phys. 8~(2) (2012)
  164--167.

\bibitem{biroli2008thermodynamic}
G.~Biroli, J.~P. Bouchaud, A.~Cavagna, T.~S. Grigera, P.~Verrocchio,
  Thermodynamic signature of growing amorphous order in glass-forming liquids,
  Nat. Phys. 4~(10) (2008) 771--775.

\bibitem{charbonneau2012geometrical}
B.~Charbonneau, P.~Charbonneau, G.~Tarjus, Geometrical frustration and static
  correlations in a simple glass former, Phys. Rev. Lett. 108~(3) (2012)
  035701.

\bibitem{yaida2016point}
S.~Yaida, L.~Berthier, P.~Charbonneau, G.~Tarjus, Point-to-set lengths, local
  structure, and glassiness, Phys. Rev. E 94~(3) (2016) 032605.

\bibitem{parisi2020theory}
G.~Parisi, P.~Urbani, F.~Zamponi, Theory of simple glasses: exact solutions in
  infinite dimensions, Cambridge University Press, 2020.

\bibitem{alder1957phase}
B.~J. Alder, T.~E. Wainwright, Phase transition for a hard sphere system, J.
  Chem. Phys. 27~(5) (1957) 1208--1209.

\bibitem{nelson2002defects}
D.~R. Nelson, {Defects and Geometry in Condensed Matter Physics}, Cambridge
  Univ. Press, 2002.

\bibitem{Tanaka2000}
H.~Tanaka, {Simple physical model of liquid water}, J. Chem. Phys. 112 (2000)
  799--809.

\bibitem{xu2005relation}
L.~Xu, P.~Kumar, S.~V. Buldyrev, S.~H. Chen, P.~H. Poole, F.~Sciortino, H.~E.
  Stanley, {Relation between the Widom line and the dynamic crossover in
  systems with a liquid--liquid phase transition}, Proc. Natl. Acad. Sci. USA
  102~(46) (2005) 16558--16562.

\bibitem{shi2018origin}
R.~Shi, J.~Russo, H.~Tanaka, Origin of the emergent fragile-to-strong
  transition in supercooled water, Proc. Natl. Acad. Sci. U. S. A. 115~(38)
  (2018) 9444--9449.

\bibitem{shi2018common}
R.~Shi, J.~Russo, H.~Tanaka, Common microscopic structural origin for water’s
  thermodynamic and dynamic anomalies, J. Chem. Phys. 149~(22) (2018) 224502.

\bibitem{ito1999thermodynamic}
K.~Ito, C.~T. Moynihan, C.~A. Angell, Thermodynamic determination of fragility
  in liquids and a fragile-to-strong liquid transition in water, Nature
  398~(6727) (1999) 492.

\bibitem{saika2001fragile}
I.~Saika-Voivod, P.~H. Poole, F.~Sciortino, Fragile-to-strong transition and
  polyamorphism in the energy landscape of liquid silica, Nature 412~(6846)
  (2001) 514.

\bibitem{zhang2010fragile}
C.~Zhang, L.~Hu, Y.~Yue, J.~C. Mauro, Fragile-to-strong transition in metallic
  glass-forming liquids, J. Chem. Phys. 133~(1) (2010) 014508.

\bibitem{zhou2015structural}
C.~Zhou, L.~Hu, Q.~Sun, H.~Zheng, C.~Zhang, Y.~Yue, {Structural evolution
  during fragile-to-strong transition in CuZr (Al) glass-forming liquids}, J.
  Chem. Phys. 142~(6) (2015) 064508.

\bibitem{wei2015phase}
S.~Wei, P.~Lucas, C.~A. Angell, {Phase change alloy viscosities down to $T_{\rm
  g}$ using Adam-Gibbs-equation fittings to excess entropy data: A
  fragile-to-strong transition}, J. Appl. Phys. 118~(3) (2015) 034903.

\bibitem{Tanaka2003}
H.~Tanaka, {A new scenario of the apparent fragile-to-strong transition in
  tetrahedral liquids: water as an example}, J. Phys.: Condens. Matter 15
  (2003) L703--L711.
\newblock \href {http://dx.doi.org/10.1088/0953-8984/15/45/L03}
  {\path{doi:10.1088/0953-8984/15/45/L03}}.

\bibitem{shi2018impact}
R.~Shi, H.~Tanaka, Impact of local symmetry breaking on the physical properties
  of tetrahedral liquids, Proc. Natl. Acad. Sci. U. S. A. 115~(9) (2018)
  1980--1985.
\newblock \href {http://dx.doi.org/10.1073/pnas.1717233115}
  {\path{doi:10.1073/pnas.1717233115}}.

\bibitem{shi2019distinct}
R.~Shi, H.~Tanaka, Distinct signature of local tetrahedral ordering in the
  scattering function of covalent liquids and glasses, Sci. Adv. 5~(3) (2019)
  eaav3194.

\bibitem{wei2017structural}
S.~Wei, M.~Stolpe, O.~Gross, W.~Hembree, S.~Hechler, J.~Bednarcik, R.~Busch,
  P.~Lucas, {Structural evolution on medium-range-order during the
  fragile-strong transition in Ge$_{15}$Te$_{85}$}, Acta Mater. 129 (2017)
  259--267.

\bibitem{wei2017glass}
S.~Wei, G.~J. Coleman, P.~Lucas, C.~A. Angell, {Glass transitions,
  semiconductor-metal transitions, and fragilities in Ge- V- Te (V= As, Sb)
  liquid alloys: The difference one element can make}, Phys. Rev. Applied 7~(3)
  (2017) 034035.

\bibitem{lucas2020liquid}
P.~Lucas, S.~Wei, C.~A. Angell, Liquid-liquid phase transitions in
  glass-forming systems and their implications for memory technology, Int. J.
  Appl. Glass Sci. 11~(2) (2020) 236--244.

\bibitem{kawasaki2010formation}
T.~Kawasaki, H.~Tanaka, Formation of a crystal nucleus from liquid, Proc. Natl
  Acad. Sci. USA 107~(32) (2010) 14036--14041.

\bibitem{dupuy1982controlled}
J.~Dupuy, J.~F. Jal, C.~Ferradou, P.~Chieux, A.~F. Wright, R.~Calemczuk, C.~A.
  Angell, Controlled nucleation and quasi-ordered growth of ice crystals from
  low temperature electrolyte solutions, Nature 296~(5853) (1982) 138--140.

\bibitem{kadiyala1984separation}
R.~K. Kadiyala, C.~A. Angell, Separation of nucleation from crystallization
  kinetics by two step calorimetry experiments, Colloids and Surfaces 11~(3-4)
  (1984) 341--351.

\bibitem{angell1986crystallization}
C.~A. Angell, Y.~Choi, Crystallization and vitrification in aqueous systems, J.
  Microscopy 141~(3) (1986) 251--261.

\bibitem{angell1988structural}
C.~A. Angell, Structural instability and relaxation in liquid and glassy phases
  near the fragile liquid limit, J. Non-Crystal. Solids 102~(1-3) (1988)
  205--221.

\bibitem{russo2012microscopic}
J.~Russo, H.~Tanaka, The microscopic pathway to crystallization in supercooled
  liquids, Sci. Rep. 2 (2012) 505.
\newblock \href {http://dx.doi.org/https://doi.org/10.1038/srep00505}
  {\path{doi:https://doi.org/10.1038/srep00505}}.

\bibitem{russo2016crystal}
J.~Russo, H.~Tanaka, Crystal nucleation as the ordering of multiple order
  parameters, J. Chem. Phys. 145~(21) (2016) 211801.

\bibitem{tan2014visualizing}
P.~Tan, N.~Xu, L.~Xu, Visualizing kinetic pathways of homogeneous nucleation in
  colloidal crystallization, Nature Phys. 10~(1) (2014) 73.

\bibitem{li2020revealing}
M.~Li, Y.~Chen, H.~Tanaka, P.~Tan, Revealing roles of competing local
  structural orderings in crystallization of polymorphic systems, Sci. Adv.
  6~(27) (2020) eaaw8938.

\bibitem{russo2012selection}
J.~Russo, H.~Tanaka, {Selection mechanism of polymorphs in the crystal
  nucleation of the Gaussian core model}, Soft Matter 8~(15) (2012) 4206--4215.

\bibitem{russo2014new}
J.~Russo, F.~Romano, H.~Tanaka, New metastable form of ice and its role in the
  homogeneous crystallization of water, Nat. Mater. 13~(7) (2014) 733--739.

\bibitem{tanaka2001interplay}
H.~Tanaka, Interplay between wetting and phase separation in binary fluid
  mixtures: roles of hydrodynamics, J. Phys.: Condens. Matter 13 (2001) 4637.

\bibitem{arai2017surface}
S.~Arai, H.~Tanaka, Surface-assisted single-crystal formation of charged
  colloids, Nat. Phys. 13~(5) (2017) 503--509.

\bibitem{tanaka1998simple}
H.~Tanaka, Simple physical explanation of the unusual thermodynamic behavior of
  liquid water, Phys. Rev. Lett. 80~(26) (1998) 5750.

\bibitem{tanaka2003possible}
H.~Tanaka, {Possible resolution of the Kauzmann paradox in supercooled
  liquids}, Phys. Rev. E 68~(1) (2003) 011505.

\bibitem{ediger2008crystal}
M.~D. Ediger, P.~Harrowell, L.~Yu, Crystal growth kinetics exhibit a
  fragility-dependent decoupling from viscosity, J. Chem. Phys. 128~(3) (2008)
  034709.

\bibitem{hu2020physical}
Y.-C. Hu, H.~Tanaka, Physical origin of glass formation from multicomponent
  systems, Sci. Adv. 6~(50) (2020) eabd2928.

\bibitem{kanno1977homogeneous}
H.~Kanno, C.~A. Angell, Homogeneous nucleation and glass formation in aqueous
  alkali halide solutions at high pressures, J. Phys. Chem. 81~(26) (1977)
  2639--2643.

\bibitem{angell2000glass}
C.~A. Angell, Glass formation and the nature of the glass transitions,
  Insulating and Semiconducting Glasses 17 (2000) 1--51.

\bibitem{egami1997universal}
T.~Egami, Universal criterion for metallic glass formation, Mater. Sci. Eng.: A
  226 (1997) 261--267.

\bibitem{johnson1999bulk}
W.~L. Johnson, Bulk glass-forming metallic alloys: science and technology, MRS
  Bull. 24~(10) (1999) 42--56.

\bibitem{inoue2000stabilization}
A.~Inoue, Stabilization of metallic supercooled liquid and bulk amorphous
  alloys, Acta Mater. 48~(1) (2000) 279--306.

\bibitem{jakse2003local}
N.~Jakse, A.~Pasturel, Local order of liquid and supercooled zirconium by ab
  initio molecular dynamics, Phys. Rev. Lett. 91~(19) (2003) 195501.

\bibitem{kelton_first_2003}
K.~F. Kelton, G.~W. Lee, A.~K. Gangopadhyay, R.~W. Hyers, T.~J. Rathz, J.~R.
  Rogers, M.~B. Robinson, D.~S. Robinson, First {X}-ray scattering studies on
  electrostatically levitated metallic liquids: demonstrated influence of local
  icosahedral order on the nucleation barrier, Phys. Rev. Lett. 90~(19) (2003)
  195504.

\bibitem{xi_correlation_2007}
X.~K. Xi, L.~L. Li, B.~Zhang, W.~H. Wang, Y.~Wu, Correlation of atomic cluster
  symmetry and glass-forming ability of metallic glass, Phys. Rev. Lett. 99~(9)
  (2007) 095501.

\bibitem{hirata_geometric_2013}
A.~Hirata, L.~J. Kang, T.~Fujita, B.~Klumov, K.~Matsue, M.~Kotani, A.~R.
  Yavari, M.~W. Chen, Geometric frustration of icosahedron in metallic glasses,
  Science 341~(6144) (2013) 376--379.

\bibitem{luo2004icosahedral}
W.~K. Luo, H.~W. Sheng, F.~M. Alamgir, J.~M. Bai, J.~H. He, E.~Ma, Icosahedral
  short-range order in amorphous alloys, Phys. Rev. Lett. 92~(14) (2004)
  145502.

\bibitem{sheng2006atomic}
H.~W. Sheng, W.~K. Luo, F.~M. Alamgir, J.~M. Bai, E.~Ma, Atomic packing and
  short-to-medium-range order in metallic glasses, Nature 439~(7075) (2006)
  419--425.

\bibitem{shen_icosahedral_2009}
Y.~T. Shen, T.~H. Kim, A.~K. Gangopadhyay, K.~F. Kelton, Icosahedral order,
  frustration, and the glass transition: evidence from time-dependent
  nucleation and supercooled liquid structure studies, Phys. Rev. Lett. 102~(5)
  (2009) 057801.

\bibitem{cheng2011atomic}
Y.~Q. Cheng, E.~Ma, Atomic-level structure and structure--property relationship
  in metallic glasses, Progress in materials science 56~(4) (2011) 379--473.

\bibitem{ding2014full}
J.~Ding, Y.-Q. Cheng, E.~Ma, Full icosahedra dominate local order in {Cu64Zr34}
  metallic glass and supercooled liquid, Acta Mater. 69 (2014) 343--354.

\bibitem{gonzalez2017competition}
B.~Gonzalez, S.~Bechelli, I.~Essafri, V.~Piquet, C.~Desgranges, J.~Delhommelle,
  Competition between crystalline and icosahedral order during crystal growth
  in bimetallic systems, J. Cryst. Growth 478 (2017) 22--27.

\bibitem{tanaka1998simple_g}
H.~Tanaka, A simple physical model of liquid-glass transition: Intrinsic
  fluctuating interactions and random fields hidden in glass-forming liquids,
  J. Phys.: Condens. Matter 10~(14) (1998) L207.

\bibitem{desgranges2014unraveling}
C.~Desgranges, J.~Delhommelle, Unraveling the coupling between demixing and
  crystallization in mixtures, J. Am. Chem. Soc. 136~(23) (2014) 8145--8148.

\bibitem{puosi2018dynamical}
F.~Puosi, N.~Jakse, A.~Pasturel, Dynamical, structural and chemical
  heterogeneities in a binary metallic glass-forming liquid, J. Phys.: Condens.
  Matter 30~(14) (2018) 145701.

\bibitem{desgranges2019can}
C.~Desgranges, J.~Delhommelle, Can ordered precursors promote the nucleation of
  solid solutions?, Phys.Rev. Lett, 123~(19) (2019) 195701.

\bibitem{ingebrigtsen2019crystallization}
T.~S. Ingebrigtsen, J.~C. Dyre, T.~B. Schr{\o}der, C.~P. Royall,
  Crystallization instability in glass-forming mixtures, Phys. Rev. X 9~(3)
  (2019) 031016.

\bibitem{wuttig2007phase}
M.~Wuttig, N.~Yamada, Phase-change materials for rewriteable data storage, Nat.
  Mater. 6~(11) (2007) 824--832.

\bibitem{lencer2011design}
D.~Lencer, M.~Salinga, M.~Wuttig, Design rules for phase-change materials in
  data storage applications, Adv. Mater. 23~(18) (2011) 2030--2058.

\bibitem{greer_new_2015}
A.~L. Greer, New horizons for glass formation and stability, Nat. Mater. 14~(6)
  (2015) 542--546.

\bibitem{wei2019phase}
S.~Wei, P.~Lucas, C.~A. Angell, Phase-change materials: The view from the
  liquid phase and the metallicity parameter, MRS Bull. 44~(9) (2019) 691--698.

\bibitem{persch2021potential}
C.~Persch, M.~J. M{\"u}ller, A.~Yadav, J.~Pries, N.~Honn{\'e}, P.~Kerres,
  S.~Wei, H.~Tanaka, P.~Fantini, E.~Varesi, et~al., The potential of chemical
  bonding to design crystallization and vitrification kinetics, Nat. Commun.
  12~(1) (2021) 1--8.

\bibitem{sun2018mechanism}
G.~Sun, J.~Xu, P.~Harrowell, The mechanism of the ultrafast crystal growth of
  pure metals from their melts, Nat. Mater. 17~(10) (2018) 881--886.

\bibitem{sun2020displacement}
G.~Sun, A.~Hawken, P.~Harrowell, The displacement field associated with the
  freezing of a melt and its role in determining crystal growth kinetics, Proc.
  Natl. Acad. Sci. 117~(7) (2020) 3421--3426.

\bibitem{gao2021fast}
Q.~Gao, J.~Ai, S.~Tang, M.~Li, Y.~Chen, J.~Huang, H.~Tong, L.~Xu, L.~Xu,
  H.~Tanaka, et~al., Fast crystal growth at ultra-low temperatures, Nat. Mater.
  (2021) 1--9\href
  {http://dx.doi.org/https://doi.org/10.1038/s41563-021-00993-6}
  {\path{doi:https://doi.org/10.1038/s41563-021-00993-6}}.

\bibitem{yanagishima2017common}
T.~Yanagishima, J.~Russo, H.~Tanaka, Common mechanism of thermodynamic and
  mechanical origin for ageing and crystallization of glasses, Nat. Commun.
  8~(1) (2017) 1--10.

\bibitem{tong2020emergent}
H.~Tong, S.~Sengupta, H.~Tanaka, Emergent solidity of amorphous materials as a
  consequence of mechanical self-organisation, Nat. Commun. 11~(1) (2020)
  1--10.

\bibitem{yanagishima2021towards}
T.~Yanagishima, J.~Russo, R.~P.~A. Dullens, H.~Tanaka,
  \href{https://link.aps.org/doi/10.1103/PhysRevLett.127.215501}{Towards
  glasses with permanent stability}, Phys. Rev. Lett. 127 (2021) 215501.
\newblock \href {http://dx.doi.org/10.1103/PhysRevLett.127.215501}
  {\path{doi:10.1103/PhysRevLett.127.215501}}.
\newline\urlprefix\url{https://link.aps.org/doi/10.1103/PhysRevLett.127.215501}

\bibitem{spellings2018machine}
M.~Spellings, S.~C. Glotzer, Machine learning for crystal identification and
  discovery, AIChE J. 64~(6) (2018) 2198--2206.

\bibitem{boattini2019unsupervised}
E.~Boattini, M.~Dijkstra, L.~Filion, Unsupervised learning for local structure
  detection in colloidal systems, J. Chem. Phys. 151~(15) (2019) 154901.

\bibitem{leoni2021nonclassical}
F.~Leoni, J.~Russo, Nonclassical nucleation pathways in stacking-disordered
  crystals, Phys. Rev. X 11 (2021) 031006.
\newblock \href {http://dx.doi.org/10.1103/PhysRevX.11.031006}
  {\path{doi:10.1103/PhysRevX.11.031006}}.

\bibitem{becker2021unsupervised}
S.~Becker, E.~Devijver, R.~Molinier, N.~Jakse, Unsupervised topological
  learning for atomic structures identification, arXiv preprint
  arXiv:2109.08126.

\bibitem{davis1965two}
C.~M. Davis~Jr, T.~A. Litovitz, Two-state theory of the structure of water, J.
  Chem. Phys. 42~(7) (1965) 2563--2576.

\bibitem{angell1971two}
C.~A. Angell, Two-state thermodynamics and transport properties for water from
  "bond lattice" model, J. Phys. Chem. 75 (1971) 3698--3705.

\bibitem{ponyatovsky1998metastable}
E.~G. Ponyatovsky, V.~V. Sinitsyn, T.~A. Pozdnyakova, The metastable {T- P}
  phase diagram and anomalous thermodynamic properties of supercooled water, J.
  Chem. Phys. 109~(6) (1998) 2413--2422.

\bibitem{urquidi1999origin}
J.~Urquidi, S.~Singh, C.~H. Cho, G.~W. Robinson, Origin of temperature and
  pressure effects on the radial distribution function of water, Phys. Rev.
  Lett. 83~(12) (1999) 2348.

\bibitem{Speedy1976}
R.~J. Speedy, C.~A. Angell, {Isothermal compressibility of supercooled water
  and evidence for a thermodynamic singularity at -45$^\circ$C}, J. Chem. Phys.
  65 (1976) 851.
\newblock \href {http://dx.doi.org/10.1063/1.433153}
  {\path{doi:10.1063/1.433153}}.

\bibitem{azouzi2013coherent}
M.~E.~M. Azouzi, C.~Ramboz, J.-F. Lenain, F.~Caupin, A coherent picture of
  water at extreme negative pressure, Nat. Phys. 9~(1) (2013) 38--41.

\bibitem{caupin2015escaping}
F.~Caupin, Escaping the no man's land: Recent experiments on metastable liquid
  water, J. Non-Cryst. Solids 407 (2015) 441--448.

\bibitem{Poole1992}
P.~H. Poole, F.~Sciortino, U.~Essmann, H.~E. Stanley, {Phase behavior of
  metastable water}, Nature 360~(6402) (1992) 324--328.

\bibitem{mishima1985apparently}
O.~Mishima, L.~D. Calvert, E.~Whalley, An apparently first-order transition
  between two amorphous phases of ice induced by pressure, Nature 314~(6006)
  (1985) 76--78.

\bibitem{Mishima1998}
O.~Mishima, H.~E. Stanley, {The relationship between liquid, supercooled and
  glassy water}, Nature 396~(6709) (1998) 329--335.
\newblock \href {http://dx.doi.org/10.1038/24540} {\path{doi:10.1038/24540}}.

\bibitem{AngellwaterR}
C.~A. Angell, Insights into phases of liquid water from study of its unusual
  glass-forming properties, Science 319 (2008) 582--587.

\bibitem{poole1994effect}
P.~H. Poole, F.~Sciortino, T.~Grande, H.~E. Stanley, C.~A. Angell, Effect of
  hydrogen bonds on the thermodynamic behavior of liquid water, Phys. Rev.
  Lett. 73~(12) (1994) 1632.

\bibitem{tanaka1982acoustic}
H.~Tanaka, Y.~Wada, H.~Nakajima, Acoustic anomaly in a critical binary mixture
  of aniline and cyclohexane at low and ultrasonic frequencies, Chem. Phys.
  68~(1-2) (1982) 223--231.

\bibitem{tanaka1985theoretical}
H.~Tanaka, Y.~Wada, Theoretical consideration on the acoustic anomaly of
  critical binary mixtures, Phys. Rev. A 32~(1) (1985) 512.

\bibitem{Tanaka2000a}
H.~Tanaka, {Thermodynamic anomaly and polyamorphism of water}, Europhys. Lett.
  50 (2000) 340--346.
\newblock \href {http://dx.doi.org/10.1209/epl/i2000-00276-4}
  {\path{doi:10.1209/epl/i2000-00276-4}}.

\bibitem{nilsson2015structural}
A.~Nilsson, L.~G.~M. Pettersson, The structural origin of anomalous properties
  of liquid water, Nat. Commun. 6~(1) (2015) 1--11.

\bibitem{johari2015thermodynamic}
G.~P. Johari, J.~Teixeira, {Thermodynamic analysis of the two-liquid model for
  anomalies of water, HDL--LDL fluctuations, and liquid--liquid transition}, J.
  Phys. Chem. B 119~(44) (2015) 14210--14220.

\bibitem{walrafen1986raman}
G.~E. Walrafen, M.~S. Hokmabadi, W.~H. Yang, Raman isosbestic points from
  liquid water, J. Chem. Phys. 85~(12) (1986) 6964--6969.

\bibitem{Errington2001}
J.~R. Errington, P.~G. Debenedetti, {Relationship between structural order and
  the anomalies of liquid water}, Nature 409~(January) (2001) 318--321.

\bibitem{tanaka2000general}
H.~Tanaka, General view of a liquid-liquid phase transition, Phys. Rev. E
  62~(5) (2000) 6968.

\bibitem{takae2020role}
K.~Takae, H.~Tanaka, Role of hydrodynamics in liquid--liquid transition of a
  single-component substance, Proc. Natl. Acad. Sci. USA 117~(9) (2020)
  4471--4479.

\bibitem{Holten2012}
V.~Holten, M.~A. Anisimov, {Entropy-driven liquid-liquid separation in
  supercooled water.}, Sci. Rep. 2 (2012) 713.
\newblock \href {http://dx.doi.org/10.1038/srep00713}
  {\path{doi:10.1038/srep00713}}.

\bibitem{Singh2016two}
R.~S. Singh, J.~W. Biddle, P.~G. Debenedetti, M.~A. Anisimov, {Two-state
  thermodynamics and the possibility of a liquid-liquid phase transition in
  supercooled TIP4P/2005 water}, J. Chem. Phys. 144~(14) (2016) 144504.

\bibitem{Biddle2017}
J.~W. Biddle, R.~S. Singh, E.~M. Sparano, F.~Ricci, M.~A. Gonz{\'{a}}lez,
  C.~Valeriani, J.~F. Abascal, P.~G. Debenedetti, M.~A. Anisimov, F.~Caupin,
  {Two-structure thermodynamics for the TIP4P/2005 model of water covering
  supercooled and deeply stretched regions}, J. Chem. Phys. 146~(3) (2017)
  034502.
\newblock \href {http://dx.doi.org/10.1063/1.4973546}
  {\path{doi:10.1063/1.4973546}}.

\bibitem{caupin2019thermodynamics}
F.~Caupin, M.~A. Anisimov, Thermodynamics of supercooled and stretched water:
  Unifying two-structure description and liquid-vapor spinodal, J. Chem. Phys.
  151~(3) (2019) 034503.

\bibitem{angell2016potential}
C.~A. Angell, V.~Kapko, {Potential tuning in the S--W system.(i) Bringing T c,
  2 to ambient pressure, and (ii) colliding T c, 2 with the liquid--vapor
  spinodal}, J. Stat. Mech. Theory Exp. 2016~(9) (2016) 094004.

\bibitem{russo2018water}
J.~Russo, K.~Akahane, H.~Tanaka, Water-like anomalies as a function of
  tetrahedrality, Proc. Natl. Acad. Sci. USA 115~(15) (2018) E3333--E3341.

\bibitem{shi2020direct}
R.~Shi, H.~Tanaka, Direct evidence in the scattering function for the
  coexistence of two types of local structures in liquid water, J. Amer. Chem.
  Soc. 142~(6) (2020) 2868--2875.

\bibitem{shi2020anomalies}
R.~Shi, H.~Tanaka, The anomalies and criticality of liquid water, Proc. Natl.
  Acad. Sci. 117~(43) (2020) 26591--26599.

\bibitem{Russo2014}
J.~Russo, H.~Tanaka, {Understanding water's anomalies with locally favoured
  structures.}, Nat. Commun. 5 (2014) 3556.
\newblock \href {http://dx.doi.org/10.1038/ncomms4556}
  {\path{doi:10.1038/ncomms4556}}.

\bibitem{soper2000structures}
A.~K. Soper, M.~A. Ricci, Structures of high-density and low-density water,
  Phys. Rev. Lett. 84~(13) (2000) 2881--2884.

\bibitem{shiratani1998molecular}
E.~Shiratani, M.~Sasai, Molecular scale precursor of the liquid--liquid phase
  transition of water, J. Chem. Phys. 108~(8) (1998) 3264--3276.

\bibitem{cuthbertson2011mixturelike}
M.~J. Cuthbertson, P.~H. Poole, Mixturelike behavior near a liquid-liquid phase
  transition in simulations of supercooled water, Phys. Rev. Lett. 106~(11)
  (2011) 115706.

\bibitem{shi2018Microscopic}
R.~Shi, H.~Tanaka, {Microscopic structural descriptor of liquid water}, J.
  Chem. Phys. 148 (2018) 124503.
\newblock \href {http://dx.doi.org/doi: 10.1063/1.5024565} {\path{doi:doi:
  10.1063/1.5024565}}.

\bibitem{angell1976density}
C.~A. Angell, H.~Kanno, Density maxima in high-pressure supercooled water and
  liquid silicon dioxide, Science 193~(4258) (1976) 1121--1122.

\bibitem{huang2004amorphous}
L.~Huang, J.~Kieffer, Amorphous-amorphous transitions in silica glass. i.
  reversible transitions and thermomechanical anomalies, Phys. Rev. B 69~(22)
  (2004) 224203.

\bibitem{loerting2006amorphous}
T.~Loerting, N.~Giovambattista, Amorphous ices: experiments and numerical
  simulations, J. Phys.: Condens. Matter 18~(50) (2006) R919.

\bibitem{bachler2021experimental}
J.~Bachler, J.~Giebelmann, T.~Loerting, Experimental evidence for glass
  polymorphism in vitrified water droplets, Proc. Natl. Acad. Sci. 118~(30).

\bibitem{palmer2014metastable}
J.~C. Palmer, F.~Martelli, Y.~Liu, R.~Car, A.~Z. Panagiotopoulos, P.~G.
  Debenedetti, Metastable liquid-liquid transition in a molecular model of
  water, Nature 510~(7505) (2014) 385--388.

\bibitem{debenedetti2020second}
P.~G. Debenedetti, F.~Sciortino, G.~H. Zerze, Second critical point in two
  realistic models of water, Science 369~(6501) (2020) 289--292.

\bibitem{gartner2020signatures}
T.~E. Gartner, L.~Zhang, P.~M. Piaggi, R.~Car, A.~Z. Panagiotopoulos, P.~G.
  Debenedetti, Signatures of a liquid--liquid transition in an ab initio deep
  neural network model for water, Proc. Natl. Acad. Sci. 117~(42) (2020)
  26040--26046.

\bibitem{palmer2018advances}
J.~C. Palmer, P.~H. Poole, F.~Sciortino, P.~G. Debenedetti, Advances in
  computational studies of the liquid--liquid transition in water and
  water-like models, Chem. Rev. 118~(18) (2018) 9129--9151.

\bibitem{handle2017supercooled}
P.~H. Handle, T.~Loerting, F.~Sciortino, Supercooled and glassy water:
  Metastable liquid (s), amorphous solid (s), and a no-man’s land, Proc. Natl.
  Acad. Sci. 114~(51) (2017) 13336--13344.

\bibitem{kim2017maxima}
K.~H. Kim, A.~Sp{\"a}h, H.~Pathak, F.~Perakis, D.~Mariedahl, K.~Amann-Winkel,
  J.~A. Sellberg, J.~H. Lee, S.~Kim, J.~Park, et~al., Maxima in the
  thermodynamic response and correlation functions of deeply supercooled water,
  Science 358~(6370) (2017) 1589--1593.

\bibitem{kim2020experimental}
K.~H. Kim, K.~Amann-Winkel, N.~Giovambattista, A.~Sp{\"a}h, F.~Perakis,
  H.~Pathak, M.~L. Parada, C.~Yang, D.~Mariedahl, T.~Eklund, et~al.,
  Experimental observation of the liquid-liquid transition in bulk supercooled
  water under pressure, Science 370~(6519) (2020) 978--982.

\bibitem{pathak2021enhancement}
H.~Pathak, A.~Sp{\"a}h, N.~Esmaeildoost, J.~A. Sellberg, K.~H. Kim, F.~Perakis,
  K.~Amann-Winkel, M.~Ladd-Parada, J.~Koliyadu, T.~J. Lane, et~al., Enhancement
  and maximum in the isobaric specific-heat capacity measurements of deeply
  supercooled water using ultrafast calorimetry, Proc. Natl. Acad. Sci.
  118~(6).

\bibitem{caupin2018comment}
F.~Caupin, V.~Holten, C.~Qiu, E.~Guillerm, M.~Wilke, M.~Frenz, J.~Teixeira,
  A.~K. Soper, {Comment on ``Maxima in the thermodynamic response and
  correlation functions of deeply supercooled water''}, Science 360~(6390)
  (2018) eaat1634.

\bibitem{kim2018response}
K.~H. Kim, A.~Sp{\"a}h, H.~Pathak, F.~Perakis, D.~Mariedahl, K.~Amann-Winkel,
  J.~A. Sellberg, J.~H. Lee, S.~Kim, J.~Park, et~al., {Response to Comment on
  ``Maxima in the thermodynamic response and correlation functions of deeply
  supercooled water''}, Science 360~(6390) (2018) eaat1729.

\bibitem{murata2012liquid}
K.~Murata, H.~Tanaka, Liquid-liquid transition without macroscopic phase
  separation in a water-glycerol mixture, Nat. Mater. 11 (2012) 436--443.

\bibitem{angell2000water}
C.~A. Angell, R.~D. Bressel, M.~Hemmati, E.~J. Sare, J.~C. Tucker, Water and
  its anomalies in perspective: Tetrahedral liquids with and without
  liquid--liquid phase transitions. invited lecture, Phys. Chem. Chem. Phys.
  2~(8) (2000) 1559--1566.

\bibitem{angell2003hyperquenching}
C.~A. Angell, L.-M. Wang, Hyperquenching and cold equilibration strategies for
  the study of liquid--liquid and protein folding transitions, Biophys. Chem.
  105~(2-3) (2003) 621--637.

\bibitem{zhao2016apparent}
Z.~Zhao, C.~A. Angell, Apparent first-order liquid--liquid transition with
  pre-transition density anomaly, in water-rich ideal solutions, Angew. Chem. -
  Int. Ed. 55~(7) (2016) 2474--2477.

\bibitem{woutersen2018liquid}
S.~Woutersen, B.~Ensing, M.~Hilbers, Z.~Zhao, C.~A. Angell, {A liquid-liquid
  transition in supercooled aqueous solution related to the HDA-LDA
  transition}, Science 359~(6380) (2018) 1127--1131.

\bibitem{sastry2003liquid}
S.~Sastry, C.~A. Angell, Liquid--liquid phase transition in supercooled
  silicon, Nat. Mater. 2~(11) (2003) 739--743.

\bibitem{lascaris2014search}
E.~Lascaris, M.~Hemmati, S.~V. Buldyrev, H.~E. Stanley, C.~A. Angell, Search
  for a liquid-liquid critical point in models of silica, J. Chem. Phys.
  140~(22) (2014) 224502.

\bibitem{xu2009monatomic}
L.~Xu, S.~V. Buldyrev, N.~Giovambattista, C.~A. Angell, H.~E. Stanley, A
  monatomic system with a liquid-liquid critical point and two distinct glassy
  states, J. Chem. Phys. 130~(5) (2009) 054505.

\bibitem{murata2013general}
K.~Murata, H.~Tanaka, General nature of liquid-liquid transition in aqueous
  organic solutions, Nat. Commun. 4 (2013) 2844.

\bibitem{poole1997comparison}
P.~H. Poole, M.~Hemmati, C.~A. Angell, Comparison of thermodynamic properties
  of simulated liquid silica and water, Phys. Rev. Lett. 79~(12) (1997) 2281.

\end{thebibliography}
\end{document}